\documentclass[11pt]{iopart}

\usepackage{amssymb} 
\usepackage{braket}
\usepackage{bm}
\usepackage[english]{babel} 	
\usepackage{graphicx} 
\usepackage{xstring} 
\usepackage[usenames,dvipsnames,svgnames,table]{xcolor}
\usepackage[colorlinks=true,
            linkcolor=blue,citecolor=blue,menucolor=.]{hyperref}
\usepackage{booktabs}
\usepackage{cleveref}

\graphicspath{{./figures/}}

\newcommand{\Scheq}{Schr{\"o}dinger equation }

\begin{document}

\title{Combined few-body and mean-field model for nuclei}

\author{D Hove$^1$, 
E Garrido$^2$\footnote{Author to whom any correspondence should be addressed}, P.~Sarriguren$^2$, D.V. Fedorov$^1$, H.O.U. Fynbo$^1$, A.S.~Jensen$^1$, N.T.~Zinner$^{1,3}$}
\address{$^1$ Department of Physics and Astronomy, Aarhus University, DK-8000 Aarhus C, Denmark}
\address{$^2$ Instituto de Estructura de la Materia, IEM-CSIC, Serrano 123, E-28006 Madrid, Spain}
\address{$^3$ Aarhus Institute of Advanced Studies, Aarhus University, DK-8000 Aarhus C, Denmark}
\ead{e.garrido@csic.es}

\begin{abstract}
The challenging nuclear many-body problem is discussed along with
classifications and qualitative descriptions of existing methods and
models.  We present detailed derivations of a new method where cluster
correlations co-exist with an underlying mean-field described
core-structure.  The variation of an antisymmetrized product of
cluster and core wave functions and a given nuclear interaction,
provide sets of self-consistent equations of motion.

First we test the technique on the neutron dripline nucleus $^{26}$O,
considered as $^{24}$O surrounded by two neutrons.  
We choose Skyrme effective interactions between all pairs of nucleons.  
To ensure correct asymptotic behavior we modify the valence
neutron-neutron interaction to fit the experimental scattering
length in vacuum. This is an example of necessary considerations both
of effective interactions between in-medium and free pairs, and
renormalizations due to restrictions in allowed Hilbert space.

Second, we investigate the heavier neutron dripline nucleus $^{72}$Ca,
described as $^{70}$Ca plus two neutrons.  We continuously vary the
strength of the Skyrme interaction to fine-tune the approach to the
dripline.  Halo structure in the $s$-wave is observed followed by the
tendency to form Efimov states.  Occurrence of Efimov states are
prevented by the exceedingly unfavorable system of two light and one
heavy particle.  Specifically the neutron-neutron scattering length is
comparable to the spatial extension of a possible Efimov state, and
scaling would place the next of the states outside our galaxy.

Our third application is on the proton dripline nucleus $^{70}$Kr,
described as $^{68}$Se plus two protons, which is a prominent
waiting point for the astrophysical $rp$-process.  We
calculate radiative capture rates and discuss the capture
mechanism as being either direct, sequential, virtual sequential or an
energy dependent mixture of them.  We do not find
any $1^-$ resonance and therefore no significant $E1$ transition.
This is consistent with the long waiting time, since both $E2$ and
background transitions are very slow.

After the applications on dripline nuclei we discuss
perspectives with improvements and applications.
In the conclusion we summarize while emphasizing the merits of
consistently treating both short- and large-distance properties, few-
and many-body correlations, ordinary nuclear structure, and concepts
of halos and Efimov states.
\end{abstract}

\submitto{\jpg}
\maketitle

\section{Introduction\label{sec:intro}}

The present  report describes methods formulated
for applications on the low-energy nuclear many-body problem.  Many
techniques have over the years been developed for that purpose.  Some
methods were designed before computers were available, while some are
recent creations using powerful computers.  The
physical insight obtained over the years from simple analytical models
are now used in extensive numerical calculations.  We shall here first
recall a number of the challenges of the multifaceted many-body nuclear 
physics. Then get down to the somewhat more dedicated purpose of
the present work, and finally describe the report structure.

\subsection{Status of the nuclear challenge}

The constituents of nuclei are first of all neutrons and protons,
which throughout this report shall be treated as either structureless
point-like particles or equivalently with frozen inert intrinsic
structure.  Nuclear physics is concerned with the many-body nucleon
problem.  The properties are basically governed by the short-range
strong interaction \cite{sie87,boh69}, which in turn originates and is
characterized through the non-perturbative theory of low-energy
quantum chromodynamics (QCD) \cite{iof06}.  One (often minor)
complication is that the protons are charged implying modification
from the associated long-range Coulomb interaction \cite{ber94}.
Another implication is that the energy must be relatively low to avoid
intrinsic nucleonic excitations occurring at $140$~MeV when the pion
can be created.  Thus, we shall consider here low-energy
properties \cite{bro90} only.

The meaning of ``many nucleons'' is a number between $2$ and $300$ for
the more or less ordinary nuclei of interest in the present report, up
to about $1000$ if non-stable exotic toroidal systems are also
considered, and $10^{55}$ if also nuclear structure in neutron stars
are included.  Clearly then few-body physics must be part of our
complete descriptions.  In addition, our many-body treatment has to
apply for the relatively low number of a few hundred, somewhat in
contrast to the genuine many-body physics problems in solid state and
condensed matter physics.

The basic theoretical tool is quantum theory which is unavoidable due
to the microscopic characteristics. On top the fermionic nature of
neutrons and protons requires quantum statistics to obey the Pauli
exclusion principle for each of these types of nucleons \cite{sai73}.
Nuclear physics with these two non-identical, but very similarly
interacting, constituents is unique and gives rise to the concept of
isospin.  The corresponding symmetry is slightly broken in nuclei but
used abundantly in particle physics.

Many nuclear properties can surprisingly be explained by macroscopic
physics concepts, like the liquid drop model or similar elaborate
extended modifications \cite{mye69}.  The level of accuracy from these
models is roughly between $5\%$ and $10\%$, whereas the remaining few
percents are overwhelmingly decisive for both the present level of
understanding and the derived properties.  The mixture of micro- and
macroscopic properties extends even further, since nuclear densities
fall off from maximum to zero over surface widths of about $2$~fm
corresponding roughly to the range of the strong interaction.  Such
leptodermous (thin skin) systems \cite{mye69} are characteristic for
mesoscopic systems where the surface and finite size have significant
influence on the acquired properties.

The models to account for all these features can be searched for by
measuring the nucleon mean-free path at low energy within nuclei.  The
result of roughly $5$~fm is comparable with the radii of medium heavy
nuclei, and therefore almost explicitly announcing that neither
mean-field nor strongly correlated structures can alone be responsible for
nuclear properties \cite{boh69}.  Thus, efficient nuclear models must
be able to describe mean-field single particle motion,
macroscopic collective modes, and various correlated structures.
These requirements are further emphasized by noting that nuclear
reaction and decay times vary from the age of the Universe and down to
$10^{-22}$~seconds \cite{lan03}.

The status at present in the few-body limit is that full and accurate
ab-initio calculations are possible and carried out for nucleon numbers less
than $10-15$.  Even here co-existing correlations are exceedingly
difficult to describe, like cluster structure within disparate
background structure exemplified by the Hoyle state 
in $^{12}$C \cite{fed03,che07b,epe11,gar15}.  The
efficient first principle methods known from quantum chemistry
\cite{vci66,bar78} can not be directly adopted, as they are built
around the idea of a perturbative description of the relevant
effective interactions.  This is a huge difference as the strong
nuclear interaction originates in the low-energy QCD, where, however,
the inherent non-perturbativity only is the tip of the iceberg with
regards to complications.

Treating heavier nuclei with more than $20$ particles requires more
assumptions and approximations. The simplest description of average
properties is obtained by use of preferentially self-consistent
mean-field calculations.  More elaborate and perhaps more ambitious
theories are abundantly formulated, but implementations always rely on
various types of approximations, where almost all can be referred to a
restriction on the available Hilbert space.  A general description of
such approximations is that some of the particles are constrained in
frozen cluster structures while others are allowed to move freely.

It is clear that nuclear physics poses daunting challenges, and great
ingenuity is needed to make any progress within the subject. However,
as a reward for attempting to overcome these challenges, nuclear
physics also provides a unique insight into the fields of many-body
quantum physics, as well as the nature of both the strong and the weak
interaction.  Furthermore, due to the general nature of the problem,
the lessons can also be useful in other subfields of physics.  In any
case we shall in this report describe our own scheme of approximations
dedicated to our immediate goals, that is especially to describe
cluster correlations emerging from a background of uncorrelated
nucleons.

\subsection{Requirements and purpose}

Given the many facets of nuclear physics, it is not surprising that no
single theory can encompass all the complexities of the field. As such
the various methods for approaching and describing nuclear systems
more often complements, than replaces, each other.  However, one can
propose a number of characteristics that an ideal model should possess
based on the fundamental challenges within the field.  We limit
ourselves to low-energy properties where the building blocks are
assumed to be nucleons. The model should be able to describe
single-particle properties and cluster correlations, as well as being
scalable and computational efficient to allow applications on both
light and heavier nuclei.

In addition, as any observation will be at large distances compared to
the size of the system, it is also vitally important to associate the
observed and necessarily large-distance behavior reliably and
consistently with the short-distance nature of the system in question.
In other words, it is necessary to have a wave function consistently
describing from spatially compact to extended configurations.  In
particular, the long-range, observable, asymptotic behavior
(scattering length, energy levels, and so on) must be correct.
Ideally such a wave function should be found with an interaction
derived from basic nucleon-nucleon properties, but not necessarily the
same although still also obeying the decisive phenomenologically
observed properties.

In this connection, the importance of spatially extended and weakly
bound configurations are in particular apparent in relation to the
concept of universality in connection with halo formation and decay,
and the extreme of Efimov physics. Both phenomena appear in nuclei and
nuclear astrophysics, as well as in cold atomic and molecular gases.
To describe properties of such weakly bound nuclear systems is
currently one of the prevailing limitations for many models.  Not only
in relation to halo \cite{fre12} and Efimov physics \cite{bra06}, but
also in relation to dripline nuclei.  In addition, computational
requirements are also a significant constraint for many of the more
popular methods.  Of course, these problems are closely related, as
the size of the needed basis drastically increases with the nucleon
number.  The more complicated the description of the small-distance
structure, the more computationally demanding the extension to larger
distances becomes.  A desirable method would provide a fairly simple
consistent and detailed description of both small and large distances

The purpose of this report is to present the details of a new method,
which incorporates several of these requirements, and which is
particularly well suited for describing weakly bound, extended
systems. A new, efficient, and flexible method for combining few-
and many-body treatments of relative and intrinsic degrees-of-freedom
of the constituent particles in a complex system is derived in detail
and applied to several topical nuclear systems. A structure consisting
of a few potentially many-body clusters is imposed, and a wave
function for the entire system based on the individual clusters and
the overarching few-body structure is established.  By performing a
variation of the energy with respect to both cluster and three-body
wave functions, a series of coupled Schr{\"o}dinger equations can be
derived. They consist of a many-body equation for each cluster, which
can be solved self-consistently, and a few-body equation where the
traditionally phenomenological two-body potentials instead are derived
from the many-body equations combined with the related effective nucleon-nucleon
interaction.

This implementation has the advantage of being conceptually very
simple. In principle, all that is needed is to settle on a many-body
and a few-body formalism, and perform a variation.  This technical
implementation of few- and many-body formalisms including the
necessary interactions are related but not separate.  The benefit of
this, from a practical point of view, is that the flexibility and
versatility of the few-body approach is maintained, along with the
detailed insight provided by the many-body model, while the
computational complexity is not significantly greater than the sum of
the parts. In addition, the ambiguity often associated with
phenomenological few-body methods is eliminated, or at least pushed to
the next level of the hierarchy, as all interactions in principle are
produced within the same framework.

\subsection{Report structure}

The previous subsections outlined the general challenges inherent to
the field of nuclear physics, and it also gave an outline of some
desirable characteristics in a model applicable to the low-energy
nuclear many-body problem.  In addition, our general purpose for
introducing a new method was stated, along with a short overview of
how this method is formulated.  This should enable the reader to place
the present method among the many variants within the myriads of other
attempts to formulate the techniques to solve (aspects of) the
nuclear many-body problem.

Sec.~\ref{sec:many} contains classifications and qualitative
outlines of several already existing methods and models.  The focus is
on selected few methods, which are either distinguished by their
historical significance, general popularity, or accuracy and
usability.  An in-depth review of each will not be presented, instead
the key concepts and assumptions will be presented and discussed along
with the primary strengths and weaknesses.  Even such a review can never be
fully exhaustive, but the intention is that it should provide the
interested reader with a general overview, along with key references
for origin and further explorations.

Afterwards, in Sec.~\ref{sec:Theory}, the detailed derivation of our
method is presented. This is divided in two parts, where the first
part in Sec.~\ref{sec:GenTheo} contains a general derivation, where as
few choices and assumptions as possible are made.  This is continued
in Sec.~\ref{sec:TheoSky} by specific derivation with more
assumptions, where selected mean-field and three-body formalisms are
chosen along with a specific type of nucleon-nucleon interaction.

The technicalities of the implementation are presented in
Sec.~\ref{sec:implement}. This includes both the method of solving 
the main equations, along with the differences of this work from the
more traditional many-body and few-body methods.  Practical
considerations and the crucial implementation of the Pauli principle
are described. In addition, some of the subtleties related to
incorporating the new effective interactions in the three-body
formalism are discussed.

The implementation on realistic examples is crucial and used to test
applicability, accuracy, and efficiency.  In Secs.~\ref{sec:24O} to
\ref{sec:68Se} the focus is on the neutron dripline nuclei $^{26}$O
and $^{72}$Ca, as well as the proton dripline nucleus $^{70}$Kr. The
$^{26}$O-nucleus is an ideal initial test case on the neutron dripline
due to its double magic structure \cite{hof09} with a spherical
$^{24}$O-core surrounded by two valence neutrons
\cite{cae13,koh13,kon16}. This is also at the limit of what is
currently experimentally feasible, and it has received much attention
recently \cite{koh13,cae13,hov17}.

However, the method is not limited to light nuclei, and in the present
formulation is directly applicable to any system where a mean-field
calculation can produce a self-consistent solution.  The application
illustrated by $^{72}$Ca in Sec.~\ref{sec:70Ca} is of special
interest.  The $s$-state near the unbound threshold provides
optimal conditions for pronounced halo structures and possibly Efimov
states.  This potential occurrence of Efimov states, or very large
halo states, has caused the heavy calcium isotopes to receive
increased attention recently \cite{hag13}.  For such heavy systems,
traditional clusterized few-body techniques seem problematic, as they
tend to neglect the internal structure of the clusters. This is not in
any way a problem here, where the internal core structure is fully
incorporated.

Practical astrophysical applications are demonstrated in
Sec.~\ref{sec:68Se} by application to the $^{70}\textnormal{Kr} \,
(^{68}\textnormal{Se}+p+p)$ nucleus, where the primary focus is on the
two-proton radiative capture on $^{68}\textnormal{Se}$. This has a
significant influence on the effective lifetime of
$^{68}\textnormal{Se}$ in stellar environments, and is therefore
central to the understanding of the rapid proton (rp) capture process.

The specific applications described in details are supplemented with a
discussion of possible future improvements and generalizations in
Sec.~\ref{sec:expansion}. Particular attention is devoted to achieve
full consistency between core and cluster interactions as well as
improvements of both the mean-field core-description and the adiabatic
few-body expansion related to the cluster-core potential.  The many
detailed derivations and formula are relegated to appendices.
Finally, Sec.~\ref{sec:con} briefly summarizes the report and presents
our conclusions from the work.

\section{The nuclear many-body problem for pedestrians\label{sec:many}}

The theories are necessarily microscopic and quantum mechanical, that
is, based on the Schr\"odinger equation.  The properties of the systems 
are contained
in the resulting wave function and derived through expectation values.
Nuclear properties require an underlying interaction which on some
level has to be phenomenologically determined. This corresponds
immediately to an uncertainty in the initial input.  Addressing the
many-body problem implies that approximations to the ideal procedure
must be adopted.  These two points of input and method are intertwined
in the sense that a given approximate procedure dictates, or at least
constrains, the parameterization or form of the interaction.

We shall here mostly be concerned with the design of a suitable method,
while the interaction is chosen to match the method in each of the
formulations.  We shall first discuss how interactions and
degrees-of-freedom are related.  Then we shall qualitatively sketch
how the nuclear many-body problem has been attempted to be solved in
previous formulations.  Assuming the interaction is known, at least two
approaches have been tried. One is to formulate a brute force method,
which along with progress has been approximated to allow practical
implementation.  The principle example is the shell model in various
formulations and approximations.

Another approach is to focus on the physics quantities of interest and
select and treat the corresponding much fewer degrees-of-freedom.
Improvements are then achieved iteratively by adding, perhaps
approximately, more degrees-of-freedom. The two approaches (from few
towards many and vice versa) should ideally coincide in some region of
parameter space.  Combinations of (approximate) full treatment and
selected degrees-of-freedom are both possible and desirable in nuclear
physics. An example could be cluster models where relative and
intrinsic degrees-of-freedom are treated differently.

We shall sketch some methods in a little more detail. This provides
perspective on advantages and disadvantages as well as prizes for the
efforts of different methods.  We shall point out a number of places
for improvements of the methods and the related interactions.
However, first we address the crucial issue of how interaction and
Hilbert space are tied together.

\subsection{Interaction versus Hilbert space}

The focus on selected degrees-of-freedom must be the first
consideration in investigations of a physics problem.  This is almost
equivalent, or at least immediately leads, to choice of Hilbert space
in the subsequent theoretical formulation.  The next question in
low-energy hamiltonian based quantum mechanics is to find the
interactions between the constituents described by the chosen
degrees-of-freedom.

\subsubsection{Basic ideas.}

The obvious starting point is to isolate two constituents from the
environment in the total system of interest, and find the interaction
by experimental or theoretical phenomenological considerations.
Direct use of such interactions for many particles in the restricted
Hilbert space can unfortunately be a very misguided suggestion.  The
actual interactions are influenced by space restrictions as seen by
considering motion in an accelerated coordinate system or on a curved
surface in ordinary three dimensional space.  The general acceptance
of effects arising from motion in a restricted Hilbert space leads to
the concept of effective interactions \cite{fes58,har70}.  The
formulation due to Feshbach is to eliminate undesired parts of the
Hilbert space and in the remaining space obtain a more complicated
interaction perhaps even with an imaginary (non-unitary) part.

It can be mathematically demonstrated that appropriately renormalized
effective interactions in the restricted space formally provide the same
correct solution of energy and wave function as unrestricted
interactions in the complete Hilbert space \cite{fes58,har70}.  It is
then a matter of using the effective interaction corresponding
precisely to the chosen restricted Hilbert space.  The problem is that
the proper transformation between complete and restricted solutions
only is fully defined in terms of the complete (unknown) solution.
However, approximations,  guesses or iterations in the procedures
have over the years been successfully employed to get closer to the
correct answers.  The same approximately known transformation
operators must be applied to calculate observable quantities.  The
procedures are easily very elaborate and often too complicated to be
fully implemented, and truncations become unavoidable.

These rather abstract considerations can be illustrated by well known
examples like a nuclear few-body system with valence particles(s)
outside a spherical nuclear core \cite{boh69}.  The Coulomb
interaction between the core and the valence particle should be either
absent for neutrons or determined by the known charge for the protons.
However, the core is in reality polarized by the short-range nuclear
interaction.  The effect in the restricted (core-nucleon) space can be
parameterized through a finite non-negligible effective charge of the
valence neutron(s) and a modified charge of the protons \cite{boh69}.
Another well-known example is the Skyrme-type of effective
interactions used in mean-field Slater determinental restricted
solutions, as we shall see throughout this article.

\subsubsection{Effective field theory.}

The phenomenological parameterization of the effects is in many
cases replaced by formal and systematic derivations with much more
controllable errors.  The prominent example is effective field theory
\cite{bur07,epe09,mac11}, where the overall key properties again are
that the physics of interest is related to degrees-of-freedom,
approximately decoupled from all other sets.  This has always been
used throughout physics as evidenced by examples ordered in
hierarchies: molecules, atoms, nuclei, nucleons, quarks and gluons.
Each level can be approximately described by use of the related 
degrees-of-freedom and a corresponding theoretical description, that
is an effective theory.  The quantities revealing this possibility are
energies or length scales ordered in classes according to size.  The
small parameters are assumed to be zero and the larger parameters are
infinitely large.  The adjective ``field'' refers to the formulation
as a field theory.

The present level of applications is interacting nucleons in bound or
low-energy continuum nuclear states \cite{ham17}.  The length and
energy scales are about $1$~fm and $140$~MeV, corresponding
respectively to the nucleon size and the pion mass, $m_{\pi}$, in
energy units.  Thus the theory aims at describing corresponding
properties larger than the nucleon radius and of lower energies, $E$, where
the pion cannot be produced.  The theory then order interactions in
powers of energy divided by pion mass, $(E/(m_{\pi}c^2))^n$, where $c$
is the velocity of light.  The leading order, $LO$, corresponds to
$n=0$, next to leading order, $NLO$, has $n=1$, next to next to
leading order, $N^2LO$ has $n=2$, etc. as $N^nLO$.

The leading order results already give indications, but is far from
sufficient for present days applications where accuracy demands
require at least $n=2$.  One more order, $n=3$, seems to provide
converged results although indications are that $n=4$ also gives
significant contribution in accuracy calculations.  Each order
introduces a number of new terms where all strengths, except one, are
fully determined by preceding orders.  The strength of this new term
has a free parameter left for phenomenological determination.  Each
order also introduces a higher order multi-body interaction which for
higher than (or equal to) three-body character are neglected under the
expectation that it is insignificant.

The advantages are that the correct symmetries from QCD are
maintained, and the method is systematic and in principle controllable
to any order required. Clearly, practical and precise formulations
must be preceded by selection of the level in the hierarchy and the
corresponding active degrees-of-freedom.  It is in principle possible
to include more than four orders in a more complete theory, but this
may be unnecessary and certainly more difficult.  Once the form and
strengths of the interactions are derived and determined, the
applications still need to decide on a method to solve the many-body
problem of interest.  We shall sketch possibilities in the next
subsections.  Practical complications involving spin and angular
momenta are present as always in nuclear physics.  The applications
are limited to few particles and therefore to light nuclei.

\subsection{The shell model concept}

The analogs to the atomic electrons in their orbits around the nucleus
are that the nucleons also can be envisaged to move in single-particle
orbits \cite{may50,may50b,bar13}. Since an analogous attractive center
is non-existent, a potential created by the other nucleons must be
responsible for these orbits.  With the related single-particle states
and given two-, three-, and $N$-body interactions
\cite{nog06,cau05,bog02}, a method to solve the nuclear many-body
problem is easily formulated.

First choose a complete basis for the single-particle states, which
not necessarily are derived from the above hypothetical
single-particle orbits.  Then construct a complete basis for the
$N$-body nuclear system, and compute and diagonalize the corresponding
hamiltonian matrix. This is easily expressed but in practice usually
impossible because the necessary matrix size is insurmountably huge. A
variety of approximations have been employed over the years.

\subsubsection{Non-interacting shell model.}

The simplest and first attempts are to avoid calculations while
guessing the average potential responsible for the single-particle
motion \cite{seb68}. The temptation here is obviously to use
analytically solvable models, where the square well and harmonic
oscillator are traditional choices.  The square well is of finite
range extending to its radius and therefore zero at large distances as
known for the short-range nuclear forces. It is most appropriate for medium and
heavier nuclei.  The harmonic oscillator never decreases to zero at
large distances, but for bound states it is remarkably successful for
lighter nuclei due to rather accurate reproduction of the potential at
the decisive intermediate distances.

Modifications of the oscillator to allow broader applications on heavy
and deformed nuclei were also extremely successful \cite{boh69,nil55}.
As soon as computers were available, the more realistic Woods-Saxon
potential resembling the square well with smooth edges was, and still
is, abundantly used.  These non-interacting shell models are clearly
only as good as the guesses.  Nevertheless, essentially all stable
nuclear structures were discovered with such phenomenological models.
However, surprises or unexpected tendencies can not be correctly
predicted as desired for unknown and unstable nuclei.  The Pauli
principle is in all these cases accounted for by allowing only single
occupancy of levels by identical neutrons and protons.

\subsubsection{Self-consistent mean-field models.}

One major step beyond guessing forms of the potential is to assume an
interaction, usually of two-body character, and a product of neutron
and proton Slater determinants.  Minimizing the energy by varying the
single-particle wave functions produce the non-linear Hartree-Fock
equations with solutions of both average potential and corresponding
single-particle wave functions \cite{vau72}.  The solutions are
self-consistent in the sense that the particles move in orbits from a
potential created by themselves.  The predictive power of this
procedure is far better than from the potentials found by guesses
based on properties of known stable nuclei.

The product wave function has by definition no correlations and
consequently is only directly able to provide meaningful average
values of observables.  The all too obviously missing pair
correlations are therefore on top included in the BCS approximation or
better by self-consistent calculations of the pair field in the
Hartree-Fock-Boguliubov scheme \cite{sch64}.  The crucial interactions
are left as phenomenological input adjusted to reproduce a number of
essential observables for example of selected (often spherical double
magic) nuclei.

The product character of the single-particle wave functions expresses
that the particle states only are related through the average
potential. These models are therefore denoted independent particle
models.  Now-a-days the same type of calculations are instead denoted
density functional theory \cite{ben03,gor09}, where the only
difference is the procedure to choose the interactions. The same form
of the interactions are parameterized to reproduce properties much
closer to the basic nucleon-nucleon interaction, but the philosophy
remains unchanged.  This implies that the mean-field, if allowed, may
turn out to be deformed with the usual problem of violation of angular
momentum conservation.  The remedy is variation before or preferentially after
projection of angular momentum.

To keep the perspective it is worth emphasizing that the bulk
properties of nuclei are described by the continuous liquid drop, or
rather the droplet models.  The quantum features of the independent
particle models are modifications on top of this classical model.  The
effects in absolute terms are often relatively small but all decisive
for the applications, let alone the understanding.  However, this
suggests that the underlying independent particle model interactions
are constrained by bulk (nuclear matter) properties like saturation
energy and density, compressibility, asymmetry energy, surface
tension.  The semi-classical model, with extended Thomas-Fermi
approximations for the kinetic energy operator, utilizes these
properties, where the quantum shell effects can be calculated and
added as in the independent particle model.

These mean field models are numerically easy to apply, and
simultaneous calculations of all nuclei in the nuclear chart are
feasible. Then any choice of interaction can be confronted with
statistics of how well binding energies and other observables are
reproduced \cite{ben03}.

\subsubsection{Interacting shell models.}

One conceptual improvement is to allow correlations in the wave
function and interactions between single-particle states \cite{cau05}.  The
insurmountable obstacle is the huge basis necessary to obtain
convergent numerical results.  The cure is to cut down on the Hilbert
space in the basis where three different restrictions can be made.
First, only the active particles in the valence shell(s) are treated,
while an inert central core is assumed.  Second, the shell model
basis are constructed from the single-particle states in the partly
occupied valence shell(s) above the inert core.  Third, the expansion
in terms of for example an oscillator basis is restricted to a few
oscillator excitations and the lowest angular momenta.

Unfortunately, each of these restrictions are very delicately related
to convergence.  The cure, as almost always, in nuclear physics is to
use a phenomenological interaction adjusted to reproduce specific
observables.  The concept of effective interactions designed to be
used in a non-complete subspace is then of greatest importance
\cite{var73,bar73,bro10}.  The numerous calculations over the years
have provided detailed understanding of nucleonic correlations in
nuclei but of course subject to the restrictions by the allocated
Hilbert space.  It quickly turned out that rather large oscillator
quanta are needed even when the length in an oscillator basis is
optimized.  Also, one valence shell is inadequate for more than very
few valence particles.  Nevertheless, the accumulated experience are
invaluable now where larger computers allow substantial extension of
the Hilbert space.

One major development is to remove the core all-together, resulting in
the no-core shell model \cite{bar13,nav09}.  After curing initially
appearing problems, first of all related to the monopole properties of
the interactions, applications to light nuclei have been very
successful.  Clearly the prize for full treatment of the core is that
even medium heavy nuclei cannot be studied.  Also, the combination of
accurate dealings with both small and large distances remain a
problem.  The shell-model is designed to do well at small distances
whereas a complete basis, also including a fair treatment of larger
distance, quickly becomes much too expensive due to demands in the
number of basis states.  The tail or large-distance properties
influence the energy minimization much less than short-distance
properties, and incorporating both are not practically possible.

The light nuclei are the targets for the no-core shell-model.
Unfortunately, this immediately implies that dealing with the huge
variation in structure of the individual nuclear states becomes
difficult. The general concept of the shell model is very appealing
but allowing descriptions of both single-particle and cluster
structures is complicated in practice.  Extensions to include
shell-model alien cluster-components in the basis has recently been
attempted \cite{bar13}.  This means that at least two types of
non-orthogonal Hilbert spaces are active.  This may very well be
necessary until both structures can be described in the same no-core
shell-model.  At present these models become more and more like the
genuine few-body models.

\subsection{Few-body descriptions}

Instead of brute force methods, where approximations are necessary at
later stages of the computations, it can be advantageous to select and
treat the anticipated most important degrees-of-freedom.  Such a
restricted problem can be solved exactly by analytical or numerical
techniques.  This can serve as the definition of few-body physics.
The only inaccuracy is therefore the uncertainties in the input
assumptions and the parameters.
Every detail in the chosen Hilbert space are then available
but only features describable in this space.  If the solutions are
missing desired quantitative agreement or specific features, more
degrees-of-freedom can be included either fully or perturbatively.

The starting point is the two-body problem which is trivial in the
present context. The much more complicated three-body problem has on
the other hand received overwhelming attention over many years.
Central in these treatments is the formulation in terms of the Faddeev
equations \cite{fad61}, which formally is extended to
Faddeev-Yakubowsky equations \cite{yak67} applicable for more than
three particles.  However, in the practical numerical context the
difficulties increase enormously with the number of particles.  We
shall sketch some of the diverse variations in few-body methods where
different strategies are employed.

\subsubsection{Cluster models.}

The conceptually simplest is a system of a finite small number of
point-like particles, that is either without intrinsic structure or
with frozen (non-active) intrinsic degrees-of-freedom.  With nucleons
as point-like particles the systems are all the light nuclei with
nucleon number less than about $12$, e.g. deuteron, $^{3}$H, $^{3}$He,
alpha-particle, etc.  More complicated structures are mixtures of
alpha-particles and nucleons in for example $^{8}$Be, $^{6}$He,
$^{6}$Li, $^{6}$Be, $^{9}$Be, $^{9}$B, $^{12}$C.  Also more
complicated clusters may be constituents as found for example in $^{15}$B,
$^{17}$B, $^{11}$Be, $^{11}$Li, $^{12}$Be, $^{20}$C, $^{21}$C,
$^{22}$C.

The procedure is then to decide on the important degrees-of-freedom as
exemplified by $^{6}$He with an alpha-particle surrounded by two
neutrons. The alpha-core is frozen and the neutron-neutron and the
neutron-alpha interactions must be parameterized with all the necessary
spin-dependences. The three-body structure problem is then
well-defined and with many present techniques also relatively easily
solved.  It is worth emphasizing that the effect of the Pauli
principle is that the effective interactions in the limited space
easily differ wildly from those in vacuum.  For the neutron-$^{4}$He
interaction in $^{6}$He calculations, the $s$-wave is repulsive for
that reason, since the $\alpha$-core already has two $s$-neutrons.

The structure of $s$ and $p$ domination is the prototype of two- and
three-body nuclear halos where the spatial extension is unusually
large and the binding energy relatively weak \cite{jen04,rii13}.  A
number of other such nuclear halo states with the same characteristics
of large radii and small binding are established.  Their properties
are rather difficult to reproduce with shell model approximations.
Thus the cluster models are not only technically simple to use, they
are also able to describe structures inaccessible by more complicated
models.

This conclusion becomes much more profound when continuum structures,
reactions and decay are required.  Even in cluster models these
properties are not trivially obtained but still they are within reach
due to the few decisive degrees-of-freedom.  Use of these models
therefore seems to be a logical first step in investigations of
complicated systems which exhibit cluster features. Let us emphasize
that some features or states may have cluster properties while others
don't and consequently they would be inaccessible by cluster models.

The actual calculations using cluster models can be carried out in many
ways, including solving the few-body Schr\"odinger equation
numerically, performing variational calculations on the chosen
degrees-of-freedom, or expansion on a convenient basis.  Each method
has been refined over the years, where subtle details often have been
implemented into practical numerical procedures.  We shall not here go
into any such details but stop with the present qualitative overall
descriptions.

\subsubsection{Zero-range models.}

The Faddeev equations apply for three point-like particles (or
clusters) without any intrinsic structure.  Their interactions may
still be of finite range. If only large-distance properties are of
interest, an assumption can be made of interaction radii much smaller
than the radius of the solution.  This is appropriately called
zero-range interaction models, which by definition then neglect all
information carried by finite-range interactions. On the other hand,
these models exhibit universal features defined by properties
independent of any details of the interactions.

An immediate technical problem is that an unlimited Hilbert space
leads to divergent three-body wave functions, since it is advantageous
to be in the same point of space. This small-distance or large
momentum divergence has to be removed. In coordinate space by applying
a short-distance cut-off adjusted conveniently to a known measured
quantity like the energy.  Examples are the zero-range effective
field models.  In momentum space the same relevant regularization can
be achieved by subtraction of the unphysical large-momentum
dependence.  The prominent example is the zero-range momentum space
formulations of the Faddeev equations \cite{adh95,fre11}.

These models can only describe universal properties, simply because
the wave function always is located completely outside the zero-range
potential.  This implies a tendency to exaggerate occurrence of
universality, since all structures are universal in these models. An
analogy can be found in the most abundant halos of two-body structure,
which over the years have been ``seen'' in different shell models,
although a more precise description would have been single-particle
states.

The advantage of zero-range interactions is the simplicity and the
relatively moderate computing time. The results depend only on the
relative masses and the bound or virtual state energies of the three
two-body subsystems.  Important results are the universal curves
obtained as dimensionless relations between two and three-body
properties.  These unique model relations are interaction independent
and they are therefore by definition revealing universal properties of
systems where observables are on such curves \cite{fre12}.

\subsubsection{Resonating group and generator coordinate method.}

The degrees-of-freedom are first chosen and typically divided into a
few slow and many fast changing coordinates \cite{whe37,whe37b}.  This
is very similar to the Born-Oppenheimer approximation
\cite{gri57,hil53}.  In general this division could be between
intrinsic and relative cluster coordinates \cite{tan78}.  First,
define clusters and corresponding parameters, then place them
geometrically in space by choosing suitable relative coordinates like
relative center-of-mass coordinates \cite{bri68,tho77,hes02}, and
finally quantize the slow coordinates and solve the corresponding
equations of motion.  More specific examples are the fission process
described by the slow collective deformation parameters and the fast
intrinsic nucleonic coordinates.

These procedures are inevitably violating basic conservation of the
total angular momentum which one way or another has to be restored
afterwards.  This is done either by projection of the solution onto
angular momentum eigenstates, or by minimizing the energy of the
projected state.

\subsection{Variational methods}

The variational principle is almost always underlying all the methods
to solve the many-body problem.  However, in some cases it is used
very directly as variation of parameterized wave functions which may be
either functionals, or parameter dependent functions, or a combination
of both.  The starting point could be the shell model \cite{cau05} or
Green's function Monte Carlo \cite{pie01} formulation with the huge
basis, but built up by random selecting and trying the importance of
new basis states \cite{shi12}.  In the end, only the most important
contributions to the energy are then left, which inevitably leaves all
exotic minor components inaccurately determined.  This problem of
determination of minor, but interesting components, remains in all
variational calculations where the energy is used as measure of
convergence.

\subsubsection{Gaussian stochastic variation.}

The basic ingredients are conveniently chosen as a number of gaussian
functions where their centers, widths, and strengths are the variational
parameters.  The gaussian structures allow analytical calculation of
matrix elements provided the interactions are analytical and
sufficiently simple or parameterized in terms of simple functions.  The
number of gaussians are increased using various dedicated strategies
depending on the problem until the energy is converged. Again only
significant basis states are then picked up, and also too similar
gaussians give numerical problems as the basis then is close to
overcomplete.  In any case the basis is by construction non-orthogonal
and should be handled accordingly.

The technique is very flexible but most directly efficient for ground
states and the first few excited states.  Total angular momentum
conservation is as always a problem in nuclear physics calculations.
Similar practical difficulties are also present for identical particles
with corresponding (anti)symmetrizations.

\subsubsection{Molecular dynamics model.}

Antisymmetrized molecular dynamics (AMD) \cite{his91,kan95} and
fermionic molecular dynamics (FMD) \cite{fel90} are two very similar
techniques for treating nuclear systems. FMD and AMD can be viewed as
improvements of the theory of quantum molecular dynamics
\cite{aic86}, where the antisymmetrization of the many-body wave
function is treated more explicitly.

In principle, FMD and AMD are created as quantum analogies to the
classical picture of point particles interacting by two-body
interactions. As such these methods were initially constructed to
study nuclear collisions and fragmentation reactions
\cite{kan03}. Nucleons are viewed as very localized wave packets to
avoid violating the Heisenberg uncertainty principle, and the
many-body wave function must be antisymmetrized according to the Pauli
principle \cite{fel95}. The fundamental procedure of AMD and FMD is
relatively traditional. An antisymmetric many-body wave function is
used as a trial state in conjunction with an appropriate Hamiltonian
for the system, and the relevant equations of motion are derived using
the time-dependent variational principle \cite{fel95}.

The differences between AMD and FMD are mostly technical, where the
main difference is that FMD contains more parameters. In particular,
the width of the wave packets is a complex variational parameter in
FMD which can be different for individual waves, while in AMD it is
the same for all wave packets. Also the spin orientation is treated as
a variational parameter in FMD \cite{nef04}.

Both FMD and AMD are very versatile methods, which are able to
describe nuclear collisions as well as nuclear structures
\cite{kan03}. When describing mean-field type structures FMD and AMD
are somewhat similar to ordinary Hartree-Fock methods, only including
correlations, but when describing cluster type structures FMD and AMD
are able to handle deformed or distorted cluster structures. It is
also possible to describe both stable and unstable nuclei as well as
excited states using FMD and AMD, and it is not necessary to assume
inert cores or restrictions like axial symmetry \cite{kan98}.

All of this makes AMD and FMD very useful approaches, capable of
describing many very different aspects of nuclear systems. However,
one type of system that both methods struggle with is spatially
extended and lightly bound systems, the so-called halo systems. The
wave functions in AMD and FMD represented by Gaussians wave packets
have difficulties in properly describing the extended tail of the halo
structures \cite{kan01}, which makes these systems inaccessible to AMD
and FMD. As always, there is also computational limitations to
consider, which, although not as significant as in shell model
calculations, still prevents the treatment of heavier nuclear systems
\cite{kan01}.  The use Gaussian two-body wave packets implies that
even three-body correlations must be described by combining individual
two-body structures. This is a severe shortcoming compared to few-body
models.

\subsection{Renormalized ab-initio methods}

The increased availability of high-performance supercomputers has
allowed adaptation of various \textit{ab-initio} methods
\cite{lei13}. They provide in principle a full description of the
many-body system, and allow systematic improvements at the expense of
computation time. These methods have some subtle differences, but the
main principles are very similar.  First the fundamental
interaction(s) must be chosen or rather adjusted to reproduce the
experimentally well established detailed nucleon-nucleon scattering
data.  The problem is that due to the strong interaction, the
potentials which reproduce these phase shifts include strong
short-range repulsion as well as strong short-range tensor forces.
This causes couplings to high-momentum modes that cannot be ignored,
which in turn leads to non-perturbative correlations.

There are basically only two practical possibilities for dealing with
this problem; either the high momentum modes must be removed, or a
transformation to decouple the high- and low-momentum modes must be
employed.  In either cases the Hilbert space is changed to be less
complete.  The interaction can then be adjusted to reproduce
observables within this reduced space, preferentially maintaining
basic QCD symmetries but allowing systematic improvements. The
prominent examples are effective field theories applied to deal with
the appropriate degrees-of-freedom like subnucleonic, relative nucleon
motion in nuclei, or maybe halo degrees-of-freedom.  However, the most
popular method is to employ some kind of transformation.  This
requires renormalization of the interaction to match the restricted
Hilbert space, but consistency simultaneously demands use of
multi-nucleon effective interactions, which in turn necessitate
truncation.

This is in stark contrast to quantum chemistry, where the
weak electromagnetic force leads to much simpler perturbative
correlations, which is why \textit{ab-initio} calculations within that
field have been in used for decades \cite{vci66,bar78}.

\subsubsection{Coupled cluster models.}

The coupled cluster method is an \textit{ab-initio} or first
principle method that was originally presented more than fifty years
ago \cite{coe58,coe60}, The use of cluster in the title is a little
deceiving because the clusters here mean particle-hole numbers of
excitation, that is $N$-body correlated structures within the
many-body system.  This method started receiving widespread use within
nuclear physics rather recently, and after various difficulties with
the necessary renormalization were removed \cite{bog03}, and
while computing power also increased dramatically.

The starting point is an approximate solution given in terms of a set
of single-particle states with ground state energy and a corresponding
hamiltonian.  A so-called similarity transformation is applied to the
hamiltonian.  The purpose is to decouple unwanted parts of the Hilbert
space from those of interest.  The transformation is expanded in
particle-hole numbers of excitations and the transformed non-hermitian
hamiltonian is diagonalized.  The cluster (correlation) amplitudes are
the solutions, and the wave function has minimum energy with respect
to one and two particle-hole excitations of the transformed
hamiltonian.

The interactions are not predetermined by the method, on the contrary
it is possible to implement a number of different interactions,
depending on the specific purpose. However, many of the more popular
interactions are either derived from one-boson exchange models
\cite{wir95,mac01} or from chiral effective field theory
\cite{wei90,epe98}.  The truncation eliminating higher order
particle-hole excitations implies that three- and higher-body
interactions are explicitly excluded, but possible to be circumvented
partly by use of density dependent two-body interactions \cite{hag16}.

When including singles and doubles (CCSD), the computational cost for solving 
the coupled cluster equations in the $m$-scheme is
$A^2 n^4$, where n is the number of single-particle valence states,
which naturally will be much larger than the number of particles,
$A$ (see \cite{hag14} for details). This results in a method that is computationally 
very expensive, although it is still cheaper than for instance a full no-core shell
model calculation.  The number of coupled equations are still huge but
clearly sometimes affordable.  Weakly bound and open quantum systems
are difficult to incorporate in coupled cluster methods \cite{mic08},
unless they are in the proximity of sub-shell closure \cite{hag07,hag12}.

\subsubsection{Similarity renormalization group.}

A similarity transformation has also the purpose of decoupling
unwanted parts of the Hilbert space from those of interest.  The
similarity renormalization group and the in-medium similarity
renormalization group (IM-SRG) approaches \cite{weg94,gla93,tsu11} are
very close cousins of coupled cluster models. The same overall
strategy of transforming the Hamiltonian to decouple the perturbative
and non-perturbative aspects of the interaction.  The idea is to
obtain a block-diagonalization of the initial Hamiltonian, by
employing a continuous, unitary transformation \cite{her16}.

For IM-SRG the central idea is to formulate and solve the
intrinsically non-perturbative Heisenberg-like equation which is
equivalent to diagonalizing the many-body Hamiltonian.  In practice,
the transformation is chosen to produce a normal-ordered Hamiltonian,
expressed in terms of second quantization, which can then be truncated
to only include interactions of a certain order. Usually, only up to
second order interactions are included, resulting in an approximate
evaluation of the eigenvalues of the equations. As in CC the main
point is that the transformation allows for such a truncation at an
appropriate level of complexity.

The main differences between coupled clusters and IM-SRG lie in the
exact transformation employed (the transformed Hamiltonian is
Hermitian in IM-SRG and non-Hermitian for coupled clusters), and how
the resulting equations are solved. Given that these differences are
mainly technical, while the general procedure and underlying
philosophy is identical, the methods have many of the same advantages
and limitations. For instance, the truncation again results in only an
approximate factorization into an intrinsic and a center-of-mass wave
function. As a result the same issues with translational invariance as
in coupled clusters appears here in IM-SRG.

When truncated to normal-ordered two-body operators, IM-SRG(2) scheme,
the computational effort for solving the flow equations is dominated by the 
two-body flow equation, which scales polynomially like ${\cal O}(N^6)$, where 
$N$ is the single-particle basis size \cite{her16}. This means that the IM-SRG(2) 
has a sizable scaling factor with model space size like the CCSD.

\section{Theoretical formulation\label{sec:Theory}}

We shall in this report present a new method for treating many-body nuclear systems, which complements the existing methods by focusing on areas currently difficult to treat with these methods. Among them, two areas are particularly interesting: very extended, weakly bound systems and heavier, more complicated structures. The issue is fundamentally a computational or technical one, as both very extended systems and complicated systems with many constituents require a very large basis to be evaluated properly.

The primary goal here is to present a method which consistently connects extended, clusterized structures with the underlying many-body structure, whereas simultaneously being so efficient that it can be applied to very complicated many-body systems.

In Sec.~\ref{sec:GenTheo} the theoretical framework will be presented with as few assumptions as possible, giving an intuitive overview of the approach and showing the general nature of the method. Afterwards, in Sec.~\ref{sec:TheoSky}, a number of assumptions, necessary to perform specific calculations, are made, and the details of the formulation are derived.

\subsection{General formulation\label{sec:GenTheo}}

\begin{figure}
\centering
\includegraphics[width=0.5\textwidth]{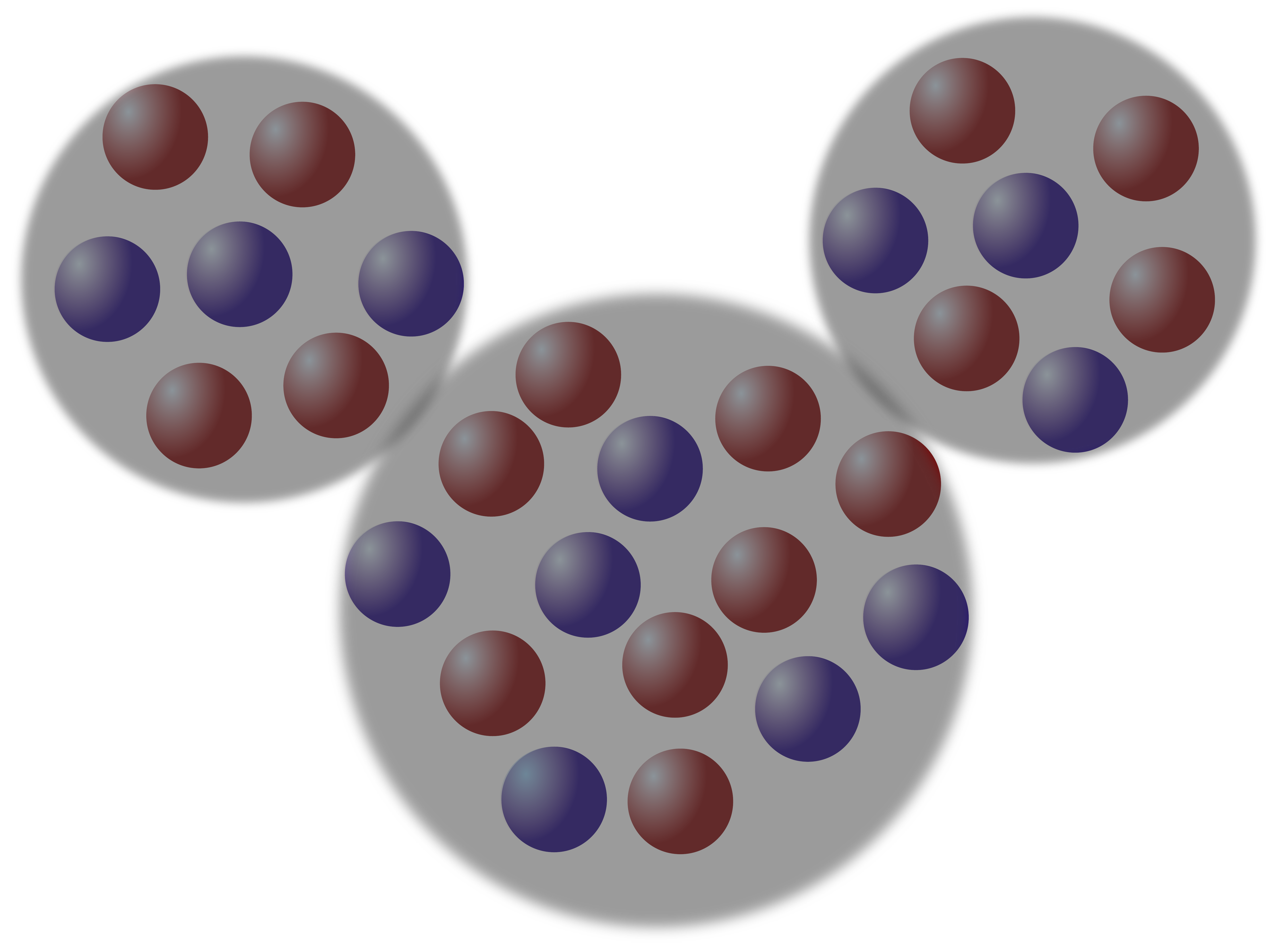}
\caption{A schematic illustration of the many-body system viewed as a three-body cluster configuration with underlying structure. It should be noted that the constituent particles are considered as point particles throughout this work, and the effect of their internal structure is included through three-body interactions. \label{fig:schIll}}
\end{figure}

Our fundamental idea is to view a many-body system as a structure consisting of few clusters, while still maintaining the full set of degrees of freedom in the description. This is illustrated in Fig.~\ref{fig:schIll}. Unless the particles in the clusters are fundamental, they will have some internal structure, and their substructure combined with two-body interactions with many particles will invariantly lead to some form of effective three-body or higher-order interactions. Here the constituent particles are considered to be point particles, but three-body interactions are included, in addition to the two-body interactions, to account for their internal structure. 

A general Hamiltonian for a many-body system consisting of $A$ particles interacting with two- and three-body interactions ($V_{ij}$ and $V_{ijk}$), in the center-of-mass frame of the system, can be written as
\begin{eqnarray}
H_{gen} = \sum_{i=1}^{A} T_i - T^{(cm)} + \frac{1}{2} \sum_{i,j}^{A} V_{ij} + \frac{1}{6} \sum_{i,j,k}^{A} V_{ijk},
\label{eq:GeneralHam}
\end{eqnarray}
where $T_i$ is the kinetic energy operator of the $i$'th particle, and $T^{(cm)}$ is the center-of-mass kinetic energy operator. It will be assumed that the system consists of three clusters, $c_1$, $c_2$, and $c_3$ with $A_1$, $A_2$, and $A_3$ particles respectively, but the generalization to $n$ clusters is straightforward. The general Hamiltonian from Eq.~(\ref{eq:GeneralHam}) can then be rewritten as
\begin{eqnarray}
H_{gen}
&=
H_{c_1} + H_{c_2} + H_{c_3} + H_{fb}, 
\label{eq:ClusterHam}
\\
H_{c_l} 
&= \sum_{i \in c_l} T_i - T_{c_l}^{(cm)} 
+ \frac{1}{2} \sum_{i,j \in c_l} V_{ij}
+ \frac{1}{6} \sum_{i,j,k \in c_l} V_{ijk},
\label{eq:CHam}
\\
H_{fb} 
&= \sum_{l=1}^3 T_{c_l}^{(cm)} - T^{(cm)}
+ \frac{1}{2} \sum_{n \neq m = 1}^3 \sum_{i \in c_n} \sum_{j \in c_m} V_{ij}
+ \frac{1}{6} \sum_{n \neq m = 1}^3 \sum_{i,j \in c_n} \sum_{k \in c_m} V_{ijk}
\nonumber
\\
&+ \frac{1}{6} \sum_{l \neq n \neq m = 1}^3 \sum_{i \in c_n} \sum_{j \in c_m} \sum_{k \in c_l} V_{ijk},
\label{eq:FBHam}
\end{eqnarray}
where $T_{c_l}^{(cm)}$ is the center-of-mass kinetic energy operator for the $l$'th cluster. For the two-body interaction in Eq.~(\ref{eq:FBHam}) both particles are never in the same cluster, and for the three-body interaction in Eq.~(\ref{eq:FBHam}) at least one particle is not in the same cluster as the others. The factors are to avoid double counting.

From Eqs.~(\ref{eq:ClusterHam}) to (\ref{eq:FBHam}) it is clear that the cluster Hamiltonians, $H_{c_l}$, are just many-body Hamiltonians like Eq.~(\ref{eq:GeneralHam}) for a system of $A_l$ particles. Also, the few-body Hamiltonian, $H_{fb}$, includes all the interactions that connect more than one cluster. If the clusters were considered as point particles, Eq.~(\ref{eq:ClusterHam}) would reduce to a regular three-body Hamiltonian.

Having settled on a conceptual approach, the following steps are in principle very simple; we have to choose a wave function, choose an interaction, and do a variation. From Eq.~(\ref{eq:ClusterHam}) the natural choice of the general wave function is an antisymmetric product of the cluster wave functions, $\Psi_{c_i}$, and the few-body wave function, $\Psi_{fb}$,
\begin{eqnarray}
\Psi_{gen} 
&= 
\mathcal{A} \left[
\Psi_{c_1}(\left\{ \bm{r}_{c_1} \right\})
\Psi_{c_2}(\left\{ \bm{r}_{c_2} \right\})
\Psi_{c_3}(\left\{ \bm{r}_{c_3} \right\})
\Psi_{fb}(\bm{r}_{R_1},\bm{r}_{R_2})
\right],
\label{eq:GenWaveFunc}
\end{eqnarray}
where $\mathcal{A}$ is the anti-symmetrization operator, $\bm{r}_{R_1}$ and $\bm{r}_{R_2}$ are the relative coordinates between the center-of-mass of the clusters, and $\left\{ \bm{r}_{c_i} \right\}$ are the $A_i$ (spin and space) coordinates for the particles in the $i$'th cluster. 

With the Hamiltonian from Eq.~(\ref{eq:ClusterHam}) and the wave function from Eq.~(\ref{eq:GenWaveFunc}) the total energy becomes
\begin{eqnarray}
E 
&= \Braket{\Psi_{gen} | H_{gen} | \Psi_{gen}} 
\nonumber
\\
&= \sum_{i=1}^3 
\Braket{\mathcal{A} \left[ \Psi_{c_i} \Psi_{fb} \right] |
 H_{c_i} 
 | \mathcal{A} \left[ \Psi_{c_i} \Psi_{fb} \right] }
+ \Braket{\Psi_{gen} | H_{fb} | \Psi_{gen}}.
\label{eq:E_gen}
\end{eqnarray}

In connection with the variation of the energy, the Lagrange multipliers $E_{c_1}$, $E_{c_2}$, $E_{c_3}$, and $E_{fb}$ are introduced as
\begin{eqnarray}
\Braket{\Psi_{gen} | H_{gen}^{\prime} | \Psi_{gen}} 
&= 
\Braket{\Psi_{gen} | H_{gen} | \Psi_{gen}} 
- \sum_{i=1}^3 E_{c_i} \int |\Psi_{c_i}(\left\{ \bm{r}_{c_i} \right\})|^2 d\left\{ \bm{r}_{c_i} \right\}
\nonumber
\\
&- E_{fb} \int |\Psi_{fb}(\bm{r}_{R_1},\bm{r}_{R_2})|^2 d\bm{r}_{R_1} d\bm{r}_{R_2}
\end{eqnarray}

To minimize the energy both, cluster and few-body wave functions, are varied individually
\begin{eqnarray}
0 &= \frac{\delta}{\delta \Psi_{c_i}^{\ast}}
\Braket{\Psi_{gen} | H_{gen}^{\prime} | \Psi_{gen}},
\\
0 &= \frac{\delta}{\delta \Psi_{fb}^{\ast}}
\Braket{\Psi_{gen} | H_{gen}^{\prime} | \Psi_{gen}},
\end{eqnarray}
for $i \in \left\{1, 2 , 3 \right\} $.

This results in a series of coupled equations for the cluster structures and the relative structure given by
\begin{eqnarray}
E_{c_i} \Psi_{c_i}(\left\{ \bm{r}_{c_i} \right\}) 
&= \Braket{\Psi_{fb} | H_{c_i} \Psi_{c_i}  | \Psi_{fb} }
+ \Braket{\Psi_{fb} | H_{fb} \Psi_{c_i} | \Psi_{fb} }
\nonumber
\\
&+ \Braket{ \mathcal{A} \left[  \Psi_{c_j} \Psi_{c_k} \Psi_{fb} \right] | \Psi_{c_i}^{\ast} \frac{ \delta H_{gen}}{\delta \Psi_{c_i}^{\ast}} \Psi_{c_i} | \mathcal{A} \left[  \Psi_{c_j} \Psi_{c_k} \Psi_{fb} \right] }, 
\\
E_{fb} \Psi_{fb}(\bm{r}_{R_1},\bm{r}_{R_2}) 
&= 
\Braket{\mathcal{A}\left[ \Psi_{c_1}\Psi_{c_2}\Psi_{c_3} \right] | H_{fb} \Psi_{fb} | \mathcal{A}\left[ \Psi_{c_1}\Psi_{c_2}\Psi_{c_3} \right] }
\nonumber
\\
&+ \sum_{i=1}^3 \Braket{\Psi_{c_i} | H_{c_i} \Psi_{fb} | \Psi_{c_i} }
\nonumber
\\
& + \Braket{\mathcal{A} \left[ \Psi_{c_1}\Psi_{c_2}\Psi_{c_3} \right] | \Psi_{fb}^{\ast} \frac{ \delta H_{gen}}{\delta \Psi_{fb}^{\ast}} \Psi_{fb} | \mathcal{A} \left[ \Psi_{c_1}\Psi_{c_2}\Psi_{c_3} \right] },
\end{eqnarray}
for $i,j,k \in \left\{1, 2 , 3 \right\} $, and $i \neq j \neq k$. The last terms in both equations are due to the fact that the interactions might be density dependent. Otherwise, these terms are zero. It is clear that the cluster interactions will depend on the three-body wave function and vice versa \cite{hov17e}.

\subsection{Specific formulation \label{sec:TheoSky}}

So far the derivations have been very general. No specific type of system has been chosen, no specific many-body formalism has been chosen, and no specific few-body formalism has been chosen. The method outlined above could be implemented with almost any choice of few- and many-body formalisms.

Here a number of specific choices will be made. As the focus is on nuclear systems, and to keep the initial investigation simple, it is assumed that the system consists of one fairly heavy cluster with $A$ nucleons surrounded by two valence nucleons, which are also assumed to be identical. 

One main point of the method is then the core description, where almost any description could be employed. As an effort is made to make the method very computationally efficient, a self-consistent Hartree-Fock mean-field method  is used to describe the core.

To only content with one core description it is assumed that the remaining two clusters consist of single, identical nucleons. As the present application focuses on weakly 
bound systems, two identical nucleons is in any case the most obvious choice. The separation of the neutron and proton driplines makes systems with simultaneously 
weakly bound neutrons and protons uncommon and exotic.

Finally, the three-body formalism employed is the hyperspherical adiabatic expansion of the three-body Faddeev equations in coordinate space \cite{nie01}. Again, any few-body formalism could be chosen, but the hyperspherical adiabatic expansion is a powerful, flexible, and accurate method, particularly well suited for describing low-energy scattering and predicting both bound states and resonances.

Applying these choices, the Hamiltonian from Eqs.~(\ref{eq:ClusterHam}) to (\ref{eq:FBHam}) initially simplifies to:
\begin{eqnarray}
H &= 
H_c + H_{3} \;,
\label{eq:Ham}
\\ 
H_c
&= \sum_{i=1}^{A} T_i - T_{c}^{(cm)} +
\frac{1}{2} \sum_{i,j=1}^{A} V_{ij} + \frac{1}{6} \sum_{i,j,k=1}^{A} V_{ijk} \;, 
\label{eq:HamCoreGen}
\\ 
H_{3} 
&= T_{c}^{(cm)} + T_{v_1} + T_{v_2} - T_{cm} + V_{v_1 v_2}  
+ \sum_{i=1}^{A} \left(V_{i v_1} + V_{i v_2} \right) 
\nonumber
\\
&+ \sum_{i,j=1}^{A} \left(V_{ijv_1} + V_{ij v_2} \right) 
+ \sum_{i=1}^{A} V_{i v_1 v_2} \;,
\label{eq:Ham3bGen}
\end{eqnarray}
where $v_1$ and $v_2$ refers to the two valence nucleons.

\subsubsection{Skyrme interaction.\label{sec:skyrme}}

Being more specific, density dependent Skyrme forces \cite{vau72} will be used in the self-consistent Hartree-Fock mean-field calculation of the core.
This has the advantage of both extremely high computational efficiency, combined with generally very accurate predictive power, and a high degree of flexibility. In addition, it is assumed that the core is spherical and even-even with respect to neutron and proton number, as this greatly simplifies many of the equations for the initial investigation. The various generalizations are well-known and straightforward, albeit cumbersome \cite{vau73,sch10,bon07}, and their implementation is discussed in Sec.~\ref{sec:expansion}.

\begin{figure}
\centering
\includegraphics[width=0.5\textwidth]{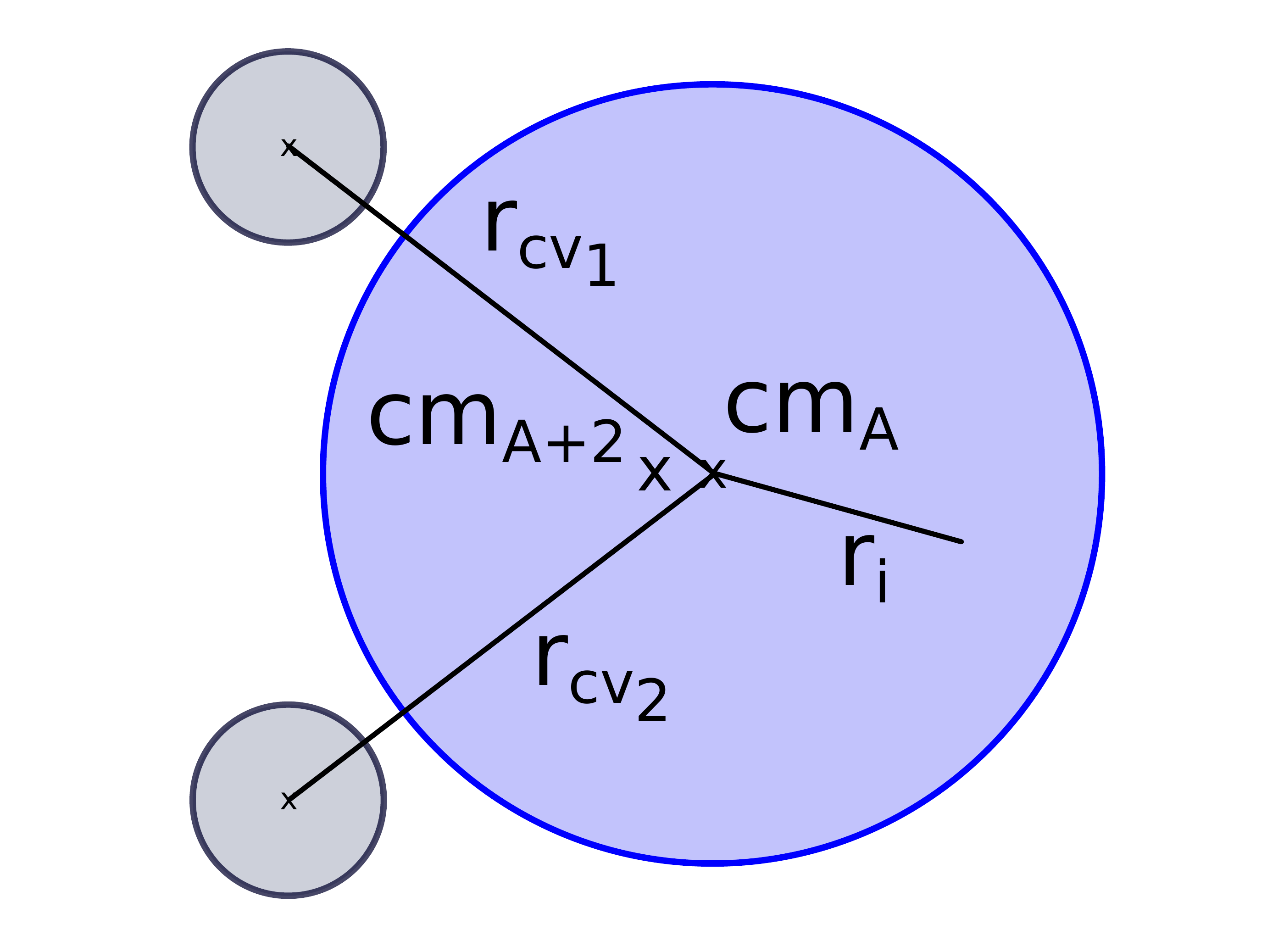}
\caption{Schematic illustration of the spatial coordinates for the wave functions in Eq.~(\ref{eq:WaveFunc}). The center of mass for the core is indicated by $cm_A$, while the total center of mass is indicated by $cm_{A+2}$. The $A$ core particles have a corresponding coordinate $\bm{r}_i$, while the valence nucleons are distinguished by $\bm{r}_{cv_1}$ and $\bm{r}_{cv_2}$, respectively. 
\label{fig:coor}}
\end{figure}

With a density dependent Skyrme force, the three-body interaction is parameterized as a density dependent two-body interaction. After introducing this parameterization, $V_{ijk}$ in Eq.~(\ref{eq:HamCoreGen}) is then included into $V_{ij}$. Likewise, in Eq.~(\ref{eq:Ham3bGen}) $V_{ij v_1}$ and $V_{ij v_2}$ are included into $V_{i v_1}$ and $V_{i v_2}$, respectively. For the moment, $\sum_i V_{i v_1 v_2}$ is kept as a regular three-body interaction, $V_{3}$. This interaction is discussed in greater detail in Sect.~\ref{sec:ValInt}. The Hamiltonians $H_c$ and $H_{3}$ then reduce to
\begin{eqnarray}
H_c
&= \sum_{i=1}^{A} T_i - T_{c}^{(cm)} +
\frac{1}{2} \sum_{i,j=1}^{A} V_{ij} \;, 
\label{eq:HamCore}
\\ 
H_{3} 
&= T_{c}^{(cm)} + T_{v_1} + T_{v_2} - T_{cm} + V_{v_1 v_2}  
+ \sum_{i=1}^{A} \left(V_{i v_1} + V_{i v_2} \right) 
+ V_{3} \;.
\label{eq:Ham3b}
\end{eqnarray}
For the simplified system the wave function (\ref{eq:GenWaveFunc}) reduces to
\begin{eqnarray}
\Psi 
&= 
\mathcal{A} \left[
\psi_{c}(\bm{r}_{1}, \dots , \bm{r}_{A})
\psi_{3b}(\bm{r}_{cv1},\bm{r}_{cv2})
\right],
\label{eq:WaveFunc}
\end{eqnarray}
where the core wave function is a Slater determinant of single-particle wave functions, i.e., $\psi_c = \det(\{ \psi_i \})$ where $i$ runs over all the core nucleons, and $\psi_{3b}$ is
the three-body wave function. As Skyrme interactions have been adjusted to Slater determinants, this is the natural choice of the core wave function, unless the Skyrme interaction is reparameterized. The coordinates are illustrated in Fig.~\ref{fig:coor}. As mean-field calculations break translational invariance, this problem is inherited here. However, the discrepancy decreases with core mass, and it does not have a significant effect on the present calculations.

As indicated by Eq.~(\ref{eq:E_gen}) the simplest expression for the energy is 
\begin{eqnarray}
E = \Braket{\Psi | H | \Psi }
= \Braket{\Psi | H_c | \Psi} + \Braket{\Psi | H_3 | \Psi}.
\end{eqnarray}
As a result the Lagrange multipliers $\epsilon_i$ and $E_3$ can be introduced by
\begin{eqnarray}
\Braket{\Psi | H^{\prime} | \Psi}
= E 
- \sum_{i=1}^{A} \epsilon_i \int | \psi_i(\bm{r})|^2 \, d \bm{r}
- E_3 \int |\psi_{3b}(\bm{r}_{cv1},\bm{r}_{cv2})|^2 \, d\bm{r}_{cv_1} d\bm{r}_{cv_2} \;.
\end{eqnarray}

To minimize the energy both, core and three-body wave functions, are again varied individually as
\begin{eqnarray}
0 = \frac{\delta}{\delta \psi_i^{\ast}}
\Braket{\Psi | H^{\prime} | \Psi},
&&
0 = \frac{\delta}{\delta \psi_{3b}^*}
\Braket{\Psi | H^{\prime} | \Psi}.
\label{eq:vari}
\end{eqnarray}

The Skyrme force used has the form
\begin{eqnarray}
V_{jk}
&= 
t_0 \left( 1 + x_0 P_{\sigma} \right) \delta(\bm{r}_j - \bm{r}_k) 
\nonumber 
\\
& + \frac{t_1}{2}  \left( 1 + x_1 P_{\sigma} \right) \left( \bm{k}'^{2} \delta(\bm{r}_j - \bm{r}_k) + \delta(\bm{r}_j - \bm{r}_k) \bm{k}^{2} \right) 
\label{eq:skyrme}
\\
&+t_2 ( 1 + x_2 P_{\sigma}) \bm{k}' \delta(\bm{r}_j - \bm{r}_k) \bm{k}
+ \frac{1}{6} t_3 (1 + x_3 P_{\sigma}) \left( n_c + n_3 \right)^{\alpha} \delta(\bm{r}_j - \bm{r}_k) \nonumber \\ &
+ i W_0 \left( \bm{\sigma}_{j} + \bm{\sigma}_{k} \right) \cdot \left( \bm{k}' \times \delta(\bm{r}_j - \bm{r}_k) \bm{k} \right), 
 \nonumber
\end{eqnarray}
where $P_{\sigma} = \frac{1}{2} ( 1 + \bm{\sigma}_{j} \cdot \bm{\sigma}_{k} )$, $ \bm{\sigma}_{1,2,3}$ are the Pauli matrices, and $\bm{k} = \frac{1}{2 i} ( \bm{\nabla}_j - \bm{\nabla}_k)$ acting on the right and $\bm{k}' = -\frac{1}{2 i} ( \bm{\nabla}_j - \bm{\nabla}_k)$ acting on the left. The density in the $t_3$ term is the total single-particle density including both core, $n_c$, and valence, $n_3$, nucleon densities, calculated in the center-of-mass of the two particles in question. It is the $t_3$ term which necessitates the variation of the interaction with respect to the wave functions (or densities). The $t_i$, $x_i$, $W_0$, and $\alpha$ are parameters of the Skyrme interaction. To this Skyrme interaction the Coulomb interaction (see Eq.~(\ref{eq_app:coul}) or for instance Ref.~\cite{cha98}) must be added for interactions between protons.

The variation itself, Eq.~(\ref{eq:vari}), is rather lengthy, but conceptually simple. As the variations are with respect to $\psi_i$ and $ \psi_{3b}$ they will lead to a core and a three-body equation, where the effective interactions in both will be coupled by $\psi_i$ and $\psi_{3b}$. The details are presented in ~\ref{appen:vari}, and lead to Eqs.~(\ref{eq:sch3Appen}) and
(\ref{eq:schCAppen}), which are the following coupled three-body and core equations:
\begin{eqnarray}
E_3 \psi_{3b}(\bm{r}_{cv1},\bm{r}_{cv2})
&= \left[
T_x + T_y
+ V_{cv}(\bm{r}_{cv_1}) + V_{cv}(\bm{r}_{cv_2})
+ V_{v_1 v_2}(\bm{r}_{cv_1} , \bm{r}_{cv_2})
\right.
\nonumber
\\
&+
\left.
V_{3}(\bm{r}_{cv_1} , \bm{r}_{cv_2})
\right] \psi_{3b}(\bm{r}_{cv1},\bm{r}_{cv2}), \label{eq:sch3}
\\
\epsilon_{iq} \psi_{iq}(\bm{r})
&= 
\left[- \bm{\nabla} \cdot \frac{\hbar^{2}}{2m^{\ast}_q(\bm{r})}  \bm{\nabla}
+ U_q(\bm{r}) 
- i \bm{W}_q(\bm{r}) \cdot ( \bm{\nabla} \times \bm{\sigma} )
\right.
\nonumber
\\
&-
\left.  \bm{\nabla} \cdot \frac{1}{m_q^{\prime \ast}(\bm{r})}  \bm{\nabla}
+ 
U^{\prime}_q(\bm{r}) 
- i \bm{W}^{\prime}_q(\bm{r}) \cdot ( \bm{\nabla} \times \bm{\sigma} )
\right] \psi_{iq} (\bm{r}). \label{eq:schC}
\end{eqnarray}
The details regarding the three-body equation are found following Eq.~(\ref{eq:sch3Appen}), and will be discussed in Sec.~\ref{sec:3bEq} . In short, $T_x$ and $T_y$ are the three-body kinetic energy operators, $V_{cv}$ is the core-valence nucleon interaction given in Eq.~(\ref{vcore}) and the equations below, $V_{v_1 v_2}$ is the valence nucleon-nucleon interaction, and $V_3$ is the three-body interaction. The full derivation of $V_{cv}$ is included in \ref{appen:vari}, whereas $V_{v_1v_2}$ and $V_3$ are discussed in Sects.~\ref{sec:ValInt} and \ref{sec:V3b}, respectively. It should be noted, that $V_{cv}$ contains effective mass terms, resulting from the gradients in the Skyrme interaction, in addition to regular central- and spin-orbit terms.

The details regarding the core equation are found following Eq.~(\ref{eq:schCAppen}), and will be discussed in Sec.~\ref{sec:coreEq}. The three first terms, without primes, are the terms found in a regular Skyrme-Hartree-Fock Schr{\"o}dinger equation \cite{vau72}, while the prime indicates the contribution from the valence nucleons, and $q$ stands for either neutrons or protons.

The technical implementation along with the practical methods for solving these equations are discussed in Sec.~\ref{sec:coreEq}, including the proper inclusion of the Pauli principle in Sec.~\ref{sec:pauli}.

\section{Technical implementation\label{sec:implement}}

In practice Eqs.~(\ref{eq:sch3}) and (\ref{eq:schC}) are solved iteratively. First, Eq.~(\ref{eq:schC}) is solved for the core in isolation, without including the new contributions from the valence nucleons. The details of the solution to Eq.~(\ref{eq:schC}) are included in Sec.~\ref{sec:coreEq}. This calculation gives rise to the density functions (\ref{densities}), and therefore
to the effective potential $V_{cv}(\bm{r})$ in Eq.~(\ref{vcore}), which is used to solve the three-body equation (\ref{eq:sch3}). The details of the modifications to the traditional hyperspherical expansion of the Faddeev equations are presented in Sec.~\ref{sec:3bEq}.

The valence nucleon densities (\ref{eq:def_n3}), (\ref{eq:def_tau3}) and (\ref{eq:def_J3}), produced in the three-body calculation, are then used in a new core calculation, which provides the potentials for a new three-body calculation. The process is repeated until convergence in energy is reached. Unless otherwise stated, convergence is here in the numerical calculations defined as the three-body energy of two consecutive iterations differing by less than $0.02$~MeV.  This criterion can of course be changed to fit the purpose, but in any
case the structure must also converge.

The new terms in  Eqs.~(\ref{eq:sch3}) and (\ref{eq:schC}) do not add much computational complexity, yet this process is still more computationally demanding than both regular Skyrme-Hartree-Fock calculations and regular hyperspherical adiabatic Faddeev calculations because of the iterations. However, convergence usually happens within 3 to 5 iterations, so the added computation time is roughly a factor of 3 to 5 on the solution of the adiabatic expansion, which is the bottleneck of the calculation. As a result, the computation time is only on the order of a few days on a single, regular processor core.

\subsection{The mean-field core equation\label{sec:coreEq}}

As seen from Eq.~(\ref{eq:schC}), the core Schr{\"o}dinger equation is very similar to the traditional Skyrme-Hartree-Fock Schr{\"o}dinger equation \cite{vau72,vau73,dec75,flo73}. There are kinetic, effective mass, central,  spin-orbit terms, and nothing else. In fact, the traditional Skyrme-Hartree-Fock equation is given by Eq.~(\ref{eq:schC}) if all "primed" potentials are omitted, and using the expressions from Eq.~(\ref{eq cOrd par}), only with $n_3=0$. The influence of the valence nucleons amounts to including the primed potentials and $n_3$, and also using the expressions  from either Eq.~(\ref{eq nval para}) or Eq.~(\ref{eq cval para}), depending on whether or not the valence nucleons are of the same type as the core nucleon in question. As only the expressions change, and no fundamentally new terms are included in the interaction, the method for solving Eq.~(\ref{eq:schC}) is identical to the solution method known from traditional Skyrme-Hartree-Fock \cite{vau72,dec75}. 

The new three-body densities, $n_3$, $\tau_3,$ and $\bm{J}_3$ from Eqs.~(\ref{eq:def_n3}), (\ref{eq:def_tau3}), and (\ref{eq:def_J3}), respectively, are produced by the three-body calculation, Eq.~(\ref{eq:sch3}), as functions of distance, and they enter directly into the calculations. The core densities,  $n_c$, $\tau_c$, and $\bm{J}_c$ from Eqs.~(\ref{eq:def_nc}), (\ref{eq:def_tauc}), and (\ref{eq:def_Jc}), respectively, are produced in the traditional self-consistent manner, as the Hartree-Fock equations are always solved iteratively (not to be confused with the iteration between our core and three-body equations).

For a spherical, even-even nucleus the single-particle wave function can be separated into spherical coordinates as
\begin{eqnarray}
\psi_i(\bm{r}, \sigma) = \rho_{\alpha}(r) y_{\beta}(\hat{r}, \sigma),
\end{eqnarray}
where $\rho$ contains the radial part, whereas $y$ contains the angular (and spin) part. In $\alpha$ the relevant quantum numbers are contained, i.e. the nucleon type, $q$, the principal quantum number, $n$, the orbital angular momentum, $l$, and the total angular momentum, $j$, of the state, while $\beta$ contains the $l$, $j$, and $q$ quantum numbers, as well as the magnetic quantum number $m$.

Due to the spherical symmetry of the core, the core densities can be expressed in terms of $R_{\alpha}(r)$, where $\rho_{\alpha}(r) = R_{\alpha}(r)/r$, as
\begin{eqnarray}
n(r) 
&= \frac{1}{4 \pi r^2} \sum_{\alpha} (2j_{\alpha} + 1) R_{\alpha}^2(r), 
\\
\tau(r) 
&= \frac{1}{4 \pi} \sum_{\alpha} (2j_{\alpha} + 1) \left( \left( \frac{d \rho_{\alpha}}{dr} \right)^2 + \frac{l_{\alpha}(l_{\alpha}+1)}{r^2} \rho_{\alpha}^2 \right) ,
\\
\bm{J}(\bm{r})
& = \frac{\hat{r}}{4 \pi r^2} \sum_\alpha (2j_{\alpha} +1) \left( j_{\alpha} (j_{\alpha}+1) - l_{\alpha}(l_{\alpha}+1) - \frac{3}{4} \right) R_{\alpha}^2(r).
\end{eqnarray}

Initially, a first approximation to the unknown radial part of the wave function, $R_{\alpha}$, is produced using harmonic oscillator wave functions. Using the expressions above, first approximations of the core densities are produced, which again is used to calculate the potentials and effective masses in Eq.~(\ref{eq:schC}). This is used to produce a new radial wave function $R_{\alpha}$, and the process continues until convergence in single-particle energies is reached.

As the parameters of the Skyrme forces used here are fitted using the Hartree-Fock equations where the spin density in the spin-orbit part of the potential has been omitted, the spin densities are also here omitted from the spin-orbit parts of Eq.~(\ref{eq:schC}). They could be included at the expense of a reparameterization, but they are in any case of very minor importance.

\subsection{The three-body equation\label{sec:3bEq}}

As mentioned in Sect.~\ref{sec:TheoSky} our chosen three-body formalism is the hyperspherical adiabatic expansion of the Faddeev equations in coordinate space \cite{nie01,mac68}. However, unlike the core equation, the three-body equation contains terms not usually found within this formalism, and more attention will be devoted to this. Specifically, the almost-local derivatives in the effective mass terms require special attention.

\begin{figure}
\centering
\includegraphics[width=0.9\textwidth]{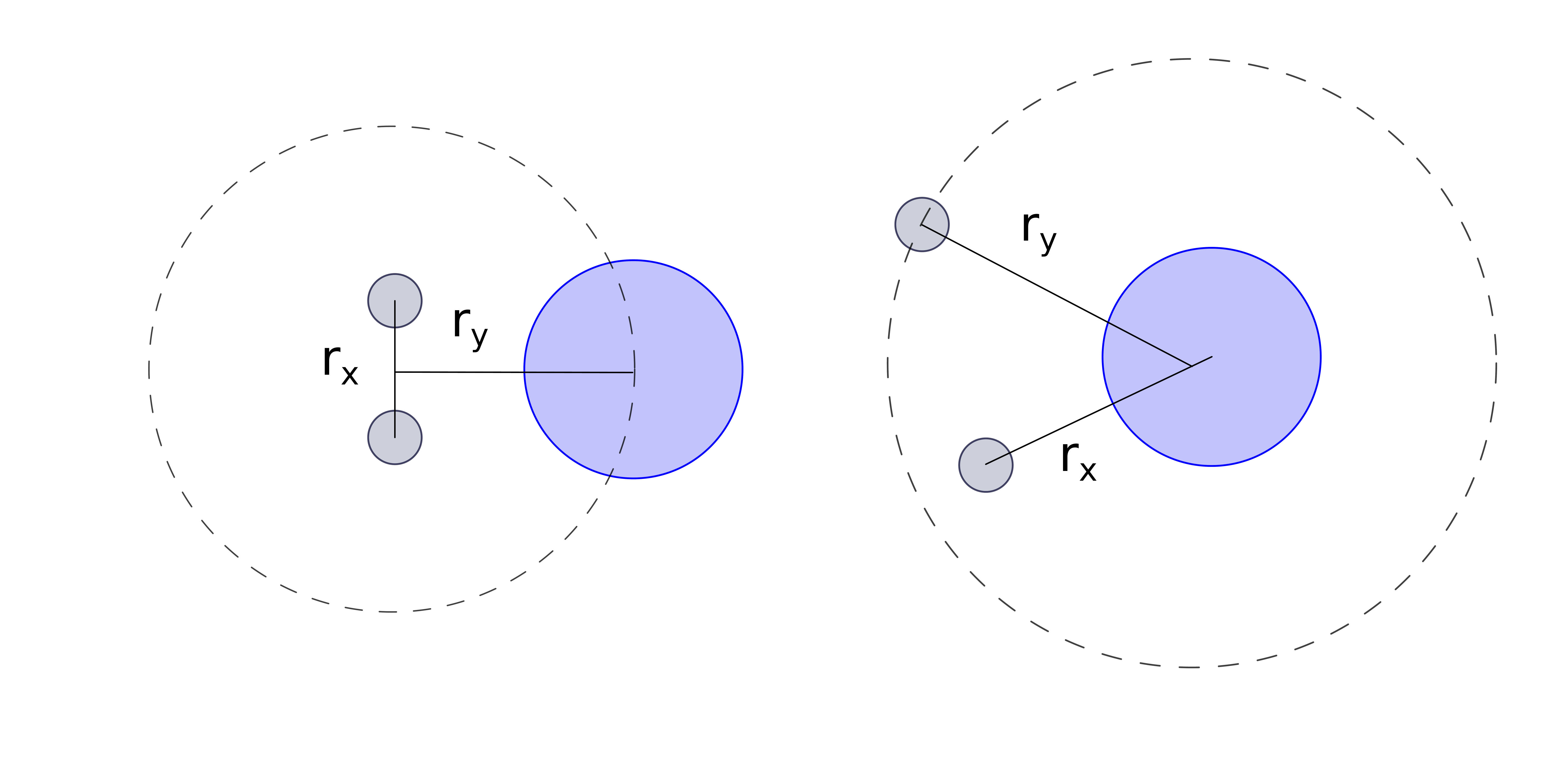}
\caption{The two relative set of coordinates for a three-body system with a large, heavy core surrounded by two identical particles, with coordinate set 1 on the left and coordinate set 2 on the right. Also illustrated is the rotational symmetry of the two coordinate sets. 
\label{fig:3Bcoor}}
\end{figure}

First, the well known three-body Jacobi coordinates are introduced as
\begin{eqnarray}
\bm{x}_i 
= \sqrt{\frac{\mu_{jk}}{m}} (\bm{r}_j - \bm{r}_k ) 
= \sqrt{\frac{\mu_{jk}}{m}} \bm{r}_x , 
\label{eq:JacCoorx}
\\ 
\bm{y}_i 
=\sqrt{\frac{\mu_{jk,i}}{m}}  \left(\bm{r}_i - \frac{m_j \bm{r}_j + m_k \bm{r}_k}{m_j + m_k} \right) 
=  \sqrt{\frac{\mu_{jk,i}}{m}} \bm{r}_y, 
\label{eq:JacCoory}
\\
\mu_{jk} 
= \frac{m_j m_k}{m_j+m_k} \;, 
\\
\mu_{jk,i} 
= \frac{m_i (m_j+m_k)}{m_i+m_j+m_k} \; ,
\label{eq:JacCoor}
\end{eqnarray}
where $\bm{r}_i$, $\bm{r}_j$, and $\bm{r}_k$ are the center-of-mass coordinates of the three clusters, with associated masses $m_i$, $m_j$, and $m_k$, while $m$ is an arbitrary normalization mass. In a three-body system with two identical particles there are two possible sets of relative coordinates, as illustrated in Fig.~\ref{fig:3Bcoor}.

The Jacobi coordinates are used to define the hyperspherical coordinates, which consist of two pairs of directional angles, $(\Omega_{x_i} , \Omega_{y_i})$, for $\bm{x}_i$ and $\bm{y}_i$, as well as the hyperangle, $\alpha_i$, and the hyperradius, $\rho$, coordinates defined by
\begin{equation}
\alpha_i=\arctan\frac{x_i}{y_i}, \hspace*{5mm} \mbox{and } \hspace*{3mm} \rho=\sqrt{x_i^2+y_i^2},
\label{coord}
\end{equation}
from which one can easily get that $x_i = \rho \sin \alpha_i$ and  $y_i = \rho \cos \alpha_i$.

When introducing these five angular and one radial coordinates the three-body kinetic energy operator separates into a hyperradial and a hyperangular part
\begin{eqnarray}
T 
&= T_x + T_y
= T_{\rho} 
+ \frac{\hbar^2}{2 m \rho^2} \Lambda^2, 
\\
T_{\rho} 
&= - \frac{\hbar^2}{2 m} \left( \frac{\partial^2}{\partial \rho^2} + \frac{5}{\rho} \frac{\partial}{\partial \rho} \right) \;,
\\
\hat{\Lambda}^2 
&= -\frac{\partial^2}{\partial \alpha_i^2} -4 \cot (2\alpha_i) \frac{\partial}{\partial \alpha_i} + \frac{\hat{l}^2_{x_i}}{\sin^2 \alpha_i} + \frac{\hat{l}^2_{y_i} }{\cos^2 \alpha_i},
\end{eqnarray}
where $\hat{l}_{xi}$ and $\hat{l}_{yi}$ are the angular momentum operators related to $\bm{x}_i$ and $\bm{y}_i$, and $\hat{\Lambda}^2$ is the hyperangular momentum operator whose eigenfunctions are the hyperspherical harmonics \cite{nie01}.

The fundamental idea behind the hyperspherical adiabatic expansion is to introduce a set of $\rho$-dependent angular wave functions, $\phi_n(\rho,\alpha)$, on which the three-body wave function can be expanded. The angular part of the Schr{\"o}dinger (or Faddeev) equation is then solved for all values of $\rho$, and the solution is used to solve the radial part of the equation. The expanded wave function is
\begin{eqnarray}
\psi_{3b}(\bm{x},\bm{y})
= \frac{1}{\rho^{5/2}}\sum_n f_n(\rho) \phi_n(\rho,\alpha),
\label{eq:3bExpanded}
\end{eqnarray}
and the Schr{\"o}dinger equation is
\begin{eqnarray}
\left[
T_{\rho}
+ \frac{\hbar^2}{2 m \rho^2} \Lambda^2 
+ V_{cv_1} + V_{cv_2} + V_{v_1v_2} + V_3
\right]\psi_{3b}(\bm{x},\bm{y})
=
E_3 \psi_{3b}(\bm{x},\bm{y}).
\label{eq:3Bsch}
\end{eqnarray}

First introduced in 1968 by J. H. Macek \cite{mac68}, the hyperspherical adiabatic representation has a unique position in the theory of multiparticle fragmentation reactions, namely, it is the only representation that maps complex, multiparticle, fragmentation theory onto a set of coupled-channel differential equations identical to those familiar from the theory of two-body reactions \cite{mac02}. 
In addition, pure three-body continuum channels are asymptotically completely separated from bound state channels, and, at the same time, all the channels are asymptotically correct \cite{mac02}. These calculations converge very quickly, and usually only the few lowest angular wave functions in the expansion (\ref{eq:3bExpanded}) are needed to produce an accurate calculation, as the coupling to higher wave functions is very modest. This is in particular important for weakly bound or very extended states, where the adiabatic expansion method is both very efficient and accurate.

Combining the adiabatic expansion with the Faddeev formalism has the added benefit of treating all Jacobi coordinate sets identically. For systems with more than one internal two-body bound or low-lying resonance state, all are treated in their natural coordinate system. When solving a regular \Scheq  the necessary rotation between coordinate systems reduces the accuracy.

We choose the angular functions $\phi_n$ to be the eigenfunctions of the angular part of the Faddeev equations, where, as mentioned, the two-body interactions include almost-local gradient terms (see Eqs.~(\ref{vcore}) and (\ref{vgrad})). In \ref{appen:3bEq} it is shown how these gradient terms can be separated into radial and angular parts. Letting $\tilde{V}$ contain the usual central and spin-orbit terms as well as the angular part of the new gradient terms, the angular Faddeev equation becomes, Eq.~(\ref{b10}), 
\begin{eqnarray}
\left(\hat{\Lambda}^2 
+\frac{2m\rho^2}{\hbar^2}(\tilde{V}_{cv_1}+\tilde{V}_{cv_2}+\tilde{V}_{v_1 v_2})
\right) \phi_n(\rho,\Omega) 
= \lambda_n(\rho) \phi_n(\rho,\Omega),
\label{eq:ang3b}
\end{eqnarray}
where $\lambda_n$ is the ($\rho$-dependent) angular eigenvalues.   Multiplying Eq.~(\ref{eq:3Bsch}), or more precisely (\ref{eq:3BFullSch}), from the left by $\phi_m^*(\rho,\Omega)$ and integrating over $\Omega = (\Omega_{x_i}, \Omega_{y_i})$ we get the set of coupled hyperradial equations
\begin{eqnarray}
0 
&=
(1-C_{nn}) \frac{\partial^2 f_n}{\partial \rho^2} 
- \frac{\lambda_n+\frac{15}{4}}{\rho^2} f_n + \frac{2m(E - V_3)}{\hbar^2}f_n
\nonumber
\\
&+ 2 \sum_m \left( P_{nm} + P_{nm}^{\prime} \right) \frac{\partial f_m}{\partial \rho}
+\sum_m \left( Q_{nm} + Q_{nm}^{\prime} \right) f_m,
\label{eq:rad3b}
\end{eqnarray}
where 
\begin{equation}
P_{nm}(\rho)=\langle \phi_n(\rho,\Omega)|\frac{\partial}{\partial\rho}|\phi_m(\rho,\Omega)\rangle_\Omega
\end{equation}
and 
\begin{equation}
Q_{nm}(\rho)=\langle \phi_n(\rho,\Omega)|\frac{\partial^2}{\partial\rho^2}|\phi_m(\rho,\Omega)\rangle_\Omega
\end{equation}
are the usual coupling terms arising from the standard adiabatic expansion method \cite{nie01} ($\langle|\rangle_\Omega$ indicates integration over the 
hyperangles only). The remaining coupling terms, $C_{nm}$, $P_{nm}^{\prime}$ and $Q_{nm}^{\prime}$ are new couplings coming from the hyperangular part of the gradient terms in Eq.~(\ref{eq:sch3}). They are defined in Eq.~(\ref{eq:Cnm}) and (\ref{eq:PpQp}).

Solving first Eq.~(\ref{eq:ang3b}) and then Eq.~(\ref{eq:rad3b}) provides the full solution for the three-body system. From there anything that could be calculated with ordinary three-body formalism, can also be calculated here. The two main differences between the traditional adiabatic expansion, and the method presented are the origin of the potentials in Eq.~(\ref{eq:ang3b}) and the new couplings in Eq.~(\ref{eq:rad3b}).

Traditionally, the two-body potentials in Eq.~(\ref{eq:ang3b}) are completely phenomenological in nature \cite{hov14b}. The solution to the Faddeev equations involves a partial wave expansion, where the two-body interactions for the individual partial waves can differ. It is known that the couplings to higher partial angular momentum in each Faddeev component is of second order in the potentials \cite{nie01}, but a number of different partial angular momentum values are often still needed for medium heavy systems. In addition, each two-body interaction would include central and spin-orbit parts, at the very least, and possible spin-spin or tensor parts as well. Even with a simple interaction like a double Gaussian, there are easily between 30 and 50 free parameters, depending on the system. As a result, more or less any result could be obtained with a traditional hyperspherical expansion of the three-body Faddeev equations, unless heavily restrained by experimental information. 

For the method presented here, a crucial aspect is that the two-body interactions are completely determined by the core description. Specifically, the Skyrme parameterization dictates the produced core potential, which translates into the two-body interactions in the three-body equation. The only important, remaining degrees of freedom lie in the choice of Skyrme parameterization and possibly in the three-body interaction to be discussed in Sec.~\ref{sec:V3b}.

The new couplings in Eq.~(\ref{eq:rad3b}) are caused by the gradient terms in the Skyrme interaction in Eq.~(\ref{vgrad}), and it is therefore a result of the specific core description chosen, and not a unavoidable necessity of the method. However, as shown in \ref{appen:3bEq} they are neither particularly problematic nor time consuming to include. On the contrary, they represent an interesting new development within three-body Faddeev formalism in their own right.

\subsection{The Pauli principle\label{sec:pauli}}

Implementation of the Pauli principle has the two sides when facing either the core or the three-body calculations.  The Slater determinant in the mean-field formulation guaranties orthogonality between occupied orbits, and the exchange terms in the corresponding Schr\"{o}dinger equation only allow interaction contributions from identical fermions in different states.

The core only requires information about the valence nucleons through ordinary and kinetic energy densities, and terms of minor importance also depending on currents.  The Hartree-Fock iteration procedure usually applied to reach self-consistency then only has information about the states occupied by the valence nucleons from these densities, which enter in Eq.~(\ref{eq:schC}) through Eqs.(\ref{eq cOrd par}), (\ref{eq nval para}), and (\ref{eq cval para}). The choice of occupied states is usually the lowest possible consistent with one for each indistinguishable nucleon.

In principle, another choice of occupied states could be made where two of these core-occupied states are left unoccupied for the valence nucleons while two higher-lying instead are occupied.  The Hartree-Fock solution obtained with the fixed valence densities must in a chosen approximation span a space orthogonal to the space of the two valence nucleons.  To which extent this is correct can be tested by calculation of the overlap integrals,
\begin{eqnarray}
\int \psi_i^{*}(\bm{r}_{cv_1}) \psi_k^{*}(\bm{r}_{cv_2}) \psi_{3b}(\bm{r}_{cv_1},\bm{r}_{cv_2}) \; d \bm{r}_{cv_1} d \bm{r}_{cv_2}, \label{eq:overlap1}
\end{eqnarray}
for any pair, $(i,k)$, of mean-field occupied states. Even independent of $r$ there should be vanishing overlap for each of the occupied states, $\psi_i$,
\begin{eqnarray}
   \int \psi_i^{*}(\bm{r}_{cv_1}) \psi_{3b}(\bm{r}_{cv_1},\bm{r}) d  \bm{r}_{cv_1} . \label{eq:overlap2}
\end{eqnarray}

However, for weakly bound systems, valence nucleons occupying states above the core occupied states are the most natural assumption, and it is the assumption made here.

The three-body calculation receives input from the Hartree-Fock solution in the form of various effective potentials, Eqs.~(\ref{vcen}), (\ref{vso}) and (\ref{vgrad}), which by definition are the potentials acting on one nucleon from all core-nucleons. These are deep potentials, and the solution to such a nucleon-core problem would produce a number of bound states corresponding to at least the number of occupied mean-field states.  This follows from the fact that these potentials are precisely those entering the Hartree-Fock calculation with a bound core.  In other words, the already occupied single-particle states in the core must be excluded from the space employed in the three-body calculation. This problem has been solved by three rather different methods added to the fundamental three-body approach \cite{nie01}. 

The first method is to adjust the two-body potential parameters to allow only the desired number of core-valence nucleon bound states, and at the same time reproduce the low-energy scattering data through a correct scattering length. In effect, the deep potentials are replaced with shallow potentials only supporting the Pauli allowed (highest lying) states \cite{gar97}. The main problem with this is that, since the potential is provided by the mean-field calculation, there is not the necessary freedom in the potential in order to produce the desired effect.

The second method redefines the two-body potential such that the initial and the new potential have precisely the same scattering properties for all energies.  The only difference is that the number of bound two-body states is reduced by precisely the number of Hartree-Fock occupied single-particle states. The new potential is called a phase equivalent potential \cite{gar99phase}. Any bound state, and not necessarily the lowest lying, could be removed using such phase equivalent potentials, without affecting the other states above or below.

A third method of dealing with the Pauli principle is to go through
the hyperspherical adiabatic expansion procedure, and in the
calculations of the radial wave functions, Eq.~(\ref{eq:rad3b}),
explicitly omitting the lowest adiabatic potentials corresponding to
the core occupied states.  The remaining adiabatic potentials are
fully exploited with their complete angular wave functions. This is a
unique feature of the adiabatic method, where bound states
asymptotically correspond to a distinct $\lambda$-function. However,
this is only asymptotically, and excluded and included adiabatic
potentials are not always completely decoupled at small and
intermediate distances, which in turn translates to the subsequent
radial calculations. It is therefore not always strictly correct
directly to exclude states by dividing the space in this way.

The Pauli principle is accounted for with the first or second
construction, since the deeper lying states are excluded as bound
two-body states in the new potentials.  The emerging solutions then
appear as ground states in the new potential, and consequently with no
nodes in the wave functions even at small distances.  The
orthogonality is not achieved through a larger number of nodes inside
the core as for the ideal structure in ideal calculations.  However,
the resulting observables obtained by use of this type of averaging
can be very accurate if either the three-body wave function is located
outside the nodal region or the observable averaging eliminate the
influence of the nodal structure.

With the third construction this is not an issue, as the wave functions are not reconstructions made to reproduce the asymptotic behavior. Instead, the allowed wave functions are represented within the unoccupied space considered as separate from the occupied space. The Pauli principle is then obeyed provided the mixing is negligibly small between these two spaces in a full calculation. Some of the defining features of the adiabatic expansion is the separation of bound and continuum states, and the tendency to asymptotic decoupling between the angular wave functions. Most of the occupied core states should decouple completely, while only the highest-lying core states could couple to the unoccupied, valence allowed states at intermediate distances, making the assumed separation of spaces exact in many cases.

In practice, the most efficient and accurate method to account for the Pauli principle is a combination of the second and the third method. All the completely decoupled states are eliminated as described in the third method, but if any of the highest lying core states couple to the unoccupied states, these are removed using phase equivalent potentials. This implementation of the Pauli principle is both unusually simple and very close to being completely exact, and therefore this is the method used in this work.

\subsection{Finer points of the three-body equation}

The fact that the Hilbert space is intrinsically connected to the choice of the interaction is often not considered explicitly. Unless the Hilbert space is infinite, the interaction must be adjusted to the truncated space. Naturally, this is also true for the method presented here, where the Hilbert space for a many-body core is expanded to encompass the additional valence nucleons. This is particularly important for the interaction between valence nucleons, which is considered in the following.

It is known that a three-body system with zero-range two-body interactions produces an infinite number of strongly bound states, an effect often referred to as the Thomas effect \cite{tho35}. This is closely related to the Efimov effect \cite{efi70}, and is in fact given by the same limit $|a/r_0| \rightarrow \infty$, where $a$ is the $s$-wave scattering length and $r_0$ is the range of the potential. Irrespective of the $s$-wave scattering length the contact interaction in the Skyrme force will lead to a Thomas effect in the three-body system, which makes it a delicate process to include properly.

\subsubsection{The valence nucleon-nucleon interaction.\label{sec:ValInt}}

When it comes to the interaction between the valence nucleons there are three possibilities. The first and probably most obvious choice is to use exactly the same interaction between the valence nucleons as between all other nucleons. The main advantage from an aesthetic point of view is that this is very consistent, but this is also the most problematic. As all kinds of effective Hartree-Fock interactions are in-medium interactions, they are not directly applicable to free nucleons without some kind of renormalization. In particular, a Skyrme interaction would produce a bound two-body system when applied to two free nucleons, where the density of the core nucleons is (approaching) zero. As one of the main reasons for developing this combination of few- and many-body formalisms is to produce a wave function with a consistent evolution from compact to extended configurations, having incorrect long range, asymptotic properties would be detrimental to our purpose.

The second and definitely most simple choice is to use an ordinary phenomenological nucleon-nucleon interaction \cite{gar99} between the valence nucleons. This lacks the pleasant consistency of the first option, but as no new implementations are needed this has the advantage of being very simple. More importantly, the long-distance, asymptotic behavior is known to be correct by construction.

The third and initially most appealing choice is to use a combination of the two options above. More specifically, to use the same effective Hartree-Fock interaction between the valence nucleons as between all other nucleons, when the valence nucleons are close to the core, but then transition to a phenomenological interaction, with the correct asymptotics, as the valence nucleons move away from the core. The valence nucleons would then be treated consistently in compact configurations, but would regain the important correct asymptotic behavior with regards to scattering lengths and internal two-body energies for extended configurations and break-up reactions. The only slightly arbitrary choice would be the method of transitioning from one interaction to the other.

More specifically, the transition could be achieved by 
\begin{eqnarray}
V_{v_1 v_2} &= W_{sk} V_{sk} + W_{ph} V_{ph}, 
\label{eq:valInt}
\end{eqnarray}
where $V_{sk}$ is the Skyrme interaction, and $V_{ph}$ is the phenomenological interaction, while $W_i$ is the associated weight, with $W_{sk} + W_{ph} = 1$. A simple choice of weights would be a Woods-Saxon shape depending on the spatial extension of the system.

The transition is in the first Jacobi coordinate system where $\bm{x}$ is between the two valence nucleons. If $R_c$ is the radius of the core, then the limits for both valence nucleons being either inside or outside the core are given by
\begin{eqnarray}
R_c
&\leq \left| \bm{r}_y \pm \frac{1}{2} \bm{r}_x \right|
= \left| \frac{1}{\mu_y} \bm{y} \pm \frac{1}{2 \mu_x} \bm{x} \right|,
\\
R_c
&\geq \left| \bm{r}_y \pm \frac{1}{2} \bm{r}_x \right|
= \left| \frac{1}{\mu_y} \bm{y} \pm \frac{1}{2 \mu_x} \bm{x} \right| .
\end{eqnarray}
The transition point, where it changes from one interaction to the other being the most favored, is at $R_c = r_{y}$, and a sensible transition width would be half the corresponding $r_{x}$, see Fig.~\ref{fig:3Bcoor}. As the extent of the core is not sharply defined a diffuseness parameter, $d_c$, could be added to the transition width. Using that $x= \rho \sin \alpha$ and $y = \rho \cos \alpha$, the weights become
\begin{eqnarray}
W_{sk}(r_x,r_y)
&= \left(
1 + \exp \left(  \frac{r_y - R_c}{\frac{1}{2} r_{x}+d_c} \right)
\right)^{-1}
\nonumber
\\
&= \left(
1 + \exp \left( 
\frac{\sqrt{\frac{m}{\mu_{jk,i}}} \rho \cos \alpha - R_c }{\frac{1}{2} \sqrt{\frac{m}{\mu_{jk}}} \rho \sin \alpha + d_c }
\right)
\right)^{-1},
 \label{wsk} \\
W_{ph}(r_x,r_y) = 1 - W_{sk} &=
\left(
1 + \exp \left(  \frac{R_c - r_y}{\frac{1}{2} r_{x}+d_c} \right)
\right)^{-1}
\nonumber
\\
&=
\left(
1 + \exp \left( 
\frac{R_c - \sqrt{\frac{m}{\mu_{jk,i}}} \rho \cos \alpha}{\frac{1}{2} \sqrt{\frac{m}{\mu_{jk}}} \rho \sin \alpha + d_c }
\right)
\right)^{-1}.
\label{wph}
\end{eqnarray}
This transition description only involves $\alpha$ in the first Jacobi set with $\bm{x}$ between valence nucleons, and it treats the limits appropriately. For $\rho \rightarrow 0$, $W_{sk} \rightarrow \left( 1 + \exp(-R_c/d_c) \right)^{-1} \rightarrow1$ for $R_c/d_c \rightarrow \infty$. Likewise, for $\rho \rightarrow \infty$, $W_{sk} \rightarrow \left(1 + \exp(2 \mu_x \cos \alpha / ( \mu_y \sin \alpha ) ) \right)^{-1} \rightarrow 0$ for $\alpha \rightarrow 0$ as would be the case for a contact valence nucleon-nucleon interaction.

The implementation of the potential in Eq.~(\ref{eq:valInt}) within the hyperspherical adiabatic expansion, and in particular the solution of the angular eigenvalue problem (\ref{eq:ang3b}), 
requires calculation of the matrix elements of the potential within the hyperspherical harmonics. The calculation of such matrix elements for a contact Skyrme interaction,
 although not complicated from the conceptual point of view, is rather lengthy and cumbersome. For this reason, and specially as a result of the discussion below, we skip here the details
 and focus just on the final result. 
 
\begin{figure}[t]
\centering
\includegraphics[width=0.7\textwidth]{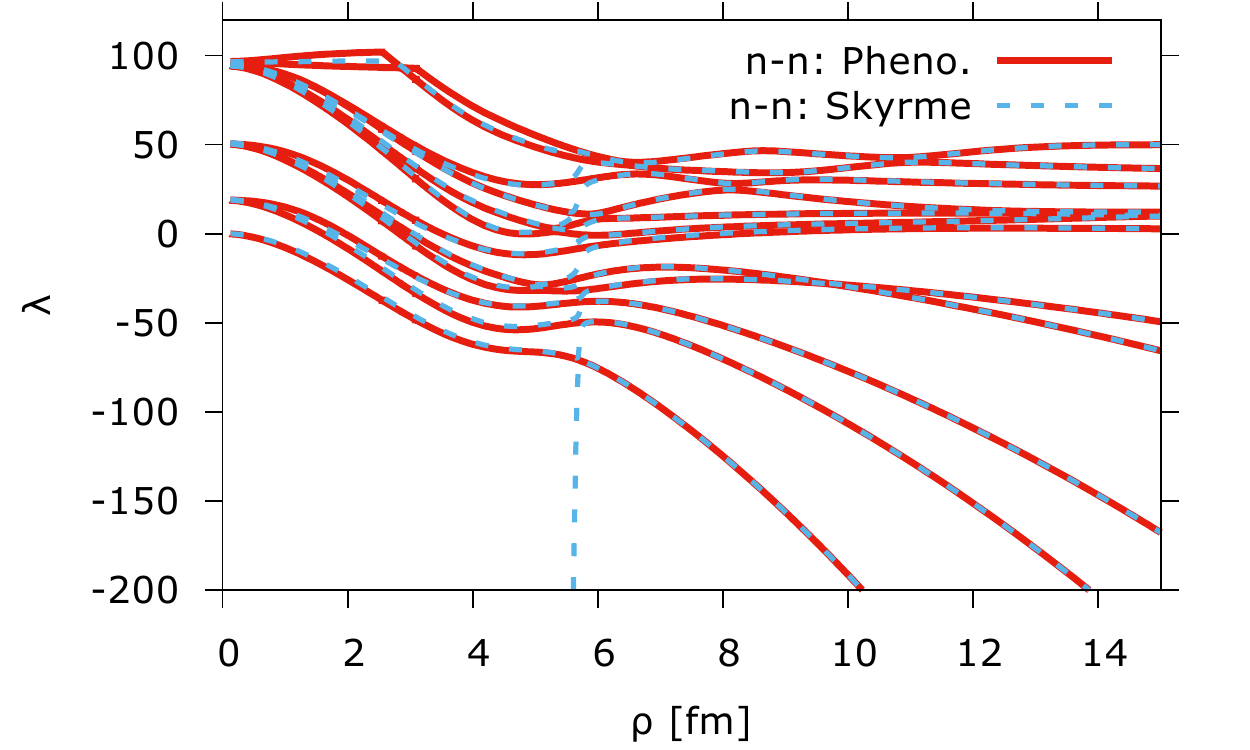}
\caption{Comparison between the raw $\lambda$ spectrum after the first iteration for a phenomenological valence neutron-neutron interaction (solid) and a more consistent interaction which transitions from Skyrme to phenomenological as described in Eq.~(\ref{eq:valInt}) (dashed). The system in question is $^{26}\textnormal{O}$ viewed as $^{24}\textnormal{O}+n+n$.
\label{fig:ValInt}}
\end{figure}

 In particular, we show in Fig.~\ref{fig:ValInt} the $\lambda_n$ spectrum obtained after solving Eq.~(\ref{eq:ang3b}) with potential (\ref{eq:valInt}) and using the transition described in Eqs.~(\ref{wsk}) and (\ref{wph}), with $d_c=0.5$~fm and $R_c = 3.0$~fm. The Skyrme potential $V_{sk}$ has been taken to be the SLy4 Skyrme force \cite{cha98}, whereas 
 the phenomenological nucleon-nucleon potential $V_{ph}$ has been taken from Ref.~\cite{gar99}. The result shown in the figure by the dashed curves corresponds to $^{26}\textnormal{O}$, viewed as $^{24}\textnormal{O}+n+n$, after the first iteration. Also included for comparison in Fig.~\ref{fig:ValInt} is the $\lambda_n$ spectrum obtained from an identical calculation, where the valence neutron-neutron interaction is just the traditional phenomenological interaction \cite{gar99} (solid curves). The spectra are seen to be identical except for a single unphysical $\lambda$ tending towards $-\infty$ as $\rho \rightarrow 0$ for a Skyrme interaction between the valence neutrons.

The infinitely deep $\lambda$ is an example of the Thomas effect caused by the contact interactions in the Skyrme force.  Even though these infinitely many bound states are here clearly supported by the lowest $\lambda$, they are difficult to exclude systematically. The lowest $\lambda$ crosses all other $\lambda$'s at intermediate distances, which makes it impossible to simply remove that specific $\lambda$ outright. The second option would be to use a phase equivalent potential, as discussed in Sec~\ref{sec:pauli}. However, as this $\lambda$ is not tied to a specific state this is not feasible either. 

For these reasons, for the moment the best option is therefore to forget the transition in Eq.~(\ref{eq:valInt}) and simply use a phenomenological interaction instead. Fortunately, as seen in Fig.~\ref{fig:ValInt}, this will have no effect on the relevant $\lambda$ spectrum. Given that the $\lambda$ spectrum dictates every other consecutive three-body calculation, using a phenomenological interaction will have no measurable effect on any of the results.

In the future this could be remedied by replacing the Skyrme force
with a finite-range force such as the Gogny force
\cite{dec80}. Without a contact interaction there would be no Thomas
effect, and the transition in Eq.~(\ref{eq:valInt}) could be
implemented without problems. Another benefit is that Gogny forces are
based on Gaussians, which would make the transition to a (Gaussian)
phenomenological interaction much less drastic.  Then all
nucleon-nucleon interactions in both valence and core space would be
the same.

\subsubsection{The three-body interaction.\label{sec:V3b}}

The final component to be discussed is the three-body interaction, $V_3$, from Eq.~(\ref{eq:sch3}). One option is to proceed as with the three-body interaction in the core, i.e. parameterize it as a density dependent two-body interaction and include it in $V_{v_1v_2}$. In fact, if a Skyrme interaction were used between the valence nucleons, the density dependent $t_3$ term would be derived from a three-body interaction between the two valence nucleons and a sum over the core nucleons. Historically, it has been found that the density dependence was needed in the Skyrme interaction to counteract the  massive and divergence causing $t_0$ term. As such it would not be sensible to parameterize $V_3$ as a density dependent two-body interaction without also using a Skyrme interaction between the valence nucleons.

It should be noted that if a finite-ranged Gogny type interaction were used for the core, it would, as mentioned, be possible to implement this interaction between the valence nucleons. In that case, the three-body interaction between the two valence nucleons and the core nucleons should be reparameterized as a density dependent two-body interaction and included in the $t_3$ term between the two valence nucleons. That would result in a method where the only freedom of any kind was in the choice of Gogny force. 

Instead, at present we use a phenomenological three-body interaction of the form 
\begin{eqnarray}
V_3(\rho) = V_0 \exp\left( - \left( \frac{\rho}{r_0} \right)^2 \right).
\label{eq:genV3}
\end{eqnarray}
This form of the three-body interaction was originally added to the three-body Faddeev formalism to better account for three-body effects related to the Pauli principle and polarization effects \cite{fed96,nob70}. As the Skyrme force is a global force, which is mainly fitted to nuclei near stability, it cannot reasonably be expected to provide predictions with an accuracy on the keV scale. This is in particular true for systems very far from stability at the dripline. The three-body interaction is then a useful free parameter, which can be used for fine-tuning the energy at the keV scale, whereas the structure and long-range behavior is kept completely unchanged.

\begin{table}[t!]
\centering
\caption{The results of choosing the three-body interaction strength $V_0$ before and after the iterations for $^{24}\textnormal{O} + n + n$. Included is the three-body energy $E_3$, the rms size of the system $\langle \rho^2 \rangle^{1/2}$, the weights of the three lowest $\lambda_n$'s in the radial solution, and the weights partial wave contributions. In both cases $r_0 = 6.0$~fm and $V_0 = -6.45$~MeV. The energies are in MeV, the sizes in fm, and the weights and contribution in percent.   \label{tab:VaryV3b}}
\begin{tabular}{*{13}{c}}
\toprule
        &              &                                & \multicolumn{3}{c}{Weights} & \multicolumn{7}{c}{Contribution} \\
\cmidrule(lr){4-6} \cmidrule(lr){7-13}
           &  $E_3$    &  $\langle \rho^2 \rangle^{\frac{1}{2}}$  &  $\lambda_1$ & $\lambda_2$  &  $\lambda_3$ & $s_{\frac{1}{2}}$  &  $p_{\frac{1}{2}}$ & $p_{\frac{3}{2}}$ & $d_{\frac{3}{2}}$ & $d_{\frac{5}{2}}$ & $f_{\frac{5}{2}}$ & $f_{\frac{7}{2}}$ \\
\midrule
Before  & -0.28  &  6.50  &  86 & 14  &  0 & 1  &  1 & 2 & 90 & 3 & 0 & 3 \\
After   & -0.27  &  6.53  &  86 & 14  &  0 & 1  &  1 & 2 & 90 & 3 & 0 & 3 \\
\bottomrule
\end{tabular}
\end{table}

The preservation of the long-range behavior is a given as the Gaussian is short-ranged. The preservation of the structure can be seen by considering the application of the three-body interaction before and after the iterations described in the beginning of Sec.~\ref{sec:implement}. This is illustrated in Table~\ref{tab:VaryV3b}, where the result of adjusting the three-body interaction before and after the iterations is seen for $^{24}\textnormal{O} + n + n$. It should be noted, that the three-body energy has not been adjusted to anything in particular, the intent is only to illustrate the effect of applying the three-body interaction. The table includes the three-body energy, the root-mean-squared (rms) size of the system, the weights of the lowest three $\lambda_n$'s, and the partial wave contributions in the core - valence neutron system. All the weights are identical both for the three $\lambda$'s and the partial wave composition. Given that convergence is here defined as consecutive iterations differing by less than $0.02$~MeV, a difference in energy of $0.01$~MeV is within the uncertainty. The difference on the second decimal of the rms size is then also to be expected.

\section{Experimental verification: \textsuperscript{24}O + n + n\label{sec:24O}}

To test the method derived in the previous sections a suitable system must be identified. This should to a good approximation be a three-body system, which means it should be possible to impose an assumed clusterization. Ideally, the clusters should be weakly bound to make the three-body formalism more viable. As stated in Sec.~\ref{sec:TheoSky} it is assumed that two of the clusters are single, identical nucleons, and that the core is much heavier than the nucleon. It is also assumed that the core is spherical, and consisting of an even number of neutrons and protons.

Our system of choice is $^{26}\textnormal{O}$ viewed as a $^{24}\textnormal{O}$ core and two valence neutrons, which is well suited for several reasons. First of all, $^{24}\textnormal{O}$ is widely believed to be doubly magical and spherical, and also at the very edge of the neutron dripline \cite{hof09}. Even though $^{26}\textnormal{O}$ is ever so slightly on the unbound side, it has recently come within experimental reach \cite{cae13,koh13,kon16}. This provides a few key observables to compare with calculations, while many questions remain unanswered.

\subsection{Fundamental considerations: Iterations and potentials}

At the heart of the hyperspheric, adiabatic expansion of the Faddeev equations lies the hyperangular eigenvalues, $\lambda_n$, from Eq.~(\ref{eq:rad3b}). They are the main connection between the hyperangular and the hyperradial equations, which are the crucial part of the adiabatic expansion. The various coupling terms ($C_{nm}, \, P_{nm}, \, P_{nm}^{\prime}, \, Q_{nm}$, and $Q_{nm}^{\prime}$ in Eq.~(\ref{eq:rad3b})) also connect the two, but should more be considered as corrections to the main quantities determined by $\lambda_n$.

\begin{figure}
\centering
\includegraphics[width=0.7\textwidth]{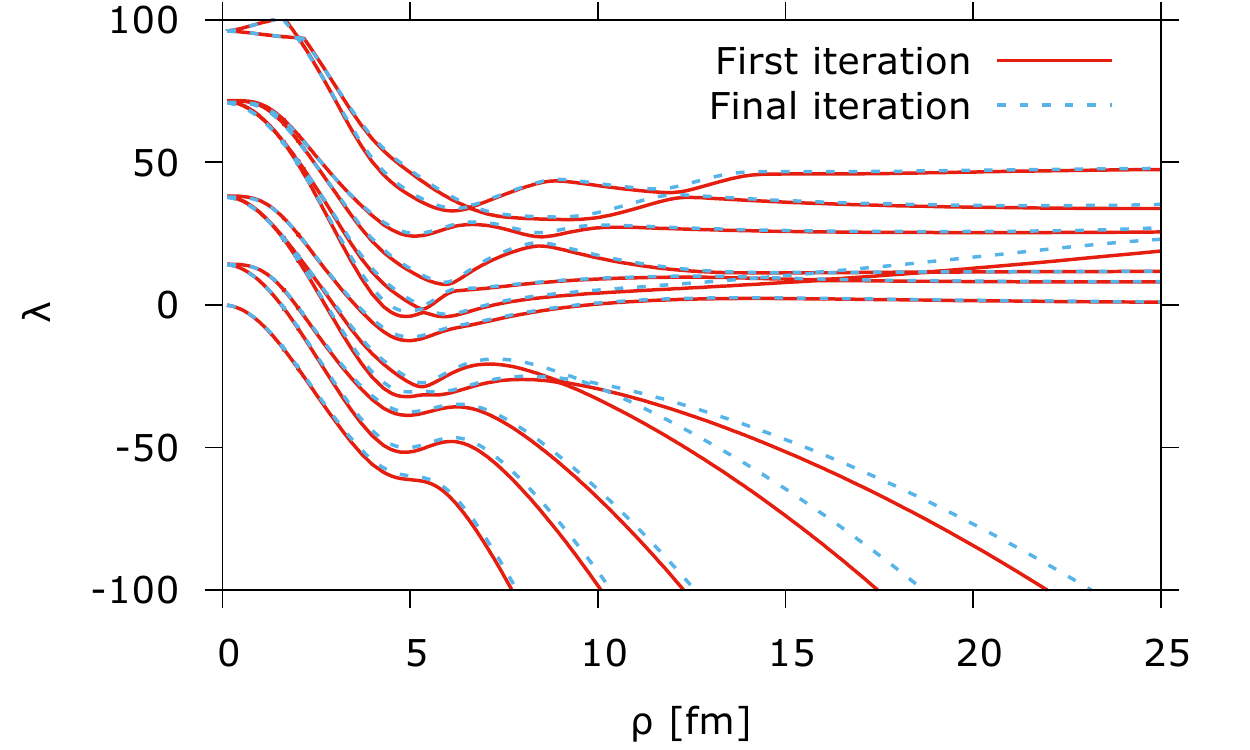}
\caption{The $\lambda$ spectrum for $^{24}\textnormal{O}+n+n$ after the first (red, solid) and last (blue, dashed) iteration using the Skyrme parameterization SkM* \cite{bar82}. The lowest five $\lambda$s correspond to core occupied states, and are therefore Pauli forbidden. Note how the iterative process changes the $\lambda$ spectrum for the occupied states, indicating a change in core structure.
\label{fig:O26Lambda}}
\end{figure}

One of the interesting apects about the method presented here is how the iterative process allows the mean-field structure to affect the three-body calculation, and vice versa. This is seen in Fig.~\ref{fig:O26Lambda}, where the spectrum of hyperangular eigenvalues is presented as a function of hyperradius for the first and last iteration. The calculation is for $^{24}\textnormal{O} + n + n$ using the Skyrme parameterization known as SkM* \cite{bar82}. The interpretation of the $\lambda$ spectrum is the same as for a regular three-body calculation, namely that $\lambda$'s which tend to $-\infty$ as $-\rho^{2}$ correspond to bound states in the two-body systems. The energy of the bound states is given by the large distance asymptotic behavior of $\lambda$ by
\begin{eqnarray}
E_{2b} = \frac{\hbar^2}{2m} \frac{\mu_{jk}}{m} \frac{\lambda}{\rho^2}
\end{eqnarray}
where $m$ is the normalizing mass, and $\mu_{ij}$ is the reduced mass.

The 16 lowest neutron levels, corresponding to the 1$s_{1/2}$, 1$p_{3/2}$, 1$p_{1/2}$, 1$d_{5/2}$, and 2$s_{1/2}$ states, are occupied by the core neutrons and therefore Pauli forbidden. The change in the $\lambda$'s corresponding to these five states in Fig.~\ref{fig:O26Lambda} reflects the change in core structure due to the valence neutrons. As discussed in Sec.~\ref{sec:pauli} the Pauli forbidden states are removed either by use of phase equivalent potentials or by simply eliminating the corresponding decoupled $\lambda$'s. As the core-occupied states clearly are decoupled, the simplest solution is here to eliminate the corresponding $\lambda$'s.

The effect of the iterative process is also seen in the changes of the three-body energy. In Table \ref{tab:O26iter} the three-body energy for each iteration in a $^{24}\textnormal{O}+n+n$ calculation is seen for three different parameterizations of the Skyrme forces known as SkM* \cite{bar82}, SLy4 \cite{cha98}, and Sk3 \cite{bei75}. The calculations are done with the three-body interaction from Eq.~(\ref{eq:genV3}) using $V_0 = -8$~MeV and $r_0 = 6$~fm in all three instances. Unless otherwise stated convergence is defined as when the three-body energies of two consecutive iterations differ by less than $0.02$~MeV, but other criteria could be chosen, depending on the focus of the investigation. Here, the energy was chosen because it is a fairly direct representation of the change between the iterations.

\begin{table}[tb]
\centering
\caption{The changes in three-body energy, $E_{3}$, in MeV between iterations for the three Skyrme forces SkM*, SLy4, and Sk3. The energy is said to have converged when the difference in three-body energy between consecutive iterations is less than 0.02~MeV. The three-body interaction strength ($V_0$) has not been adjusted to anything in particular. It is -8 MeV for all forces and iterations, with $r_0 = 6$~fm.
\label{tab:O26iter}}
\begin{tabular}{l *{4}{c} }
\toprule
                           & \multicolumn{4}{c}{$E_{3B}$}  \\
                                 \cmidrule(lr){2-5}
               &   1    &   2   &  3    &  4   \\
\midrule
SkM*           & -1.75 & -1.06 & -1.12 & -1.14 \\
Sk3            & -0.31 & -0.07 & -0.10 & -0.11 \\
SLy4           & -0.93 & -0.85 & -0.87 &       \\
\bottomrule
\end{tabular}
\end{table}

In Table \ref{tab:O26iter} it should first of all be noted how fast the convergence is, which means the added computation time from the iterations is rather limited. Secondly, there is a relatively large difference between the first and the second, compared to the later, iterations. This is because the mean-field in the first iteration by choice cannot include the effect of the valence neutrons. In the second iteration the effect of the nucleons are felt, and a significant change in energy is seen to occur.

Skyrme forces have over the years  been shown to be very successful effective two-body interactions in mean-field calculations of ground state properties of nuclei. The parameters are usually fitted to reproduce nuclear matter properties, as well as binding energies and root-mean-square charge radii of doubly magic nuclei. However, different parameterizations can satisfy these requirements, and as a result, multiple Skyrme forces are found in the literature. It is therefore not surprising that the final energies in Table \ref{tab:O26iter} vary by slightly more than 1~MeV for the three Skyrme parameterizations used here. The Sk3 parameterization is an example of one of the earliest and most simple, but also very successful parameterizations. It contains a linear dependence on the density ($\alpha = 1$) and $x_0=1$ as the only $x_i$ parameter. The SkM* parameterization is more elaborate, and originally constructed to improve $E0$ and $E1$ giant resonances, as well as fission barriers. Finally, the SLy4 parameterization is a more recent example of Skyrme forces being revisited with the aim of improving the isospin properties of nuclei away from the valley of stability. As such, it should be well-suited as a starting point for many of our investigations along the nucleon driplines.

\begin{figure}
\centering
\includegraphics[width=0.7\textwidth]{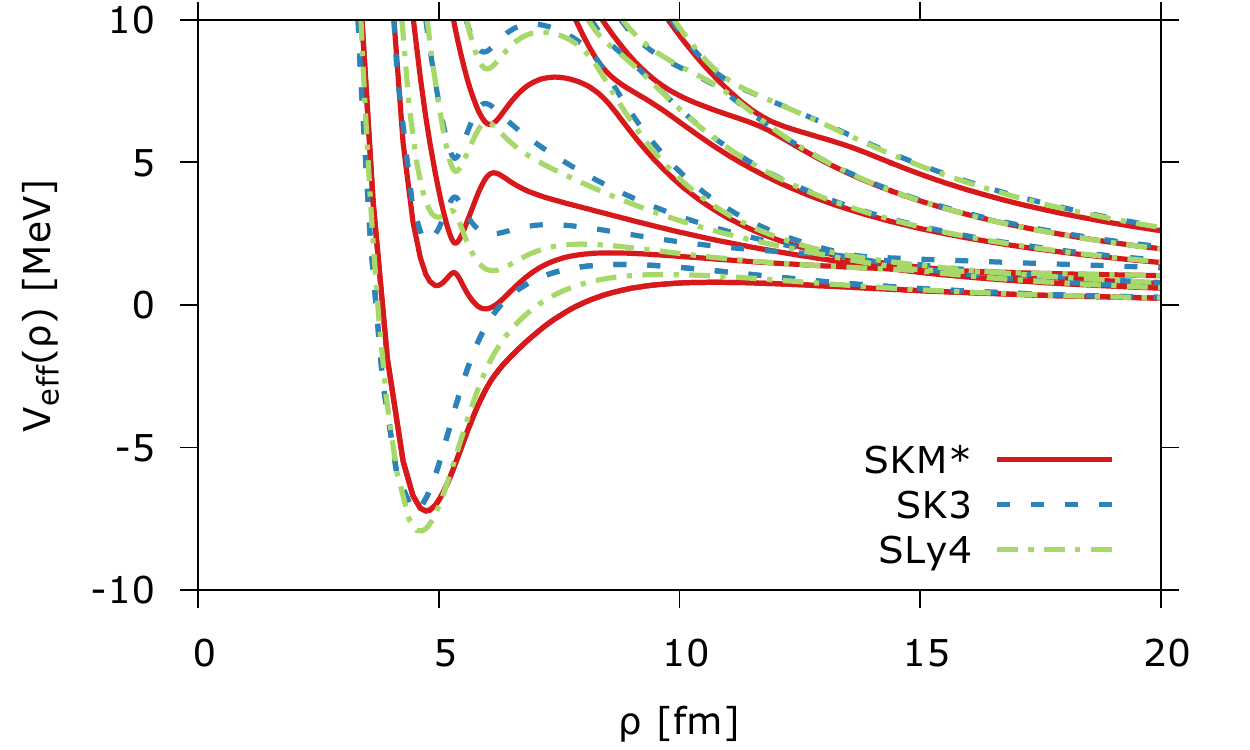}
\caption{The effective potentials for $^{24}\textnormal{O}+n+n$ using the Skyrme parameterizations known as SkM* \cite{bar82} (solid, red), Sk3 \cite{bei75} (dashed, blue), and SLy4 \cite{cha98} (dashed-dotted, green). The core occupied, Pauli forbidden states have been removed by excluding the corresponding effective potentials. This does not include three-body potentials and couplings such as $Q_{nn}$. Note the overall similarity between the potentials, where the main difference is at the second effective potential at intermediate distances.
The results for SLy4 are taken from Fig.1 of Ref.~\cite{hov17}.
\label{fig:O26EffPot}}
\end{figure}

However, the apparent differences between the results of the various Skyrme parameterizations are mainly an overall displacement and not a structural difference. This is seen in Fig.~\ref{fig:O26EffPot}, where the effective potentials based on the Skyrme parameterizations from Table \ref{tab:O26iter} are shown for $^{24}\textnormal{O}+n+n$. Neglecting coupling terms and the three-body interaction to better compare the parameterizations directly, the effective diagonal potentials can be seen from Eq.~(\ref{eq:rad3b}) to be
\begin{eqnarray}
V_{eff}(\rho) = \frac{\hbar^2}{2 m} \frac{\lambda_n(\rho) + 15/4}{\rho^2},
\end{eqnarray}
where $m$ is the normalization mass.

The $\lambda$'s corresponding to core occupied, Pauli forbidden states are not included in Fig.~\ref{fig:O26EffPot}. It is seen that all three parameterizations form a pocket at roughly the same distance and depth, capable of sustaining a three-body state. The differences in shape and width of this pocket is responsible for the variations in energies seen in Table \ref{tab:O26iter}. The higher-lying effective potentials have the same qualitative structure, but differ slightly in their exact placements. Importantly, the asymptotic behavior of the potentials are basically identical, which ensures very similar long-range structures and dynamics.

\subsection{Reproducing experimental values}

Given that $^{25}\textnormal{O}$ and $^{26}\textnormal{O}$ have recently come within experimental reach a quantitative examination and comparison can be performed. The lowest-lying resonance in $^{25}\textnormal{O}$ has been measured to be a $d_{3/2}$ resonance at $0.749(10)$~MeV with a width of $88(6)$~keV \cite{kon16}. Likewise, the ground-state of $^{26}\textnormal{O}$ is measured to be a $0^+$ state at $0.018(3)$~MeV \cite{kon16}. In the same study the half-life was deduced to be $T_{1/2} \sim 0.01 - 1$~fs.

The observed ground-state of $^{26}\textnormal{O}$ can be used to fix the three-body interaction, which is the only freedom remaining within this method, after a specific Skyrme parameterization has been chosen. Every other calculation and observable are completely determined by these choices. This is in contrast to most phenomenological models, where the two-body resonance state would also be used as an input parameter \cite{hagino16}. In Table \ref{tab:O26Energies} the predicted energy of the $d_{3/2}$ resonance in $^{25}\textnormal{O}$ is shown for the three Skyrme parameterizations, along with the energy predicted using regular mean-field Hartree-Fock (HF) calculations with the same parameterizations. In addition, the half-lifes resulting from the calculated $d_{3/2}$ energies have also been included. Finally, as the various Skyrme parameterizations produce different energies, it is necessary to use different three-body interaction strengths, to achieve total three-body energy of $0.018$~MeV. These strengths are also included in Table \ref{tab:O26Energies}.

All four rows in Table \ref{tab:O26Energies} deserve attention. First of all, it is interesting to note that the three-body interaction strengths for SLy4 and SkM* are almost identical despite the final energies in Table \ref{tab:O26iter} differing by almost $0.3$~MeV. The different response to the three-body interaction is caused by the slight difference in potential shape, as seen in Fig.~\ref{fig:O26EffPot}. Despite this, the produced $d_{3/2}$ resonance energies are basically identical for the SLy4 and the SkM* parameterization. Not only that, but they are within $0.1$~MeV of the experimental value. It is not so surprising that the Sk3 prediction is slightly further off, given that it is one of the simpler, earlier parameterizations, fitted to a much more limited data set, close to stability. It should also be noted that the fine-tuning with the three-body interaction just shifts the three-body energy slightly. It does not affect the structure or the two-body interaction, as discussed in Sec.~\ref{sec:V3b}. The very accurate reproduction of the $d_{3/2}$ energy level is therefore not a result of the three-body interaction. It is a direct consequence of the method, with no remaining freedom after the Skyrme parameterization has been chosen.

\begin{table}
\centering
\caption{This table is divided according to the three different Skyrme forces used. The three-body energy has been adjusted to the experimental value of 0.018 MeV using the three-body interaction strength $V_0$, but keeping the range constant at $r_0 = 6$~fm. All presented results are after the final iteration. Included is the energy of the first $d_{3/2}$ resonance state in $^{25}$O computed using our method and compared to the result of traditional HF calculations. This should be compared with the experimental value of $0.749(10)$~MeV \cite{kon16}. The half-life $T_{1/2}$ is calculated using a WKB approximation for the tunneling probability, and harmonic oscillator approximation for the knocking rate. The experimental value from Ref.~\cite{kon16} is $\sim$ 0.01 -- 1 fs. All energies are in MeV and all lengths are in fm.
\label{tab:O26Energies}}
\begin{tabular}{l *{3}{c} }
\toprule
                                                        &       SLy4     &      Sk3             &       SkM*        \\
\midrule
$V_0$                                           & -5.71          & -7.71        &  -5.72                \\
$E_{d_{3/2}} (Our)$             &  0.85          &  1.23        &       0.83            \\
$E_{d_{3/2}} (HF)$                      & -0.96          & -0.53        &  -1.15                \\
$T_{1/2}\,\left[ fs \right]$& 0.4                &  0.6         &   0.2                 \\
\bottomrule
\end{tabular}
\end{table}

It is then very interesting to see, that regular mean-field Skyrme calculations with the same parameterizations yield predictions that are not only far from the experimental value, but also clearly bound. Mean-field calculations are known to become inaccurate, when approaching the driplines, so an erroneous result is to be expected. What is interesting is that the two-body potential, which produces the $d_{3/2}$ resonance energy in our method, is provided by the core calculation. This indicates that the combination of the three-body approach with the self-consistent mean-field calculation provides a small, but crucial change to the effective potential.

The fourth and final row in Table \ref{tab:O26Energies} is the half-life of $^{26}\textnormal{O}$ calculated using a simple WKB approximation for the tunneling probability, and a harmonic oscillator approximation for the knocking rate. Despite the simple approximations, the calculated half-life is right in the middle of the experimental interval. Given that the half-life depends exponentially on the barrier height and thickness this is a remarkably good agreement.

\begin{figure}
\centering
\includegraphics[width=0.7\textwidth]{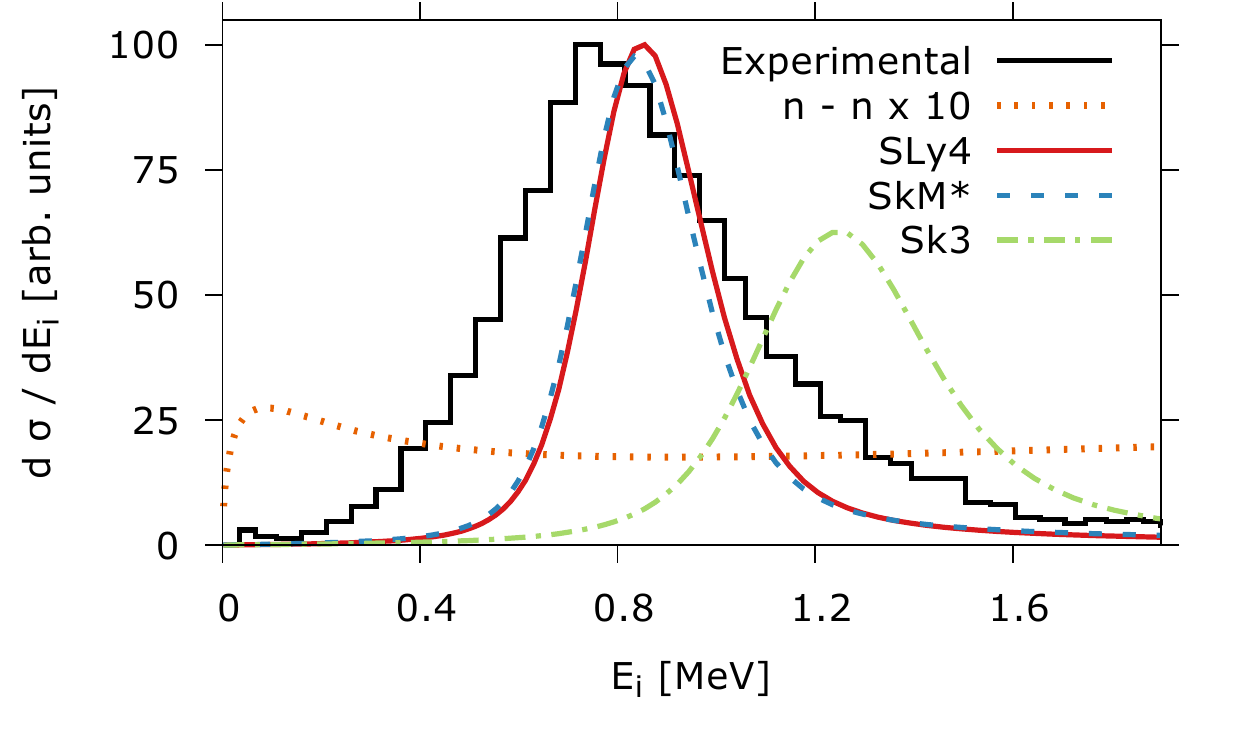}
\caption{The invariant mass spectrum of the core-neutron two-body system for $^{24}\textnormal{O}+n+n$ with the SLy4 (solid, red), SkM$^{\ast}$ (dashed, blue), and Sk3 (dashed-dotted, green) Skyrme parameterizations as a function of the two-body energy. The SLy4 neutron-neutron (dotted, orange) invariant mass spectrum is also included, scaled by a factor of 10 to make it visible. The black curve is the measurements from Ref.~\cite{kon16}. The SLy4 and SkM$^\ast$ curves are taken from Fig.4 of Ref.~\cite{hov17}.
\label{fig:O26Invar}}
\end{figure}

The predicted energy for the $d_{3/2}$ state is derived from the invariant mass spectrum seen in Fig.~\ref{fig:O26Invar}. For the core-neutron system the invariant mass, $E_{core-n}$, can be interpreted as the kinetic energy of this system in the final state (after removal of a neutron), where $\bm{p}_{core} + \bm{p}_n = 0$, with $\bm{p}_{core}$ and $\bm{p}_n$ being the core and neutron three-momentum, respectively \cite{gar97}. The resonance energy is then defined as the energy, where this spectrum has a maximum.

The invariant mass spectrum is particularly useful for two reasons. First, it is a direct calculation of the resonance energies, which can otherwise be difficult to calculate when there is no barrier to enable bound-state approximations. And second, it is a physical observable allowing for direct comparison between theory and experiment \cite{zin97}. In Fig.~\ref{fig:O26Invar} the experimental data from Ref.~\cite{kon16} is seen alongside the three Skyrme parameterizations. Unfortunately, the experimental beam profile is not available (and therefore not taken into account), which makes the calculated width slightly more narrow than the experimental one \cite{gar97}. Also included is the neutron-neutron invariant mass spectrum, which is structureless as there are no neutron-neutron resonances.

\subsection{Further predictions}

Technically, the core-nucleon invariant mass spectrum is calculated from the core-nucleon phase-shifts through the cross section under the so-called sudden approximation \cite{gar97,gar96,gar01,gar99breakup}. The sudden approximation is the assumption that a high-energy beam hits a target such that one constituent particle is removed instantaneously from the three-body system. Instantaneous here means that the reaction times are much shorter than the time for motion of the particles in the three-body system. As the focus here is on systems with very weakly bound valence nucleons this requirement is fulfilled even for moderate beam-energies.

Given that there is a final state interaction between the two remaining particles after the sudden removal of the third, the two-body wave function will be distorted. In keeping with the convention for three-body Jacobi coordinate systems the relative and total momentum of the two-body system with respect to the center-of-mass of the three-body system is $\bm{k}_x$ and $\bm{k}_y$, respectively. The probability, $P$, of finding the remaining two particles with momentum $(\bm{k}_x, \bm{k}_y)$ is proportional to the transition matrix element
\begin{eqnarray}
P^{JM}_{s_x \sigma_x}(\bm{k}_x,\bm{k}_y) \propto \braket{e^{i \bm{k}_y \cdot \bm{y}} w_{s_x \sigma_x}(\bm{k}_x,\bm{x}) | \psi^{JM}(\bm{x},\bm{y})},
\end{eqnarray}
where $J$ and $M$ are the total spin and projection of the halo nucleus, while $s_x$ and $\sigma_x$ are the spin and projection of the two-body final state, and $w$ is the distorted two-body wave function \cite{gar96}. The cross section (or momentum distribution) is then proportional to the square of the transition matrix element, averaged over initial states $(M)$, and summed over final states $(s_x, \sigma_x)$
\begin{eqnarray}
\frac{d^6 \sigma}{d \bm{k}_x d \bm{k}_y} \propto \sum_M \sum_{s_x, \sigma_x} | P^{JM}_{s_x \sigma_x}(\bm{k}_x,\bm{k}_y) |^2.
\label{eq:3b_crossSec}
\end{eqnarray}

The earlier interpretation of the invariant mass as the kinetic energy in the final two-body system can be expressed more explicitly as
\begin{eqnarray}
E_{core-n} = \left( (E_{core} + E_n)^2 + c^2 (\bm{p}_{core} + \bm{p}_n)^2 \right)^{1/2} - \left( M_{core} + M_{n} \right) c^2,
\end{eqnarray}
where $E_i$ is the energy, $\bm{p}_i$ the momentum, and $M_i$ the rest mass of the individual particles. Based on this, it can be shown \cite{gar97} that the invariant mass spectrum, defined as $d \sigma/d E_{core-n}$, is related to the cross section as
\begin{eqnarray}
\frac{d \sigma}{d E_{core-n}} = \frac{m (M_{core} + M_n)}{M_{core} M_n} \frac{E_{core} + E_n}{E_{core} E_n} \frac{1}{k_x} \frac{d \sigma}{d k_x},
\end{eqnarray}
where the cross section from Eq.~(\ref{eq:3b_crossSec}), integrated over unobserved parameters, results in $d\sigma / dk_x$.

\begin{figure}
\centering
\includegraphics[width=1\textwidth]{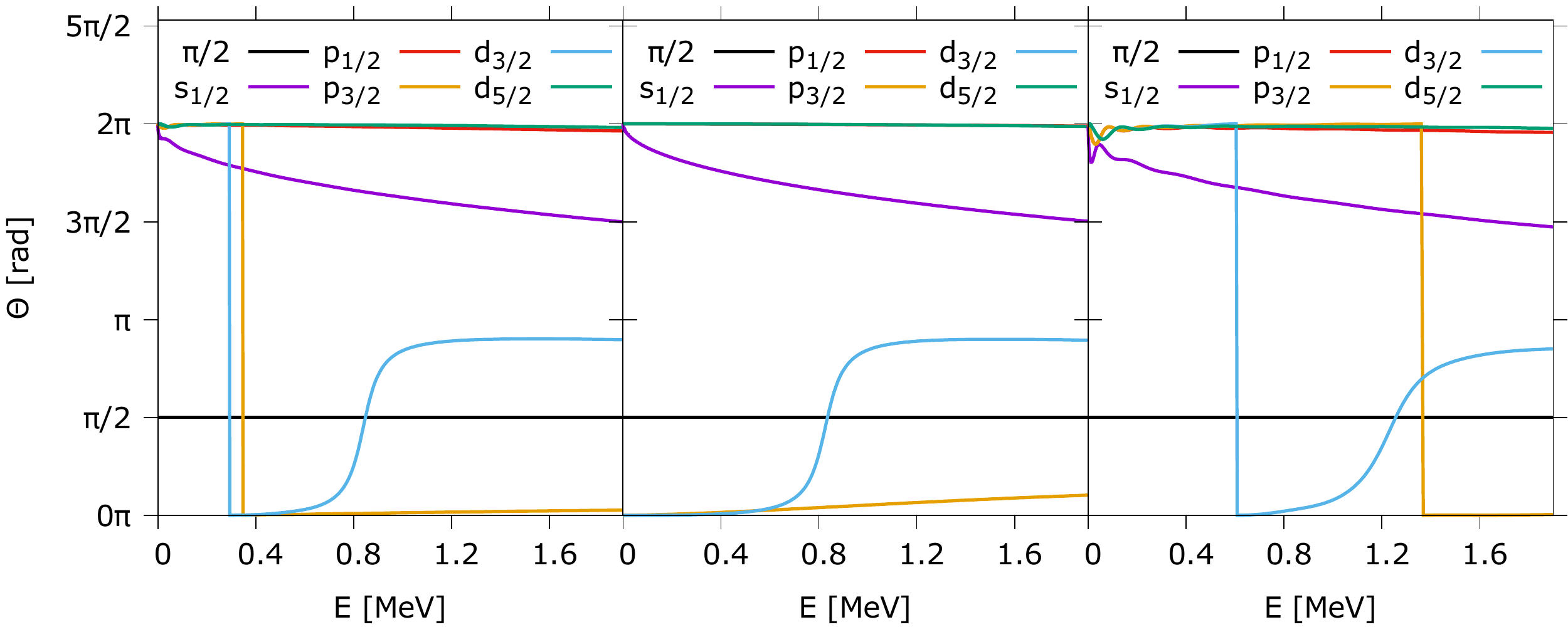}
\caption{The phase shifts for the core-neutron two-body system in $^{24}\textnormal{O}+n+n$ for the $s$-, $p$-, and $d$-states for SLy4 (left), SkM* (middle), and Sk3 (right). The horizontal black line at $\pi/2$ is included to guide the eye. The vertical transitions do not indicate a resonance, but are a result of the phase shifts being calculated in the range $0$ to $2\pi$.
\label{fig:O26PhS}}
\end{figure}

The invariant mass spectrum therefore depends on this cross section, which is determined by the distorted wave function. As the distorted wave function depends crucially on the phase shifts, the energy structure seen in Fig.~\ref{fig:O26Invar} could also be deduced from the phase shifts themselves. These are seen in Fig.~\ref{fig:O26PhS}, where the crossing of $\pi/2$ indicates a resonance state. The phase shifts are here calculated modulo $2\pi$, which means the vertical lines are not resonance states. The decomposition into individual partial waves clearly shows that there are no other possible resonances in the low energy region, and it also shows the state to be a $d_{3/2}$ state in accordance with the experimental observations.

Various other interesting observables can be extracted from the cross section in Eq.~(\ref{eq:3b_crossSec}). The momentum distribution of the fragments after a breakup reaction is experimentally a very direct look into the structure of the system. The main difficulty in interpreting such measurements is that there is often a mixture of effects stemming from both the original structure of the system and the reaction mechanism itself, and it is not easy to disentangle these contributions. However, if the third particle is removed suddenly, the structure of the remaining two particles is unaffected by the reaction mechanism.  Under the sudden approximation, the remaining particles are released with momentum distributions equal to the distribution in the initial wave function.

\begin{figure}
\centering
\includegraphics[width=0.7\textwidth]{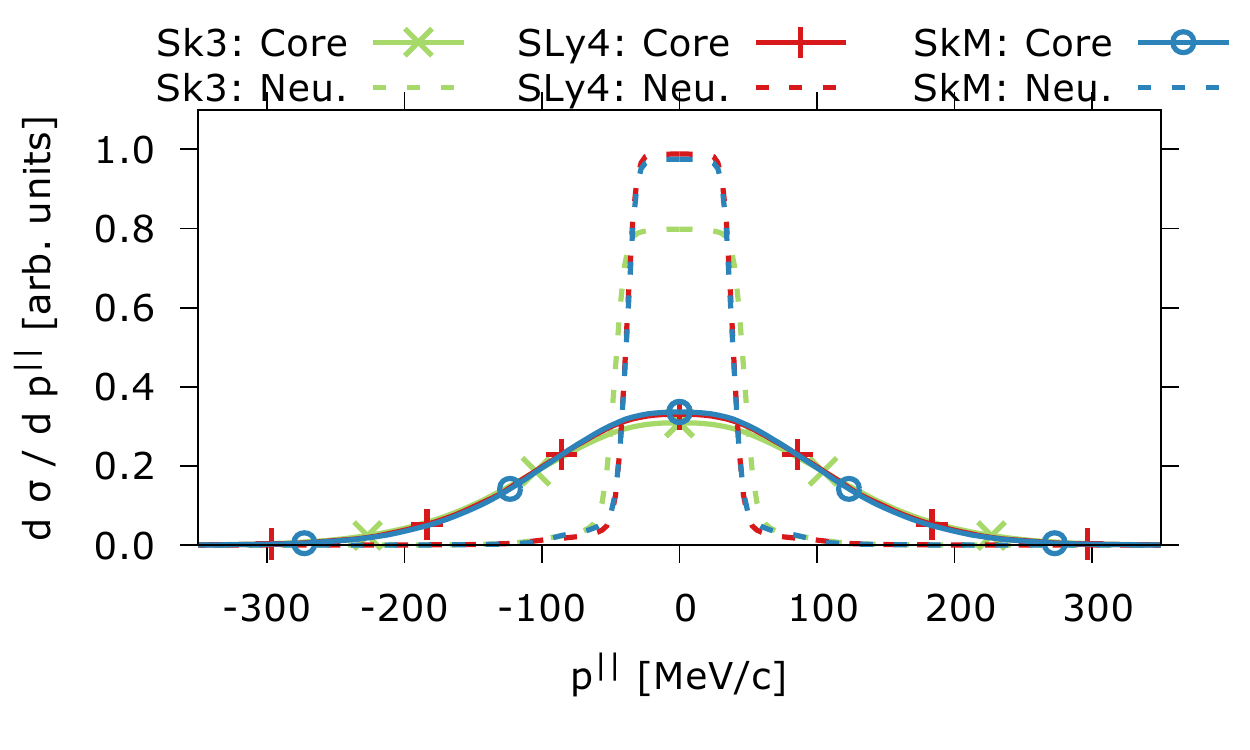}
\caption{The longitudinal momentum distributions of the $^{24}$O core (solid) and neutron (dashed) after knockout of one neutron calculated using the SLy4 (red), Sk3 (green), and SkM (blue) Skyrme parameterizations.
\label{fig:O26MomDist}}
\end{figure}

\begin{figure}
\centering
\includegraphics[width=0.7\textwidth]{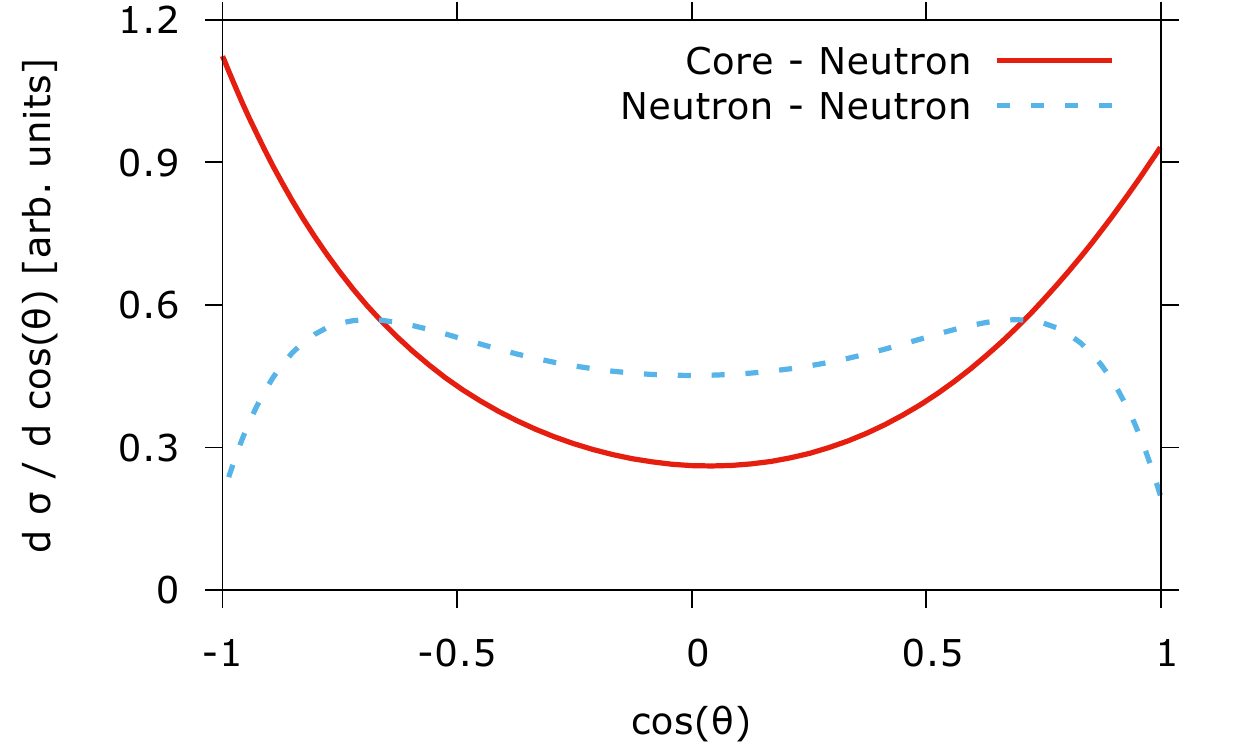}
\caption{The angular distribution of $^{24}\textnormal{O}+n+n$ between directions of neutron and core-neutron, and core and neutron-neutron momenta, respectively. The results are for the SLy4 Skyrme force.
\label{fig:O26AngDist}}
\end{figure}

The longitudinal momentum distribution for both the $^{24}\textnormal{O}$ core and the remaining neutron in $^{24}\textnormal{O} + n +n $ are shown in Fig.~\ref{fig:O26MomDist} after removal of one of the neutrons.  The full width at half maximum (FWHM) for the core distribution is about 200 MeV/c. In contrast, the FWHM for the neutron distribution is only about 100 MeV/c, which reveals the moderate neutron halo structure allowed in $d$-waves. The relatively restrained spatial configuration leads to a more extended configuration in momentum space. The almost identical distributions resulting from the various Skyrme parameterizations once again demonstrate how independent of the specific parameterization the results presented here are.

Further structural insights are provided by the angular distribution seen in Fig.~\ref{fig:O26AngDist}, where both the angular distribution after removal of a neutron (core-neutron system) and the angular distribution after removal of the core (neutron-neutron system), are shown. These angular distributions are again derived from Eq.~(\ref{eq:3b_crossSec}) by integrating over unobserved parameters. In the angular distribution for the core-neutron system the angle is between the $^{25}\textnormal{O}$ momentum and the relative momentum between neutron and core after fragmentation, while the angle is between the center-of-mass momentum and the relative momentum of the two neutrons after fragmentation. Angular distributions are very sensitive to the mixing of various partial waves, and can therefore be used to probe the structure of the system \cite{gar02}. In Fig.~\ref{fig:O26AngDist} the asymmetric parabola is again an indication of the dominating $d$-wave in the state. Neither the momentum distribution, nor the angular distribution has been measured in $^{26}\textnormal{O}$ so far, but both Figs.~\ref{fig:O26MomDist} and \ref{fig:O26AngDist} are predictions for the observables of a state that lives sufficiently long for a knockout process to be initiated.

\section{Clusters and Efimov: \textsuperscript{70}Ca + n + n\label{sec:70Ca}}

The specific parameters in the Skyrme forces are fitted to experimental data
on stable nuclei. Naturally, this makes any prediction based on such forces
more reliable closer to stability and in areas where changes from
nuclei to nuclei is more gradual \cite{hov13}. This is very clearly
reflected in the uncertainty of the predicted positions of the dripline based on various
parameterizations. This uncertainty is inherent in any mean-field
prediction reaching outside the experimentally known region. Considering
for instance the calcium isotope chain, the neutron dripline is usually
thought to be in the range $A=68-76$ \cite{erl11,bha05}. With the SLy4
parameterization \cite{cha98} $^{70}\textnormal{Ca}$ is the last bound
isotope, while $^{72}\textnormal{Ca}$ ($^{70}\textnormal{Ca}+n+n$)
can be bound in the following three-body calculation with three unbound
subsystems, depending on the three-body interaction.

The main point of interest in this section is the neutron dripline for
calcium isotopes studied from a new and systematic
perspective with the method presented here~\cite{hov17b}. Given that
Skyrme parameters are fitted to observable quantities of experimentally
well-known nuclei, existing parameterizations cannot be expected to
present accurate predictions well into experimentally unknown regions
without alterations. This fact will here be exploited to examine how
nuclear configurations evolve as the dripline is approached. Specifically,
it will be examined how halo structures and possibly even Efimov states
could appear close to the dripline, and whether this would be reflected
in observable long-distance structures.

With the traditional ordering of the single-particle levels
$^{70}\textnormal{Ca}+n+n$ is an ideal choice to study halo and
possible Efimov states. The 50 neutrons in $^{70}\textnormal{Ca}$
would traditionally exactly fill the $g_{9/2}$ state, making it
a very strong magic number, and leaving $s_{1/2}$ and $d_{5/2}$
as the nearest unoccupied states for the two valence neutrons. In
particular, the available $s_{1/2}$ is interesting as sufficiently
low binding in this state should lead to the formation of halos
\cite{jen04,rii92,mis97,zho90}. It is also a necessary requirement for the
formation of Efimov states, along with an extremely large $s$-wave
scattering length \cite{efi70,fed93,jen03,gar06,maz97,maz00,zho94,zho97}.

To allow for systematic continuous variation, a specific Skyrme
parameterization---the SLy4---is chosen and the overall $t_i$ parameters
are scaled as $t_i \rightarrow S \, t_i$, while the $x_i$ and $W_0$
parameters are left unchanged. Scaling identically the $t_i$ parameters
amounts roughly to an overall scaling of the potential itself, approaching
the dripline as needed, while the structure dictated by $x_i$ and $W_0$
is left unchanged. The SLy4 parameterization is chosen as a baseline
as it was originally designed to improve predictions for nuclei far
from stability.

Far from stability the magic numbers are less sharply defined, they
could change slightly, or there could be inversion of
the ordering in $gds$ shell \cite{hag13,men02}. However, an inversion,
placing the $g_{9/2}$ state above the $s_{1/2}$ state and making
$^{60}\textnormal{Ca}$ the last bound calcium isotope, as suggested by
some recent coupled-cluster calculations \cite{hag13}, would have a very
minor effect on any conclusion drawn here. This would change only the
neutron-core potential by the contribution from the $g_{9/2}$ state. As
this is fully occupied in the $^{70}\textnormal{Ca}$ calculation, it does
not affect strongly the valence nucleons. The essential part in both cases
is the unoccupied $s_{1/2}$ state.

\subsection{Short-distance structure}

\begin{figure} \centering
\includegraphics[width=0.7\textwidth]{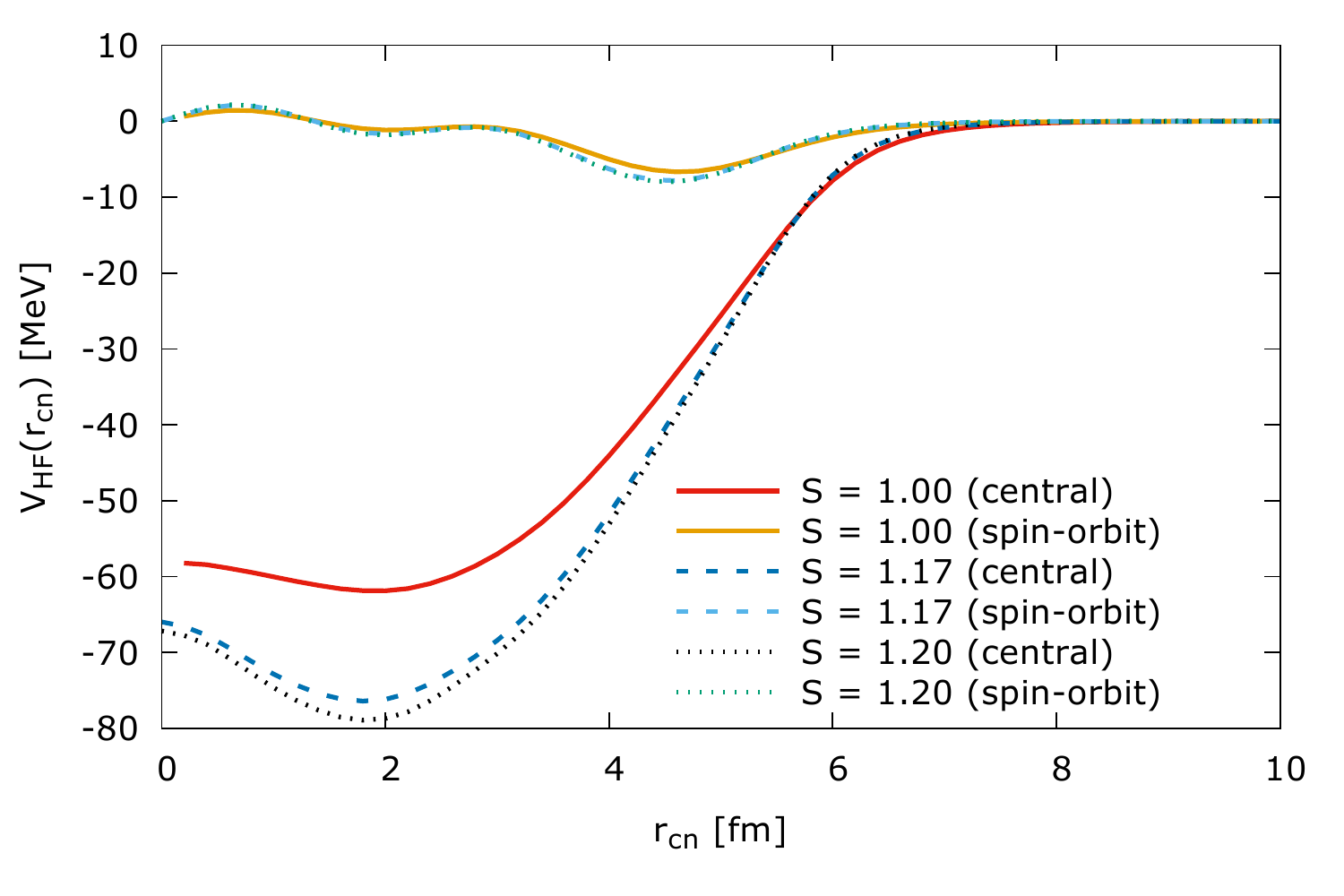} \caption{The
central and spin-orbit part of the self-consistent mean-field core-neutron
potential for $^{70}\textnormal{Ca} + n + n$ calculated for three
different scalings of the Skyrme parameters $(t_i \rightarrow S t_i)$
using the SLy4 parameterization as a baseline.  \label{fig:Ca72HFPot}}
 \end{figure}

Scaling the Skyrme parameters in a $^{70}\textnormal{Ca}+n+n$
calculation as $t_i \rightarrow S t_i$ for the SLy4 parameterization
with the factor $S=1.17$ and $1.20$, in addition to $S=1.00$, yields
the core-valence neutron central and spin-orbit potential shown in
Fig.~\ref{fig:Ca72HFPot}. These specific scaling values are chosen for
the interesting properties of the resulting potentials in relation to
halo formations and Efimov physics, which will be discussed later. The
important point is that for $S=1.17$ there is a near degeneracy
between the relevant two-body energies $E_{s_{1/2}} = 0.017$~MeV and
$E_{d_{5/2}} = -0.026$~MeV, but for $S=1.20$ the relevant two-body energies
are $E_{s_{1/2}} = 2.4 \cdot 10^{-4}$ and $E_{d_{5/2}} = -0.32$, all
in MeV. For values of $S$ below $1.17$ the ordering is opposite, and
$s_{1/2}$ is above $d_{5/2}$. 

An interesting complication to be studied here is therefore how the value
of $S$ affects the energy of the $d_{5/2}$ state relative to the $s_{1/2}$
state. The inversion between the $s$- and $d$-level with $S$ makes it
possible to study the conditions for halo structures. It
is suspected that there might be a near degeneracy among the $s$-, $d$-,
and possibly even $g$-orbitals around the very neutron-heavy end of the
calcium isotope chain \cite{hag13,men02}. Any realistic investigation
of these isotopes must therefore be able to account for
such a degeneracy, in particular with respect to halos and Efimov states
as these depend so crucially on the characteristics of the $s$-orbital.

A clear difference is seen in Fig.~\ref{fig:Ca72HFPot} for the
central part, which becomes more attractive as the potential initially
is attractive, and the factor is larger than one. However
the spin-orbit potential is basically unchanged indicating the
structure of energy levels is mostly unchanged, they are only shifted
down.
         
The structure of a three-body system is fundamentally determined by
two aspects: the two-body and the three-body interactions. Assuming
a given valence neutron-neutron interaction, here the phenomenological
interaction from Ref.~\cite{gar99}, the crucial quantities for the
formation of cluster structures are the core-valence neutron interaction
and the overall binding of the three-body system. The core-valence
neutron interaction is the SLy4 parameterization of the Skyrme force,
scaled as specified. However, the binding of the three-body system is
not completely determined by that interaction alone, as there remains a
degree of freedom in the three-body interaction.

As previously a Gaussian three-body interaction is used of the type
specified in Eq.~(\ref{eq:genV3}), where the range is kept constant at
$\rho_0 = 8$~fm, and the strength $V_0$ is used to adjust the three-body
energy. Changing the three-body energy without changing the underlying
two-body interactions makes it possible to see how and under which
circumstances halo structures and possibly Efimov states appear for a
given two-body interaction.

\begin{figure}
\centering
\includegraphics[height=0.92\textheight]{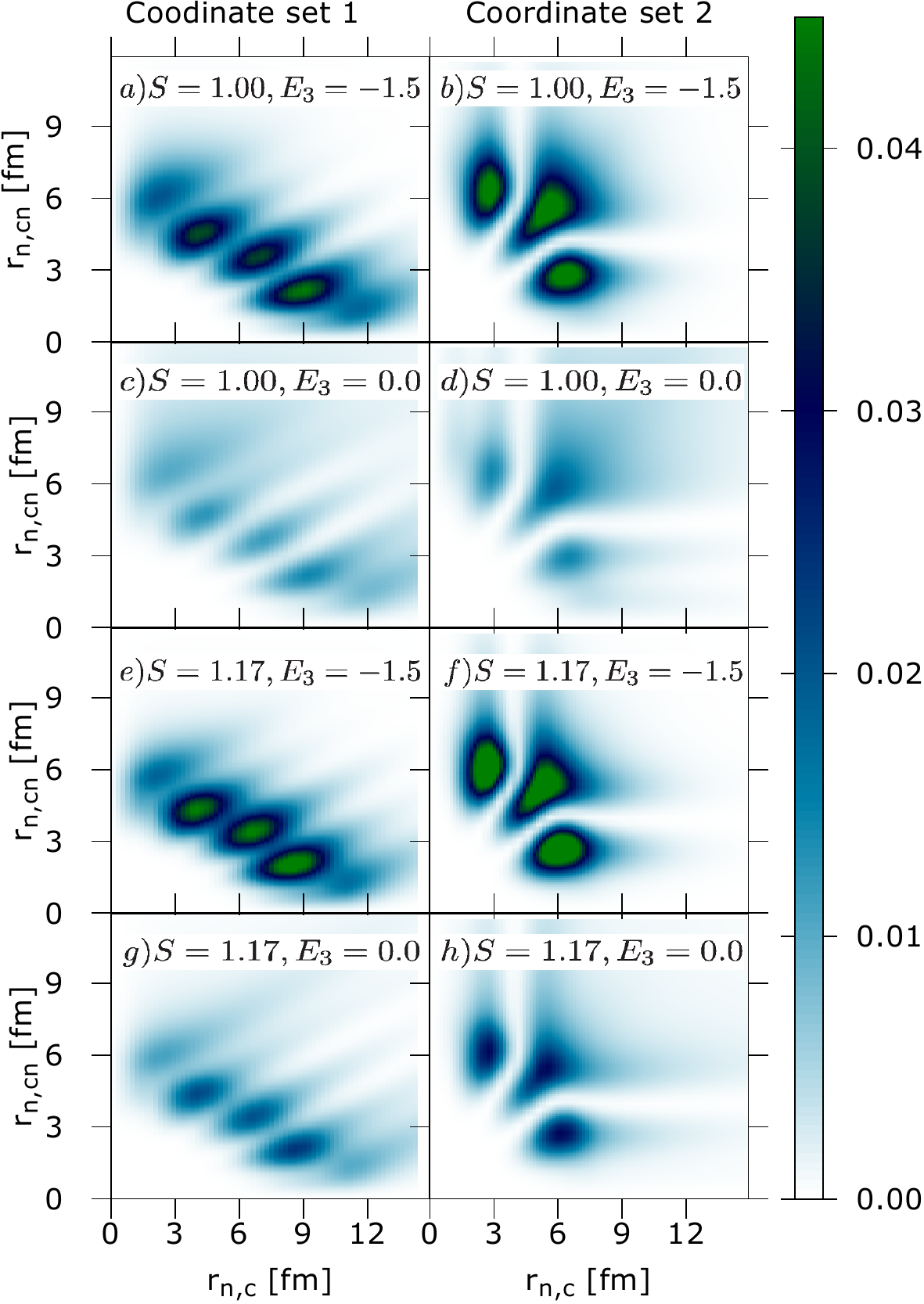}
\caption{The probability distribution from Eq.~(\ref{eq_70ca:propDist}) for the ground state $^{70}\textnormal{Ca} + n + n$ for 
both relative coordinate systems, with $S=1.00$ and $1.17$ for both $E_3=0.00$~MeV and $E_3=-1.5$~MeV. 
\label{fig:Ca72PropDist}}
\end{figure}

To gain an initial understanding of the nature of the system from
these potentials the probability distribution of the three-body wave
function can be considered in traditional relative coordinates as in
Fig.~\ref{fig:3Bcoor}.  The detailed spatial structure can be seen by
integrating the probability distribution of the three-body wave
function over the directional angles, i.e \begin{eqnarray} P(r_x,r_y)
  = \int r_x^2 r_y^2 |\psi_{3b}(\bm{r}_x, \bm{r}_y)|^2 d \Omega_x d
  \Omega_y, \label{eq_70ca:propDist}
 \end{eqnarray} where $\Omega_x$ and $\Omega_y$ are the directional
 angles from Sec.~\ref{sec:3bEq}. Using the coordinates from
Fig.~\ref{fig:3Bcoor} results in two different visual representations of
Eq.~(\ref{eq_70ca:propDist}). In addition, a specific value of $S$ and a
specific three-body energy must be chosen. Figure \ref{fig:Ca72PropDist}
 contains 8 sub-figures showing the probability distribution for the ground
  state of $^{70}\textnormal{Ca}+n+n$ for $S=1.10$ and $1.17$ as well
as for $E_3 = -1.5$~MeV and $0.00$~MeV in both relative coordinate sets.

In Fig.~\ref{fig:Ca72PropDist} it is therefore possible to directly
see the effect both of changing three-body energy and of scaling the
two-body interaction. Unsurprisingly the most unambiguous
configuration is seen in Fig.~\ref{fig:Ca72PropDist}~(f), where the
two-body interaction is most attractive and the three-body system is
most bound. Here three clear peaks are observed. Although three-body
configurations in general are complicated superpositions of many
possible particle placements, it is common to try and reduce them
qualitatively to more intuitively understandable configurations. For
Fig.~\ref{fig:Ca72PropDist}~(f) the three peaks correspond roughly to
three very distinct and understandable configurations. The first peak
at $r_{n,c} \simeq 2$~fm and $r_{n,cn} \simeq 6$~fm can be interpreted
as a configuration where one valence neutron is very close to the core
and the other is further away. The second peak at $(r_{n,c},r_{n,cn})
\simeq (6,2)$ (both in fm) can be interpreted as the opposite, where
the other neutron is close to the core. Finally, the central peak can
be interpreted as a more separated structure where both neutrons are
away from the core.

However, the configurations are not as sharply defined as the discussion
above leads to believe. As illustrated in Fig.~\ref{fig:3Bcoor}
the direction of the orientation is not specified, so a rotation
is possible for each configuration. This is demonstrated by
Fig.~\ref{fig:Ca72PropDist}~(e), which specifies exactly the same
wave function as Fig.~\ref{fig:Ca72PropDist}~(f), but it is much more
smeared out. There are five less sharply defined peaks, which roughly
correspond to range from both neutrons being closely together on the same
side of the core, to the neutrons being far apart on opposite sides of
the core. In effect, a more exact specification of the configuration
would be to say that for each of the probable configurations in
Fig.~\ref{fig:Ca72PropDist}~(f), each of the probable configurations in
Fig.~\ref{fig:Ca72PropDist}~(e) are allowed.

Fortunately, the evolution is generally as one would expect. Given
an initial configuration in the second coordinate system, as
for instance the one seen in Fig.~\ref{fig:Ca72PropDist}~(f) or
Fig.~\ref{fig:Ca72PropDist}~(b), the probability distribution tends
towards a more diluted structure when the three-body binding is weakened,
but maintains the underlying configuration. When increasing the strength
of the two-body interaction with $S$ the tendency is the same, only
less pronounced. The same is true for the first relative coordinate set,
only the structure is more muddled.

\subsection{Long distance structure}

\begin{figure}
\centering
\includegraphics[width=0.8\textwidth]{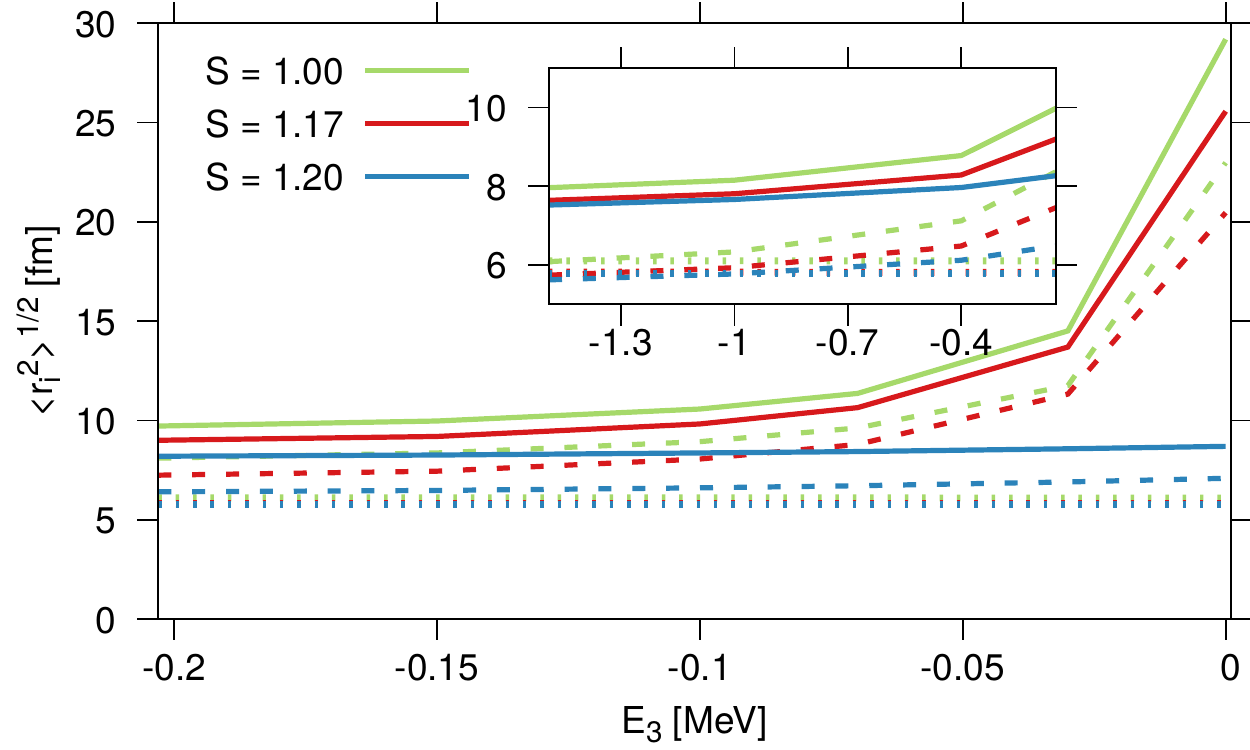}
\caption{Root-mean-square distances as functions of the three-body energy, $E_{3}$, for the ground state of $^{70}\textnormal{Ca} + n + n$. The two-body systems distances are valence neutron-neutron $\braket{r_{n,n}^2}^{1/2}$ (solid), valence neutron-core $\braket{r_{c,n}^2}^{1/2}$ (dashed), and core neutron-neutron $\braket{r_{c,c}^2}^{1/2}$ (dotted) radius for $S=1.00$ (red curves), $S=1.17$ (blue curves), and $S=1.20$ (green curves). The insert shows the behavior for larger binding. (Update of Fig.2 in Ref.~\cite{hov17b}).
\label{fig:Ca72GroundDist}}
\end{figure}

Halo structures are defined by their very large spatial extension,
which necessitates $s$-wave (or possibly $p$-wave) structures,
with very small binding, and no long-range Coulomb interactions
\cite{jen04,rii92}. As seen from Fig.~\ref{fig:Ca72PropDist} very
weakly bound extended structures are difficult to interpret directly
from the three-body wave function. Instead, simplest properties
reflecting the structure for very extended systems are the various
average nucleon-nucleon distances. At least three different distances
are of interest: the valence neutron-neutron average distance,
$\braket{r_{n,n}^2}^{1/2}$, the valence neutron-core center of mass
average distance, $\braket{r_{c,n}^2}^{1/2}$, and neutron-neutron average
distance inside the core, $\braket{r_{c,c}^2}^{1/2}$. These are seen
in Fig.~\ref{fig:Ca72GroundDist} as functions of three-body energy,
$E_{3}$, for $S=1.00, \, 1.17$, and $1.20$.

Much information about the structure is gained from
Fig.~\ref{fig:Ca72GroundDist}. First of all, for a large binding
the valence neutron-core distance (dashed lines) approaches the core
neutron-neutron distance (dotted lines), because the system becomes more
tightly bound. However, the valence neutron-core distance remains larger
reflecting the fact that the valence neutrons are located at the surface
of the core. As the threshold is approached the valence neutrons move
away from both the core and each other (solid lines), while the core
neutron-neutron distance remains unchanged, indicating the core is not
changing, and the three-body structure is preserved in the calculation.

This behavior is identical for $S=1.00$ and $S=1.17$, only slightly more slowly for $S=1.17$ due to the more attractive potentials, demonstrating how the mere availability of the $d_{5/2}$ state does not affect the emergence of halos. However, for $S=1.20$ a clearly bound $d_{5/2}$ state is produced, as mentioned above, and the corresponding adiabatic potential will by definition asymptotically approach the energy of this state. Decreasing the three-body binding energy towards this threshold will populate the $d$-wave, which will prevent an increase in spatial extension, leaving the average distances unchanged as seen in Fig.~\ref{fig:Ca72GroundDist}. The appearance of halo structures is therefore not an inherent characteristic of the method when approaching the dripline, but requires some particular conditions accounted for by the method itself.

\begin{figure}
\centering
\includegraphics[width=0.7\textwidth]{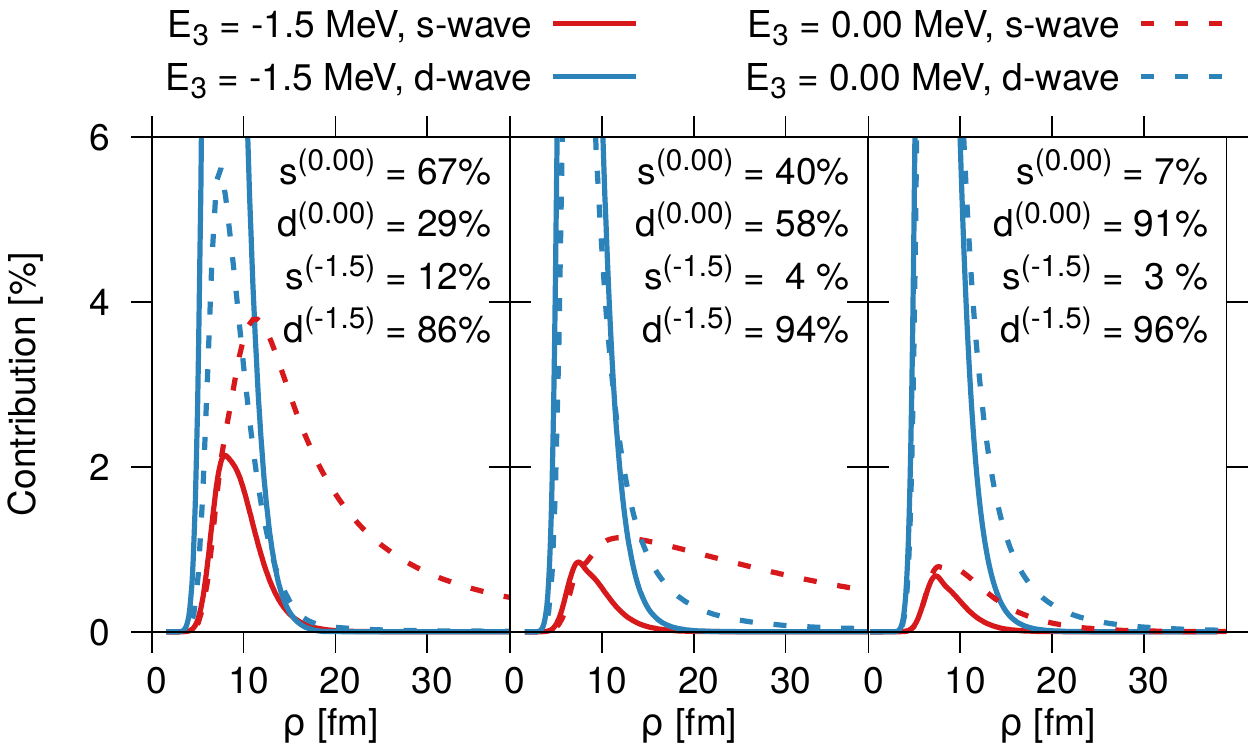}
\caption{The partial wave contributions from $s$- (red) and $d$-waves (blue) for the neutron-core subsystem in $^{70}\textnormal{Ca}+n+n$ as functions of hyper-radius, $\rho$, for $S=1.0$ (left), $S=1.17$ (center), and $S=1.20$ (right). Full and dashed lines indicate a three-body energy of $-1.5$~MeV and $0.00$~MeV, respectively. The total partial wave contribution after integration over $\rho$ is also given for both energies. (Update of Fig.3 in Ref.~\cite{hov17b}).
\label{fig:Ca72ParWaveRho}}
\end{figure}

To further explore how the two-body structure affects the formation of halos one can study the partial wave composition as function of three-body energy, $S$, and $\rho$. This is seen in Fig.~\ref{fig:Ca72ParWaveRho}, where the three panels from left to right correspond to $S=1.00, \, 1.17$, and $1.20$. For large binding the wave function is dominated by $d$-waves, but as the energy is increased the wave function is extending further out with an $s$-wave tail that becomes more favorable for $S=1.00$ and $S=1.17$. For $S=1.20$ the $d$-wave dominates so heavily that the $s$-wave tail is completely suppressed. This is also reflected in the relative weights of the $s$- and $d$-wave included in the figure, where for $S=1.20$ the overall $s$-wave contribution is insignificant even at $E_3 = 0.00$~MeV.

The partial wave composition is determined by the contributing adiabatic potentials, the structure of which is determined by the underlying valence neutron-core interaction. As a result the large distance nature of the tail of the wave function is affected by the fine-tuning of the Skyrme interaction. The characteristics of the wave function tail will later be seen to have profound effect on calculated observables.

\begin{figure}
\centering
\includegraphics[width=0.7\textwidth]{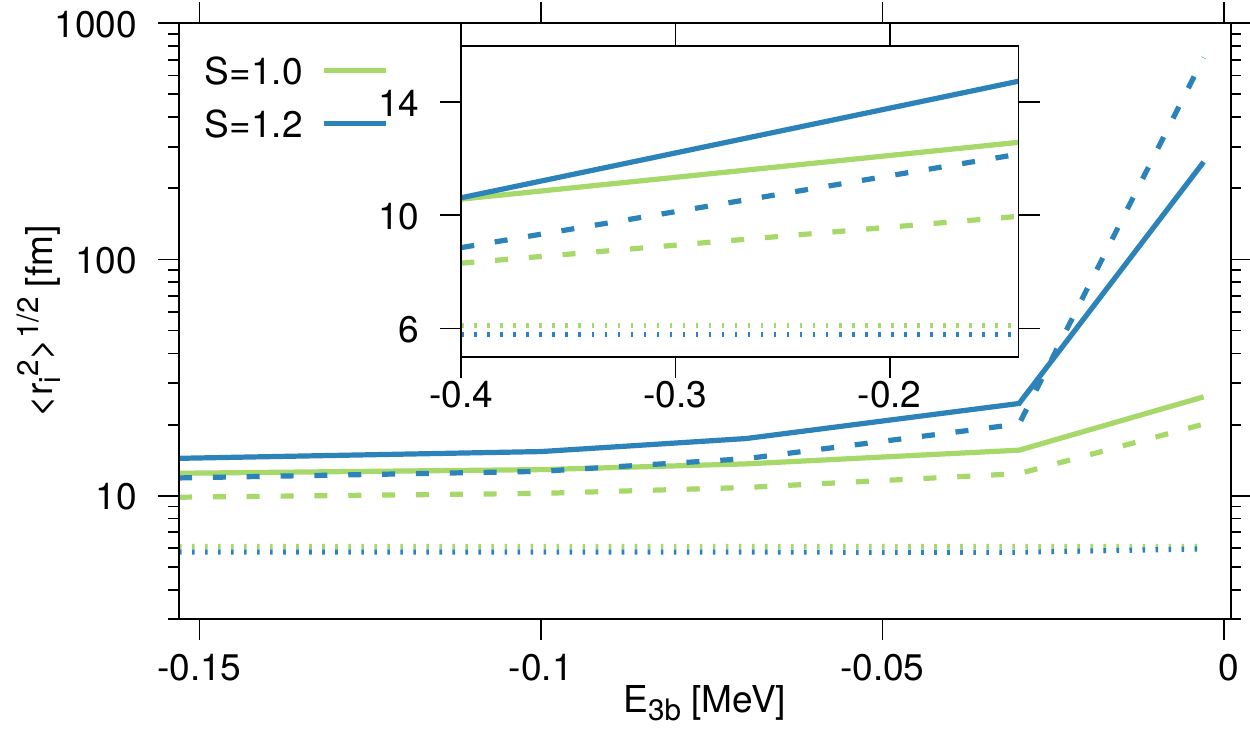}
\caption{Root-mean-square distances as functions of the three-body energy, $E_{3b}$, for the first excited state of $^{70}\textnormal{Ca} + n + n$. The two-body systems distances are valence neutron-neutron $\braket{r_{n,n}^2}^{1/2}$ (solid), valence neutron-core $\braket{r_{c,n}^2}^{1/2}$ (dashed), and core neutron-neutron $\braket{r_{c,c}^2}^{1/2}$ (dotted) radius for $S=1.00$ (green curves) and $S=1.20$ (blue curves). (Update of Fig.4 in Ref.~\cite{hov17b}).
\label{fig:Ca72ExcitDist}}
\end{figure}

It is equally possible for halo structures to form in excited states, it
is again only dependent on the energy and partial wave structure of the
state. However, these states will naturally be more unstable and are often
very delicate. In Fig.~\ref{fig:Ca72ExcitDist} the same average distances are shown
with the same line coloring and type as in Fig.~\ref{fig:Ca72GroundDist},
only for the first excited state in $^{70}\textnormal{Ca}+n+n$. Again
a halo structure is seen for $S=1.00$ at small energies, but unlike in
Fig.~\ref{fig:Ca72GroundDist}, halo structure also appears for $S=1.20$. 

\begin{figure}
\centering
\includegraphics[width=0.7\textwidth]{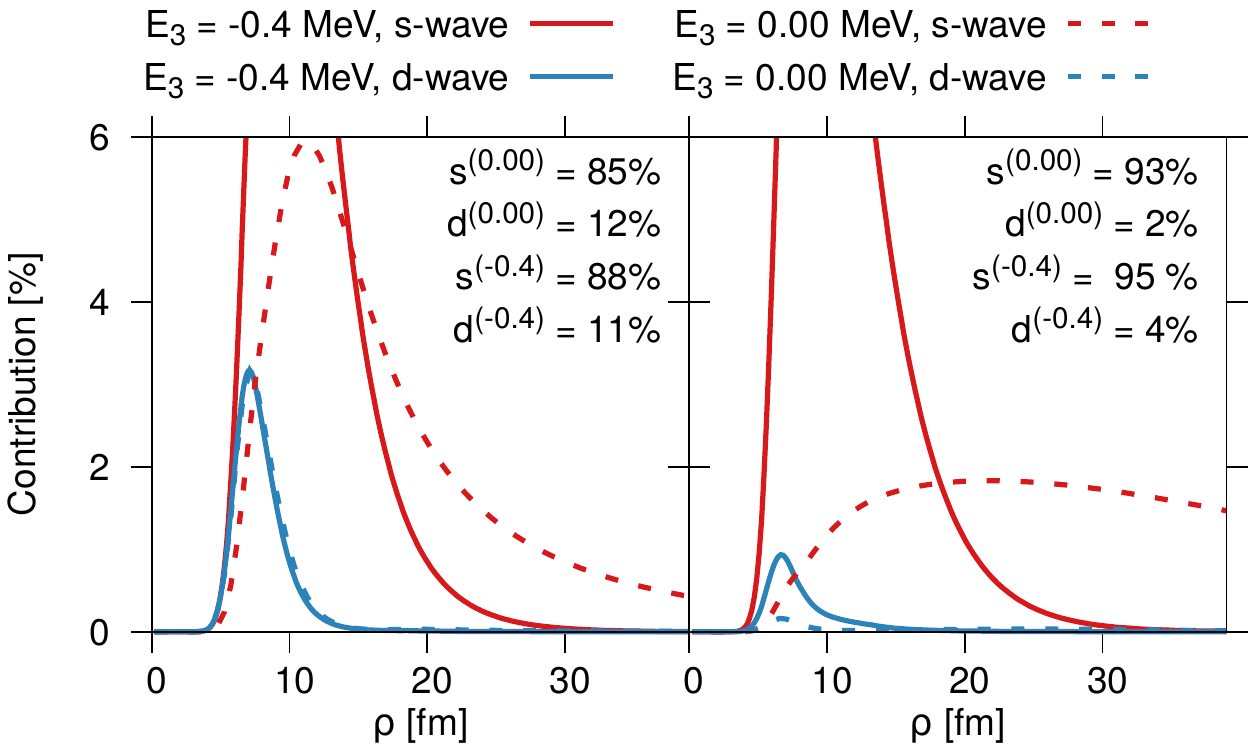}
\caption{The partial wave contributions, similar to Fig.~\ref{fig:Ca72ParWaveRho}, with $S=1.00$ (left) and $S=1.20$ (right), for the excited states shown in Fig.~\ref{fig:Ca72ExcitDist}.  \label{fig:Ca72ExcitParWave}}
\end{figure}

This is best explained by considering the partial wave composition shown in Fig.~\ref{fig:Ca72ExcitParWave}, analogous to Fig.~\ref{fig:Ca72ParWaveRho}. For $S=1.00$ (left panel) some $d$-wave contribution is seen irrespective of energy, while for $S=1.20$ (right panel) almost no $d$-wave contribution is seen. Actually, for $S=1.20$ the composition is almost the reverse of what was seen in Fig.~\ref{fig:Ca72ParWaveRho}, which means there is no barrier to confine the wave function spatially.

The formation of halo structures is sometimes superficially quoted as
the natural consequence of two requirements: the very small binding
energy, and the lack of the barrier associated with $s$-wave. However,
based on Figs.~\ref{fig:Ca72GroundDist} to \ref{fig:Ca72ExcitParWave},
the second requirement is seen to be more complicated than
that. Fractional halo structures are possible even in a state
dominated by $d$-waves, given only a relatively small $s$-wave tail in
the wave function.

Even though the wave function is not an observable, it can still
be used to calculate useful observables, that have unambiguous
interpretations. One of the most accessible non-trivial observables
is the final state single-particle energy distribution. The scattering
length is known to greatly affect the energy distribution \cite{gar06},
therefore different outcomes are to be expected based on the value of $S$.

As the method provides a consistent connection between the short-distance
bulk properties and the large-distance observable, it is even
possible to speculate which path the particle is likely to take before
observation. As quantum mechanics only treat initial and final states,
a more accurate statement would be that it is possible to deduce the
initial state related to the observed final state. In turn this can be
used to infer the general structure of the original system before decay.

It can be shown that the kinetic energy distribution of the fragments in
a three-body decay can be expressed by the square of the wave function
in coordinate space, with the important difference that the hyper-angles
are angles in momentum space \cite{gar07,gar06b}. If $\bm{k}_x$ and
$\bm{k}_y$ are the Jacobi momenta corresponding to the Jacobi coordinates
$\bm{x}$ and $\bm{y}$, then $k_y^2 \propto \cos^2 \alpha$. But $k_y$
is the momentum of the third particle relative to the center of mass
of the other two (with appropriate mass factors), which means $k_y^2
\propto \cos^2 \alpha$ is the energy of the particle relative to its
maximum possible energy \cite{gar07}. As a result the energy probability
distribution can be expressed as \begin{eqnarray} P(\rho, \cos^2 \alpha)
\propto \sin(2 \alpha) \int |\psi_{3b}(\rho,\alpha,\Omega_x,\Omega_y)|^2
d\Omega_x d\Omega_y. \end{eqnarray}

\begin{figure}
\centering
\includegraphics[width=0.7\textwidth]{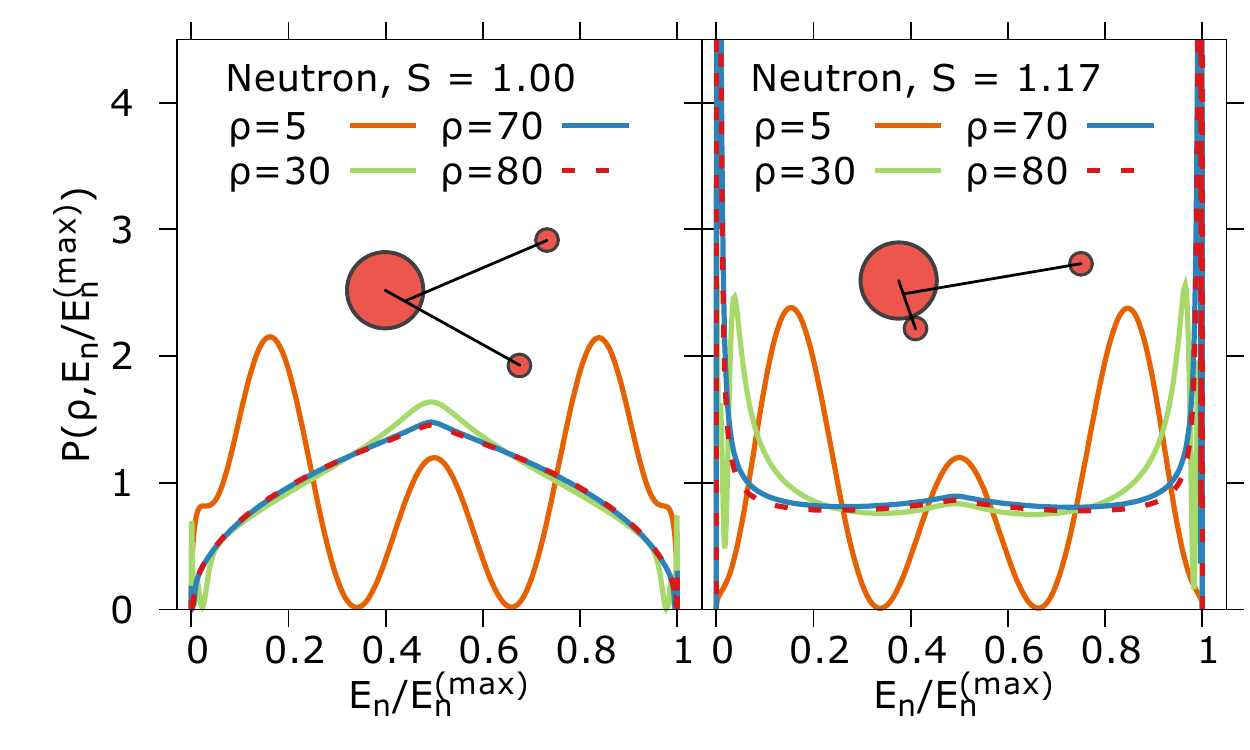}
\caption{The single-particle energy distributions for the neutron after decay of $^{70}\textnormal{Ca}+n+n$ at $\rho = 5, \, 30, \, 70$, and $80$~fm, for $S=1.0$ (left) and $S=1.17$ (right). Schematic illustrations of the large distance configurations are also included. (Fig.5 in Ref.~\cite{hov17b}).
\label{fig:Ca72EDist}}
\end{figure}     

The energy distribution for $^{70}\textnormal{Ca}+n+n$ is shown in
Fig.~\ref{fig:Ca72EDist} for selected values of $\rho$ for both $S=1.00$
and $S=1.17$. This is the "second" relative coordinate system where
$\bm{k}_y$ is the relative momentum of a valence neutron compared to
the center of mass of the core-neutron system. It has been expressed
directly in terms of $E_n/E_n^{(max)}$ instead of $\cos^2 \alpha$. Also
included is a schematic illustration of the long-distance configuration.

The structure seen in Fig.~\ref{fig:Ca72EDist} is almost identical for $S=1.00$ and $S=1.17$ at short distances. Three peaks are seen with maximums at around $E_n/E_n^{(max)} \simeq 1/6$, $3/6$, and $5/6$. However, the evolution to their respective large distance structure is remarkably different. 

For $S=1.00$ the oscillations disappear, and a single peak is seen at around $E_n/E_n^{(max)} \simeq 1/2$. Energetically, this can be interpreted as a situation where both valence neutrons are moving away from the core. This type of energy distribution indicates a decay directly into the continuum without the involvement of any two-body resonance states.

For $S=1.17$ the oscillations also disappear, but the final energy
distribution different. Two peaks are seen, instead
of one, at $E_n/E_n^{(max)}$ approaching 0 and 1, corresponding to a
situation, where either one of the neutrons are close to the core, while
the other moves away. This is a clear indication of a sequential decay,
where a two-body resonance is used to facilitate the decay. The 
most likely scenario is that the core-neutron $d$-state is being used.

This demonstrates how large-distance observables not only potentially
differ substantially from the equivalent short-distance structures,
but are also completely determined by potentially very subtle changes
in these short-range characteristics. In other words, to explore the
experimentally inaccessible short-distance nature of nuclear systems,
it is necessary to rely on theoretical interpretations. Much then hinges
on a proper connection between how the short- and long-ranged aspects
of the systems are treated.

\subsection{Possibility of Efimov behavior}
        
\begin{figure}
\centering
\includegraphics[width=0.7\textwidth]{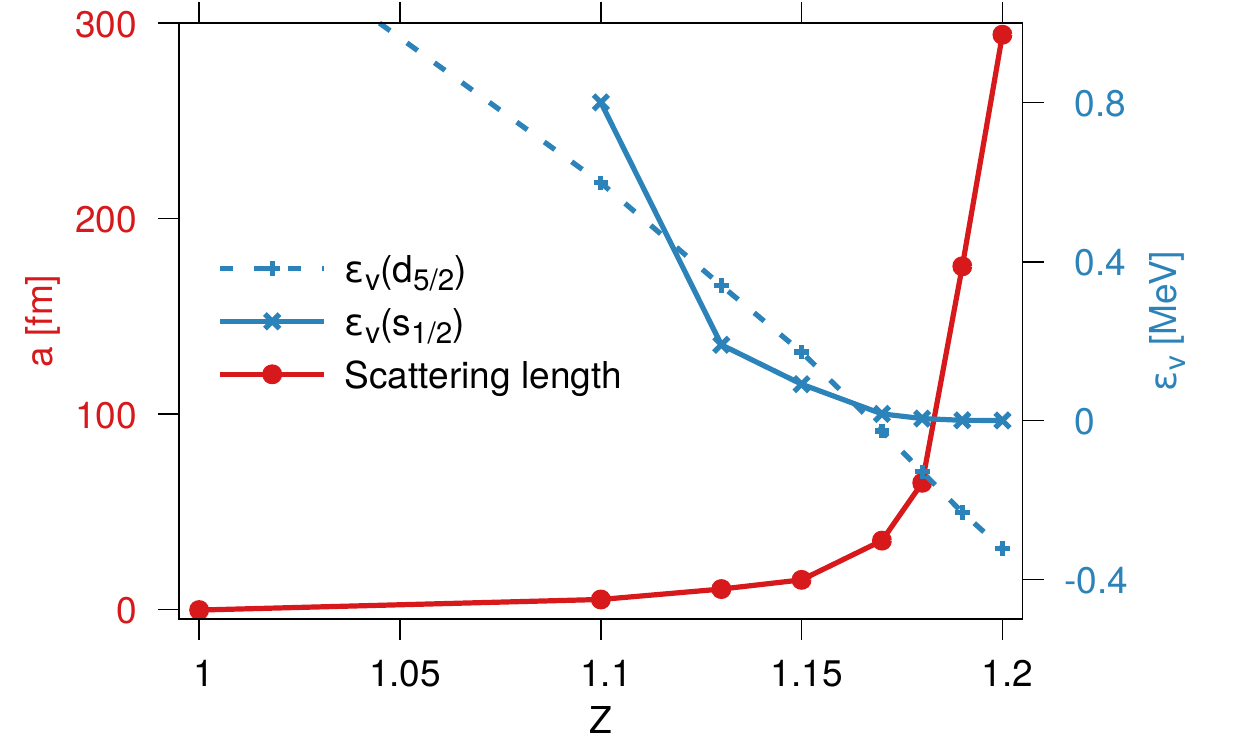}
\caption{The $s$-wave neutron-core scattering length (red, solid) as function of the scaling of the Skyrme-parameters, $S$, for $^{70}\textnormal{Ca} + n + n$. The virtual energy of $s_{1/2}$ (blue, solid) is calculated from Eq.~(\ref{eq:ca72_virEngy}), while the energy of the $d_{5/2}$ (blue, dashed) is obtained from the corresponding invariant mass spectrum, when it is unbound, and from a two-body calculation, when bound. 
\label{fig:Ca72ScatLen}}
\end{figure}

The specific values of $S$ were chosen due to their effect on the scattering length of the potential and the virtual energies of the first unoccupied states. The $s$-wave scattering length for the core-neutron potential is seen in Fig.~\ref{fig:Ca72ScatLen} as a function of $S$. Also included in Fig.~\ref{fig:Ca72ScatLen} is the virtual energy of the unoccupied $s_{1/2}$ state calculated as
\begin{eqnarray}
\epsilon_v(s_{1/2}) = \frac{\hbar^2 }{2 \mu a^2},
\label{eq:ca72_virEngy}
\end{eqnarray}
where $\mu$ is the reduced mass of the system, and $a$ is the corresponding scattering length. Where the following sign convention is adopted
\begin{eqnarray}
\lim_{k \rightarrow 0} k \cot \delta(k) = \frac{1}{a},
\end{eqnarray}
with $k$ being the wave number, and $\delta(k)$ the phase shift. Finally, the energy of the $d_{5/2}$ state, calculated from an invariant mass spectrum when unbound and from a straight forward two-body calculation when bound, is also shown in Fig.~\ref{fig:Ca72ScatLen}. 

The results in Fig.~\ref{fig:Ca72ScatLen} are shown for several
values of $S$, and not just the three specific values from
Fig.~\ref{fig:Ca72HFPot}. The values of $S=1.00, \, 1.17$, and $1.20$
are simply particularly interesting cases. For $S=1.00$ the parameters
correspond to regular, unaltered SLy4, which gives a sensible starting
point for later comparisons. At $S=1.17$ the $s_{1/2}$ and $d_{5/2}$
states are nearly degenerate. This will help illustrate how the
wave function responds to the interplay between several allowed,
and energetically equally favorable, states. Finally, for $S=1.20$ the
$s$-wave scattering length is starting to tend towards infinity. Already
for $S=1.201$ the $s$-state becomes bound with a scattering length of
$a=-251$~fm.

As an extension of the halo structures discussed above, one could
consider the possibility of Efimov states. It has been a long standing
theoretical prediction that an infinite series of bound three-body states
can be formed in a three-body systems where two or three of the two-body
subsystems have a bound or virtual $s$-wave state very close to zero
energy \cite{efi70}. The energy and mean square radii of neighboring
solutions in this series are related by the scaling $s$ defined by

\begin{eqnarray}
s^2 = \exp \left(\frac{2 \pi}{\xi} \right),
\label{eq:Efi_scat}
\end{eqnarray}

where $\xi$ is a constant related to both the underlying two-body
interactions, the effective range of the potentials, and the scattering
length of the two-body systems. The effective potential for three-body
systems in hyper-spherical coordinates will maintain the value

\begin{eqnarray}
V_{eff}(\rho) = - \frac{\hbar^2}{2 m} \left(  \frac{\xi^2 + 1/4}{\rho^2} \right),
\end{eqnarray}

(where $m$ is the normalizing mass) for a large $\rho$-interval which
is needed for the formation of Efimov states \cite{jen03}. However, this
convergence is very slow and an added requirement is that the effective
range, $r_0$, of the potential is several orders of magnitude smaller
than $\rho$, which again must be several orders of magnitude smaller
than the scattering length, i.e. $r_0 \ll \rho \ll a$. This necessitates
a scattering length well above $10^4$~fm, which generally is extremely
unlikely in nuclei.

However, to go one step further before rejecting the idea of Efimov
states in nuclei completely, one could assume a system with a sufficiently
large core-neutron scattering length, and this should then be
the foundation for a series of Efimov states. With identical masses and
large scattering lengths $\xi$ is determined by (see Ref.~\cite{jen03})
 \begin{eqnarray} 8 \sinh \left( \frac{\xi \pi}{6} \right) = \xi
\sqrt{3} \cosh \left( \frac{\xi \pi}{2} \right). \label{eq:Efi_xi3}
\end{eqnarray} This results in $\xi = 1.00624$, which in turn gives the
famous scaling value $s=22.7$. The expression becomes more complicated
if there is mass differences between the particles along with the three
large scattering lengths.

However, as the defining characteristic of an Efimov state is its relation
to the other states in the series, a state cannot reasonably be called
an Efimov state in isolation. The assumption of three simultaneously
large scattering lengths (compared with the range of the potentials)
includes the assumption that the neutron-neutron scattering length
is not only much larger than $r_0$ \cite{fre12,bra06}, but that
it is also larger than the extend of second Efimov state (at $s$
times the first state). With a neutron-neutron scattering length of
around $20$~fm, the assumption does not hold, and instead this must
be treated as a system with only two large scattering lengths. This
will always be the case for two nucleons connected to some core, due
to the inherently modest neutron-neutron scattering length. For only
two large scattering lengths, $\xi$ is determined by (see again for
instance Ref.~\cite{jen03}) \begin{eqnarray} \xi \cosh \left(\xi
\frac{\pi}{2}\right) \sin(2 \phi)  = 2 \sinh \left( \xi \left(
\frac{\pi}{2} - \phi\right)\right), \label{eq:Efi_xi2} \\ \phi  =
\arctan \left( \frac{\sqrt{(m_c(m_c+2m_n)}}{m_n} \right), \end{eqnarray}
 where $m_c$ and $m_n$ is core and neutron mass respectively.

Using this in relation with $^{72}\textnormal{Ca}$, where $m_c \simeq
70 m_n$, results in $\xi \simeq 0.01035$ and a scaling factor of $s \sim
10^{131}$, completely outside any realm of possibility. The possibility
of Efimov states has been considered in many other nuclei over the years,
such as $^{60}\textnormal{Ca}$ \cite{hag13} or $^{11}\textnormal{Li}$
\cite{jen03}. For $^{60}\textnormal{Ca}$ the mass imbalance leads to $\xi
\simeq 0.01205$ and $s \sim 10^{113}$ and for $^{11}\textnormal{Li}$
the mass imbalance leads to $\xi \simeq 0.07382$ which results in a
scaling of $s \sim 10^{18}$. This might be smaller, but is still many,
many orders of magnitude too large to be realizable.

In nuclear physics systems of three identical particles are not
realistic candidates for Efimov physics. Three neutrons do not form
a bound system, and anything heavier would include Coulomb interactions whose
long-range nature would prohibit formation of Efimov states. Furthermore,
from Eq.~(\ref{eq:Efi_xi2}) it is clear that no realistic scaling can
ever be obtained with only two identical particles. To make this even
clearer, one could look at a series expansion of Eq.~(\ref{eq:Efi_xi2})
 \begin{eqnarray} |\xi| \approx \frac{4 m_{\textnormal{light}}}{
\sqrt{3} \pi m_{\textnormal{heavy}}}. \end{eqnarray} With an exponential
dependence on the mass ratio, and the masses being discrete, there is
no hope of having a second Efimov state in a nuclear system. However,
in other areas of physics, such as cold atomic gases, where three
identical particles can be used, it has become possible in recent years
to actually detect Efimov states \cite{zac09}. Another possibility,
in other areas of physics, is to use two heavy and one light particle,
which also produces more favorable scaling conditions, and very recently
the emergence of Efimov physics has even been suggested in relation to
strongly interacting photons \cite{gul17}.

\section{Astrophysical applications\label{sec:68Se}}

In previous sections, we have focused on applications of the formalism at the neutron dripline. Clearly, since the formalism is 
of a general nature, it is perfectly well-suited to also address the questions with protons as the valence nucleons. In this section, we therefore discuss the case of proton valence particles and examine an important example along the proton dripline. In particular,
we will adapt the case of the nuclear astrophysical rapid proton capture (rp) process \cite{wal81} and cases where the process can have important contributions from three-body dynamics. The rp-process involves the rapid capture of protons by a nucleus in a stellar 
environment, forming a system further from the line of stability. At some point the proton dripline is reached, where additional proton capture would render the system unstable. 

Although, neutron capture processes are more discussed and well-known \cite{mat85,bar06}, the importance of the proton capture processes must not be underestimated. The formation of at least 40 stable and a large number of unstable nuclei is only possible due to various proton capture processes \cite{bur57,arn03,rau13,rei14}. Among them, the rp-process is thought to be most important in relatively high-temperature, hydrogen rich environments \cite{wal97}. The most likely environment is in the accretion of a close binary system containing a neutron star or white dwarf, which results in x-ray bursts \cite{sch01,sch98}. Other possibilities include inside supermassive stars or in nova and supernova bursts \cite{wal81}, but all scenarios include temperatures in the Giga-Kelvin range.

At the proton dripline, where the capture of single protons would lead to reemission, the system will initially wait for the comparably slow $\beta^+$ decay. Following a $\beta^+$ decay, the capture of one or more protons is usually possible, until the new position of the proton dripline is reached. This implies that there is an accumulation of matter at the nuclei with the longest $\beta$-decay lifetime, and these nuclei are therefore referred to as waiting points of the rp-process. 
Going beyond the light nuclei, the most important rp-process waiting points, the so-called critical waiting points, are $^{64}$Ge, $^{68}$Se, and $^{72}$Kr \cite{sch98,woh04}, which have a $\beta$-decay half-life of $64$ \cite{rob74}, $36$ \cite{bau94}, and $17$ \cite{piq03} seconds, respectively. Out of these critical waiting point nuclei, $^{68}$Se is thought to be the most important 
\cite{sch07,tu11}.

In this section we therefore focus on this particular nucleus as the prime example that demonstrates the applicability of the method to proton capture processes and the proton-rich parts of the nuclear chart. 
Different effects including nuclear pairing \cite{hov13} make the driplines become quite rugged in structure. Here the three-body formalism that we discuss offers unique opportunities in relation to the critical waiting points, as it should be possible to bridge the mass gap in the dripline through a three-body reaction. Although the addition of one proton would form a particle unstable system, two protons could be added, forming a stable, Borromean system. Our focus is on the process
$^{68}\textnormal{Se} + p + p \rightarrow ^{70}\textnormal{Kr}+\gamma$, and the rate for this two proton capture process would affect the effective lifetime of the waiting points in a stellar environment. 
Consequently, this would affect where, in a proton-rich nucleosynthesis process, matter is accumulated and in turn influences the evolution of the particular stellar system. 

The rest of this section is devoted to a discussion of the cross sections and rates for the process calculated based on the three-body framework discussed in previous sections, as well as a discussion of the decay mechanism for which knowledge can also be accessed. Lastly, we include a section that discusses the technicalities of the potential terms that need to be included and in particular of the couplings between different hyperspherical effective potential channels in relation to the physical parameters that enter the setup.

\subsection{Cross sections and reaction rates}

The process in question is the capture of two protons on a core, $c$, forming a three-body system, $A$, which then $\gamma$-decays to the $0^+$ ground state. The cross section, $\sigma_{ppc}$, for this process $c + p + p \rightarrow A + \gamma$, is related to the dissociation cross section, $\sigma_{\gamma}$, for the reverse process by \cite{gar15}
\begin{eqnarray}
\frac{\sigma_{ppc}(E)}{\sigma_{\gamma}(E_{\gamma})}
= \nu! \frac{2(2J_A + 1)}{(2J_{p_1} +1 )(2J_{p_2} + 1) (2J_c + 1)} \frac{32\pi}{\kappa^5} \left(\frac{E_{\gamma}}{\hbar c} \right)^2,
\label{eq:SeSigppc}
\end{eqnarray}
where $J_i$ is the total angular momentum of the related particle,
$\nu$ is the number of identical particles, and $\kappa$ is the
three-body momentum, which is defined as $\kappa =
\sqrt{2mE/\hbar^2}$. The mass $m$ is the normalization mass used to
define the Jacobi coordinates in Eqs.~(\ref{eq:JacCoorx}) and
(\ref{eq:JacCoory}), and the energies are related by $E_{\gamma} = E + |E_{gr}|$, where $E_{gr}$ is the $(0^+)$ ground-state energy, $E$ is the three-body energy, and $E_{\gamma}$ is the energy of the emitted photon. 

The process can occur as a resonant reaction, going through a well-defined three-body resonance, or a non-resonant reaction, going through a continuum state. In either case the dissociation cross section is a sum over contributing electric and magnetic multipole transitions of order $\ell$. Here we shall focus on the electric transitions, and the corresponding dissociation cross section is given by \cite{hov16}
\begin{eqnarray}
\sigma^{\ell}_{\gamma}(E_{\gamma}) 
=& \frac{(2 \pi)^3 (\ell +1)}{\ell ((2\ell +1)!!)^2} \left( \frac{E_{\gamma}}{\hbar c}\right)^{2\ell-1} \frac{d}{d E}\mathcal{B}(E\ell, 0 \rightarrow \ell), \label{eq:SeSiggam}
\end{eqnarray}
where $E\ell$ represents the electric multipole transition of order $\ell$ whose strength function for the $0\rightarrow \ell$ transition is 
\begin{eqnarray}
 \frac{d}{d E} \mathcal{B}(E\ell, 0 \rightarrow \ell) =
 \sum_i \left| \braket{\psi_{\ell}^{(i)} | | \hat{\Theta}_{\ell} | | 
 \Psi_{0}} \right|^2  \delta(E-E_i), 
 \label{eq:Setran}
\end{eqnarray}
where $\hat{\Theta}_{\ell}$ is the electric multipole operator, and $\psi_{\lambda}^{(i)}$ is the three-body wave function of energy $E_i$, for all bound and (discretized) three-body continuum states in the summation.

The reaction rate, $R_{ppc}$, can be expressed in terms of $\sigma^{\ell}_{\gamma}$ by \cite{die11,hov17c}
\begin{eqnarray}
 R_{ppc}(E) = &  \frac{8 \pi}{(\mu_{cp} \mu_{cp,p})^{3/2}}  \frac{\hbar^3}{c^2} 
 \left( \frac{E_{\gamma}}{E} \right)^2 \sigma^{\lambda}_{\gamma}(E_{\gamma}), 
\label{eq:SeRate}
\end{eqnarray}
where $\mu_{cp}$ and $\mu_{cp,p}$ are the reduced masses of the proton-core system and the proton to proton-core system, respectively.

We want to address reactions in an astrophysical environment described as a gas of temperature $T$, hence the reaction rate must be averaged over the (energy normalized) Maxwell-Boltzmann distribution
\begin{eqnarray}
B(E,T) = \frac{1}{2}  \frac{E^2}{T^3} \exp\left(- E/T \right),
\end{eqnarray}
resulting in an energy average reaction rate given by
\begin{eqnarray}
\braket{R_{ppc}(E)} = \int_{0}^{\infty} B(E,T) R_{ppc}(E)
 \, dE, \label{eq:SeAveRate}
\end{eqnarray}
where the temperature is in units of energy (the Boltzmann constant is set to unity).

\subsection{\textsuperscript{68}Se + p + p cross sections}

A potential problem with a traditional three-body investigation of the systems that we are considering is the phenomenological character of the approach, which makes it 
difficult to present concrete predictions \cite{hov16,hov17d}. However,  very little arbitrariness is left in the method we use here, and therefore the prediction capability increases.

For the particular case of the  $^{68}\mbox{Se}+p+p\rightarrow ^{70}\mbox{Kr}+\gamma$ reaction, 
the SLy4 Skyrme parameterization is again used, as it is well suited for describing nuclei far from $\beta$ stability. Importantly, as the reaction rate will depend exponentially on the barrier thickness at the resonance energies, accuracy in the keV range is needed. To achieve this, we recall the discussion in Sec.~\ref{sec:70Ca} and rescale the Skyrme parameters $t_i$ as $t_i \rightarrow S t_i$ to achieve this level of accuracy, leaving $x_i$ and $W_0$ unchanged.
Experimentally, the lowest lying resonance in $^{69}\textnormal{Br} \, ( ^{68}\textnormal{Se}+p)$ is known to be a $f_{5/2}$ state at $0.6$~MeV \cite{san14}. Using an unaltered SLy4 parameterization the lowest allowed state also turns out to be an $f_{5/2}$ state \cite{hov17c}, in accordance with experimental findings. To achieve the correct energy the SLy4 parameterization is scaled by a factor of $S = 0.9515$, which has been used in all the results presented in this section. With this scaling there are no $1^-$ resonance states, and the important states are therefore the $0^+$ and $2^+$ states. Notice that $S$ is very close to unity, again attesting that only a small adjustment is needed to get the experimental data correct to the necessary keV accuracy.

Once the known  $f_{5/2}$ resonance in $^{69}\mbox{Br}$ has been used to fine tune the Skyrme parameterization, the only remaining degree of freedom is the three-body interaction. 
The $0^+$ ground state is predicted from systematics to be at $-1.34$~MeV \cite{aud12}, and the three-body interaction is used to reproduce this value. The form from Eq.~(\ref{eq:genV3}) is used with a range of $r_0 = 6$~fm and a strength of $V_0 = -17.5$~MeV. To see the effect of the resonance energy the $2^+$ state is varied from $0.5$ up to $4.0$~MeV using the same range, $r_0$, but varying the strength from $-35.05$ to $-26.22$~MeV.

With a choice of Skyrme interaction and a choice of the three-body interaction the set of discretized 2$^+$ continuum states can be calculated by discretization with a box boundary condition. As the plane wave state for a free particle contains all possible angular momentum in a partial wave expansion, the continuum states are not characterized by one complete set of discrete quantum numbers \cite{hov16}. However, the bound final states are clearly defined by a set of quantum numbers, and the transition is dictated by a appropriate multipole operator, as specified in Eq.~(\ref{eq:Setran}). The transitions for the individual discretized continuum states are independent, and they can as such also be calculated and added individually. As the available single-particle levels all have odd angular momentum, it is not possible to produce negative parity final states such as $1^-$. Likewise, there are no excited $0^+$ states available, so the only relevant transition is the $E2$ transition.

\begin{figure}
\centering
\includegraphics[width=0.7\textwidth]{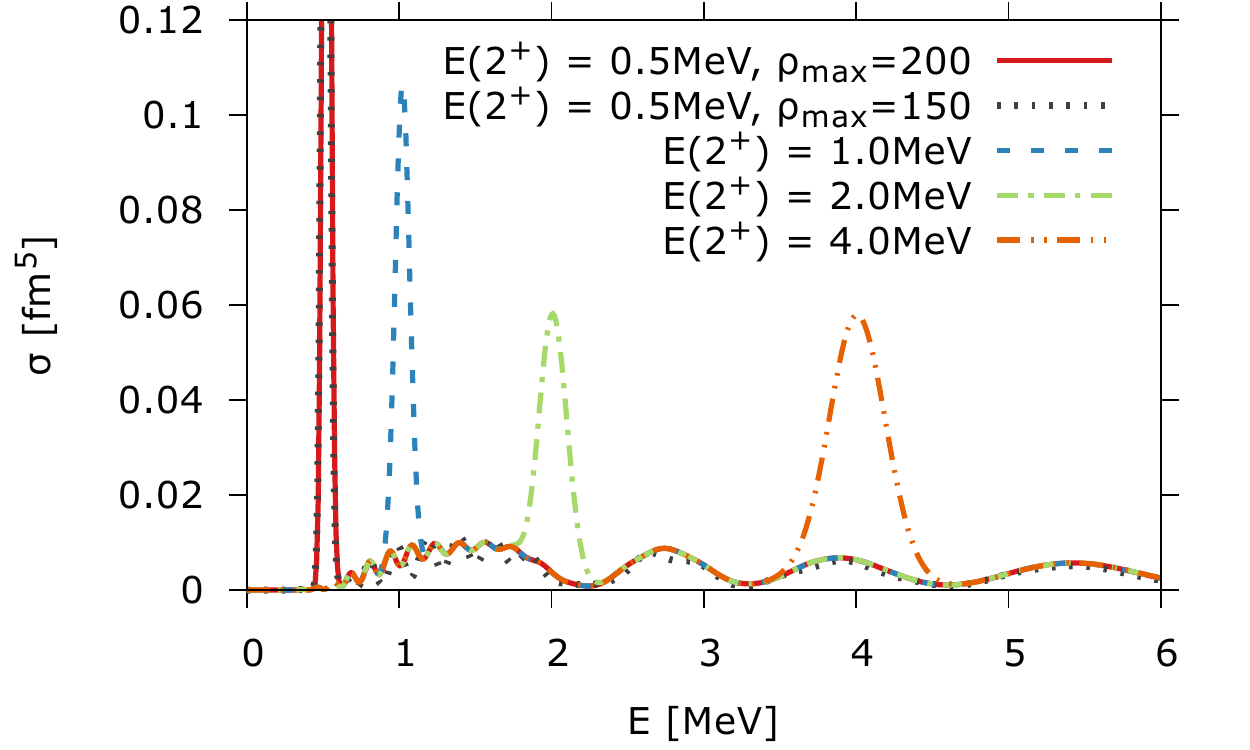}
\caption{The electromagnetic $\mathcal{E}2$ proton capture cross section, $\sigma_{ppc}(E)$, for the process, $^{68}\textnormal{Se}+p+p \rightarrow ^{70}\textnormal{Kr} + \gamma$, as a function of three-body energy.  The $0^+$ final state energy is $-1.34$~MeV and the $2^+$ resonance energies are $E=0.5, \, 1.0, \, 2.0,$ and $4.0$~MeV, respectively.  The discretized continuum states are obtained using box sizes of $\rho_{max} = 150, 200$~fm for $E=0.5$~MeV, and only $\rho_{max} = 200$~fm for all other energies.
\label{fig:SeCross}}
\end{figure}

Once we have calculated the discretized continuum spectrum, Eq.~(\ref{eq:Setran}) produces the photodissociation cross section in combination with Eq.~(\ref{eq:SeSiggam}). This is directly related to the proton capture cross section through Eq.~(\ref{eq:SeSigppc}). The proton capture cross section is shown in Fig.~\ref{fig:SeCross} for a range of different $2^+$ resonance energies and box sizes.

First of all, Fig.~\ref{fig:SeCross} illustrates that the box size used to discretize the continuum is sufficiently large. For a $2^+$ resonance energy of $0.5$~MeV the cross section is calculated using box sizes of $150$~fm and $200$~fm, and the results are identical, as they should be for large box sizes \cite{gar15b}. In addition, it is seen how the peaks at the cross section correspond to the resonance energies, while the non-resonant, background contribution between the peaks is less important, although not insignificant. This follows from Eqs.~(\ref{eq:SeSiggam}) and (\ref{eq:Setran}) where the overlap between the ground state the continuum state is seen to dictate the cross section. This background contribution is also independent of the the $2^+$ resonance energy.

\subsection{\textsuperscript{68}Se + p + p reaction rates}

\begin{figure}
\centering
\includegraphics[width=0.7\textwidth]{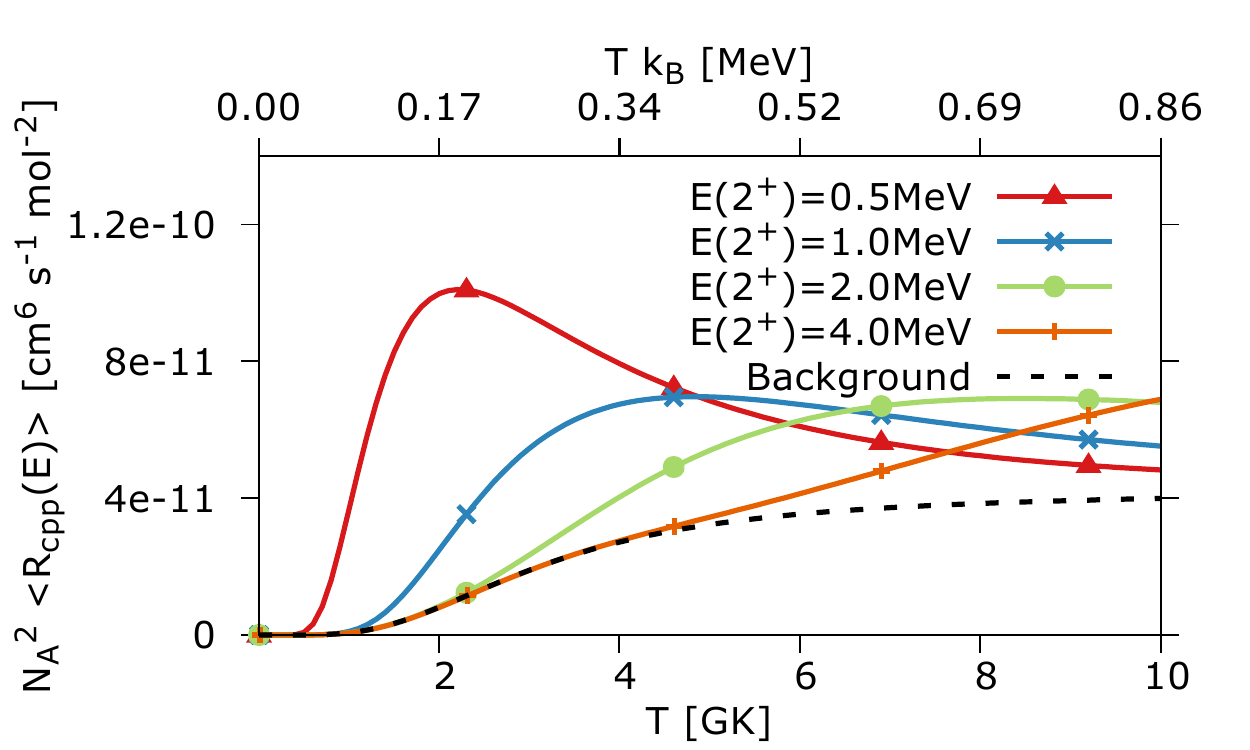}
\caption{The reaction rate for the radiative capture process 
$^{68}\textnormal{Se}+p+p \rightarrow ^{70}\textnormal{Kr} + \gamma$, as function of temperature for the 
different $2^+$ resonance energies in Fig.~\ref{fig:SeCross}. The black dashed curve is the background 
contribution. The top axis shows the temperature multiplied by Boltzmann's constant in MeV. (Fig.3 in Ref.~\cite{hov17c}).
\label{fig:SeRate}}
\end{figure}

The proton capture cross section leads to the two-proton absorption rate through Eq.~(\ref{eq:SeSigppc}) and Eqs.~(\ref{eq:SeRate}) to (\ref{eq:SeAveRate}), which is the central result in relation to astrophysical evaluations. The average rates from Eq.~(\ref{eq:SeAveRate}) are shown in Fig.~\ref{fig:SeRate}. The four solid lines correspond to the four cross section calculations from Fig.~\ref{fig:SeCross}, while the dashed line is the non-resonant, background contribution, common to all four calculations.

These average rates are smeared out due to the Maxwell-Boltzmann energy distribution, but the rates have their maximum at the Gamow peak \cite{ili15} resulting from the compromise between the decreasing temperature distribution and the tunneling probability increasing with resonance energy. The non-resonant, background contribution is clearly smaller, but by less than a factor of 2 for temperatures above $4$~GK (GigaKelvin). As the non-resonant contribution is independent of the resonance energy, this very firmly determines the scale of the reaction rate. For temperatures in the astrophysically relevant regime for the rp-process, i.e. $2$--$4$~GK, the reaction rate for the two-proton capture process is of the order $\sim 4 \cdot 10^{-11} \textnormal{cm}^6 \, \textnormal{s}^{-1} \, \textnormal{mol}^{-2} N_A^2$. This result depends on the specific shape of the potential barrier, which again depends on the charge, mass, and available single-particle states of the system. 

It should be noted that given a specific choice of Skyrme parameterization, this is a very comprehensive calculation. First of all, other types of transitions are very unlikely, as $1^-$ and $3^-$ resonance states are not possible due to the available single-particle states \cite{nic14}, while no excited $0^+$ states are found -- something also supported by the mirror nucleus \cite{nes14}. Excited $2^+$ states are fully included, as the full cross section spectrum is included in the rate calculation. Core excitations are also unlikely to be significant, as the first excited state of $^{68}\textnormal{Se}$ is at $0.854$~MeV above the ground state \cite{mcc12}. The probability of occupation at a temperature of $4$~GK would then be $\exp(-E_{c1} /T) = 0.09$, which heavily suppresses any contribution from this state \cite{hov16}. 

As a result the reaction rates presented here provide a very solid range for use in the calculations of evolutions of stellar environments. The upper limit is to a large extend given by the rate curve with a low-lying resonance, while the lower limit is given by the background contribution.

\subsection{Reaction mechanisms}

Another vital issue, which touches on the most fundamental aspects of the reaction, is to examine the reaction 
mechanism \cite{hov16,hov17c}. The two-proton capture process can proceed through a well-defined three-body resonance, or as a non-resonant reaction through a continuum state, while after penetrating the Coulomb barrier the only transition to the $0^+$ ground state is through $\gamma$-decay. However, both the resonant and the non-resonant penetration of the Coulomb barrier can be imagined as a direct, one-step process, where both protons are captured simultaneously, or a sequential, two-step process, where the protons are captured successively
one at the time. In other words, the sequential process \cite{gri01,alv08} can occur as
\begin{eqnarray}
c + p + p \rightarrow (cp) + p &\rightarrow A^{\ast} \rightarrow A + \gamma, \\
c + p + p \rightarrow (cp) + p &\rightarrow A + \gamma,
\end{eqnarray}
where $A^{\ast}$ is the well-defined resonance state. Likewise, the reaction mechanism usually referred to as direct \cite{alv08,gri05} can proceed as
\begin{eqnarray}
c + p + p &\rightarrow A^{\ast} \rightarrow A + \gamma, \\
c + p + p &\rightarrow A + \gamma.
\end{eqnarray}
These distinctions in practice mostly conceptual, as the actual reaction mechanism would be superpositions of all possibilities, all of which are accounted for in the present formulation. They are, nonetheless very important for understanding physical mechanisms. Moreover, extreme cases would be observable \cite{che07}, as the resulting energy distribution would carry distinct signatures of the mechanism. This is similar to the discussion in Sec.~\ref{sec:70Ca}. 

\begin{figure}
\centering
\includegraphics[width=0.7\textwidth]{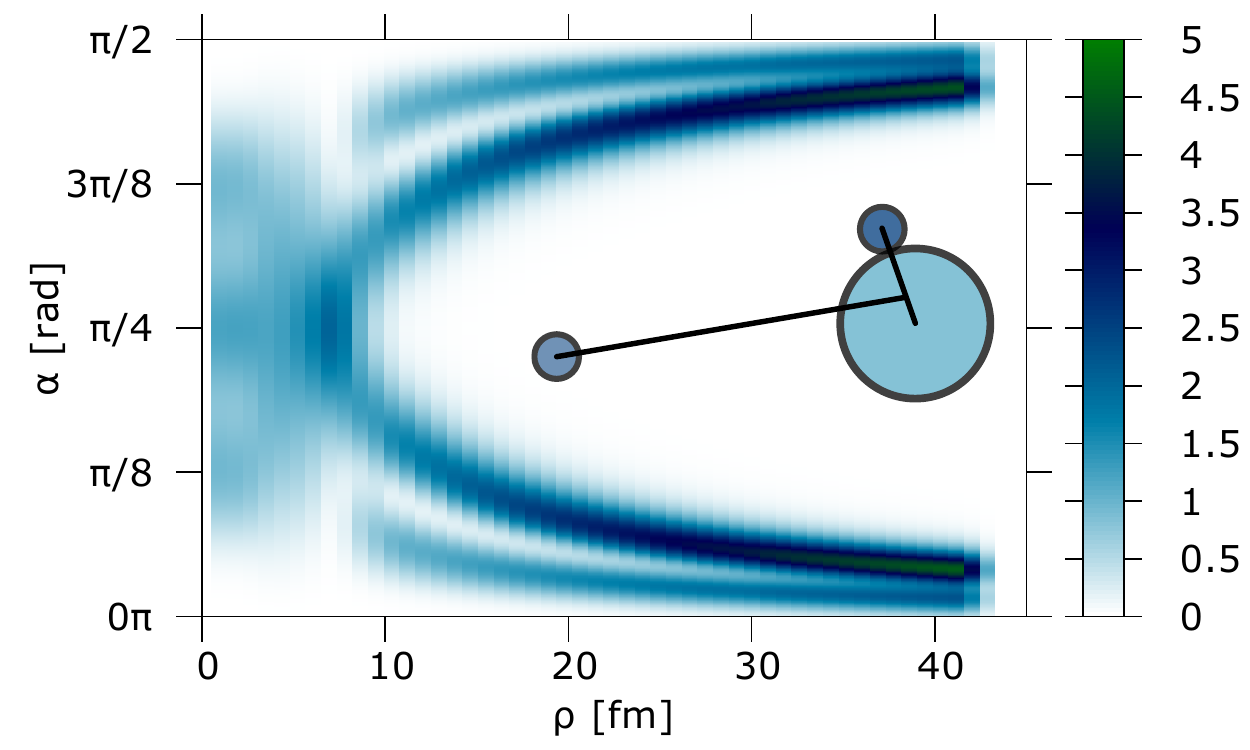}
\caption{The probability distribution, as expressed by Eq.~(\ref{eq:SeProp}), of the three-body, $^{68}\textnormal{Se}+p+p$, wave function for the lowest allowed potential in a $2^+$ state. This is given as a function of hyperradius, $\rho$, and hyperangle, $\alpha$, related to the Jacobi coordinate system where "x" is between core and proton.  (Fig.4 in Ref.~\cite{hov17c}).
\label{fig:SeAngDist}}
\end{figure}

In order to pin down specific reaction mechanisms, we turn to the angular wave function, integrated over directional angles as
\begin{eqnarray}
P(\alpha,\rho) =  \sin^2 (\alpha) \cos^2 (\alpha) 
\int |\phi_n(\alpha, \rho, \Omega_x, \Omega_y)|^2 d\Omega_x d\Omega_y,
\label{eq:SeProp}
\end{eqnarray}
where $\sin^2 \alpha$ and $\cos^2 \alpha$ are phase factors.
This probability distribution is shown in Fig.~\ref{fig:SeAngDist}
for the lowest allowed angular eigenfunction, shown in the second Jacobi coordinate system in Fig.~\ref{fig:3Bcoor}. This shows the reaction clearly proceeding  through a sequential reaction mechanism, with only two clear peaks in the spectrum beyond $\rho = 10$~fm. An angle of $\alpha = 0$ or $\alpha = \pi/2$ represent to a configuration where one proton is close to the core, while the other is located somewhere at a much larger distance. In other words, this corresponds very clearly to an extreme sequential reaction. We note that the next few, higher-lying allowed angular eigenfunctions show the same behavior.

This very unambiguous mechanism can be understood through some schematic potentials associated with the geometric configuration of the reaction \cite{gar05}. If one proton is moving away from a core-proton resonance state with energy $E_{pc}$, and it has passed outside the strongly attractive region, it would experience a Coulomb potential given by \cite{hov16}
\begin{eqnarray}
V_{seq} (\rho) = \frac{e^2 (Z_c + 1)}{\rho} + E_{pc},
\end{eqnarray}
where $Z_c$ is the charge of the core.
Alternatively, if both protons are moving away from the core symmetrically, they would experience a potential given by
\begin{eqnarray}
V_{dir}(\rho) = \frac{E^2 \left( 2 Z_c + \frac{1}{2} \right) \sqrt{2}}{\rho}.
\end{eqnarray}
These schematic potentials cross at a distance, $\rho_{crit}$, and at a potential energy, $V_{crit}$, given by
\begin{eqnarray} 
\rho_{crit} 
&= \frac{e^2 \left( (2\sqrt{2}-1)Z_c -1 + 1/\sqrt{2} \right)}{E_{pc}} \approx \frac{2.6 Z_c}{E_{pc}} \;,
\label{eq cros1}
\\ 
V_{crit} 
&=  E_{pc}\frac{(2Z_c+1/2)\sqrt{2}}{(2\sqrt{2}-1)Z_c -1 + 1/\sqrt{2}} \approx  1.6 E_{pc} \; ,
\label{eq cros2}
\end{eqnarray}
for $Z_c \gg 1$. With $E_{pc} = 0.6$~MeV and $Z_c = 34$ this results in $V_{crit} \simeq 0.94$~MeV and $\rho_{crit} \simeq 147$~fm, far outside the decisive region of interest discussed here. Given the relatively low-lying resonance, the reaction is clearly sequential.

\section{Improvements, generalizations and perspectives \label{sec:expansion}}

The spotwise demonstration of applicability and flexibility of the
method in the previous sections can now be supplemented by a number of
possible improvements and appealing generalizations.  In principle,
improvements can be related independently to the core description or
to the few-body treatment.  However, the interconnectivity of the core
and few-body descriptions means that modifications affect both
structures.  We shall not here implement any of the extensions but
discuss them in various degrees of details depending on the subject of
investigation and short- or long-term goals.

We divide this section into four categories, that is first the
immediately desirable or necessary improvements where details are
available, second choices of more complicated systems or other
modifications of the technical method, third descriptions of
applications beyond those of the previous sections, and fourth
indications of how and where to transfer the present experience to
other subfields of physics.  As we go along with this agenda the
amount of details are clearly decreasing with the time perspective for
the implementations.

\subsection{Short-range valence nucleon-nucleon interactions\label{subsec:valNN}}

Contact interactions somehow need regularization due to the divergence
at short distance.  This was realized long before the first Skyrme
interactions were used in mean-field treatments, where the Slater
determinant automatically eliminates the divergence problem.  However,
using the complete Hilbert space for the two valence nucleons revives
this problem as touched upon in Sec.~\ref{sec:ValInt}.  This
inconsistency with the use of Skyrme interactions in valence space
leads to an infinite number of spurious deep-lying bound states built
on the lowest diverging adiabatic potential described by $\lambda$,
while the relevant higher-lying $\lambda_n$'s are unaffected.  Thus,
in principle the problem disappears with removal of this $\lambda$.
Unfortunately the many crossing $\lambda$'s in practice make this
difficult although not impossible.

A better and more obvious option to rectify this divergence problem
and eliminate the spurious bound states, would be to replace the
contact interaction with a finite-range interaction. If simplicity is
still desired, a natural choice would be the finite-range Gogny
interaction \cite{dec80}, which completely would eliminateall spurious
bound states.  The original Gogny-proposal in mean-field calculations
was to replace $t_0, \, t_1,$ and $t_2$ terms in the Skyrme force with
a sum of two Gaussians, in the style of Brink and Boeker \cite{bri67},
but keeping contact $t_3$ and spin-orbit terms.  This resulted in a
density dependent two-body interaction between particles $j$ and $k$
of the form \cite{dec75}.  The spin-orbit and $t_3$-terms remain as in
Eq.~(\ref{eq:skyrme}) while $t_0$, $t_1$ and $t_2$-terms in this
equation are substituted by
\begin{eqnarray}
 V_{jk}^{Gogny} = \sum_{n=1}^2 \left[
  \exp{\left(-\frac{(r_j-r_k)^2}{\mu_n^2}\right)} \left( W_n + B_n P_{\sigma}
  - H_n P_{\tau} - M_n P_{\sigma} P_{\tau}  \right) 
 \right] \; ,
\end{eqnarray}
where $P_{\tau} = (1+\bm{\tau}_j \cdot \bm{\tau}_k)/2$ is defined in
analogy to $P_{\sigma}$ in Sec.~\ref{sec:skyrme}. They are operators
exchanging spin and iso-spin coordinates, while $\mu_n, \, B_n, \,
H_n, \, M_n, $ are parameters of the interaction.

A large number of different versions exist for the Skyrme force, while for the Gogny force the so-called "D1S" force \cite{ber84} is the most common. Other parameterizations, such as the D1N \cite{cha08} and the D1M \cite{gor09} do exist, but even today the D1S force is still the prevailing parameterization. More recent forces tackle the contact interaction in the density dependent $t_3$-term, and are usually designated with the prefix "D2" to mark the two finite-range parts. One recent example is shown in Ref.~\cite{cha15}, which demonstrate that a finite-range density dependent term generally improves the accuracy of various predicted nuclear matter properties.

To completely eliminate contact interactions the spin-orbit
interaction should also be replaced by a finite-range version. This can
be done similarly to the other terms \cite{dec75,cha15}, but the
parameters would necessarily have to be refitted to the new form of
the  force. However, the importance of the spin-orbit term is
largest at the surface, diminishing at both small- and
large-distances. It is therefore not in itself supporting
spurious bound states in our method, and a contact spin-orbit
interaction could be maintained between the valence nucleons while
still achieving full consistency.

The procedure for implementing the Gogny interaction between all
nucleons in the core equation is identical to the procedure for
implementing the Skyrme interaction. The main difference is, that the
input functions in the Sch{\"o}dinger equation would contain an
integral over the coordinate which previously was trivially eliminated
by the $\delta$-function.  This would increase the numerical
complexity slightly, but as the integrands are Gaussians, this would
only be a very minor complication. For the three-body part the
procedure would not change, as the Gogny force along with central and
spin-orbit potentials also produces effective masses, which could be
incorporated in the same process.

It should be emphasized that even if a finite-range interaction, as for
instance the Gogny interaction, is implemented between all nucleons, a
reparameterization is needed when the valence nucleons are outside the
core.  This is imperative, as global mean-field interactions are
fitted to reproduce in-medium nuclear matter properties, while the
correct asymptotic nucleon-nucleon scattering properties in vacuum
must be obeyed.  The need for reparameterization can be viewed as a
consequence of the increased Hilbert space expanded to encompass the
added valence nucleons. Such a reparameterization, in one form or
another, is always needed when changing the Hilbert space, but especially
in the present case when free-space properties are crucial.

\subsection{Adiabatic treatment of mean-field from valence space}

While the improvements in Sec.~\ref{subsec:valNN} dealt with
implementing more sophisticated and consistent nucleon-nucleon
interactions, and the extensions in Sec.~\ref{subsec:deform} shall
deal with more complicated core structures, a third improvement has to
do with the procedure for solving the coupled equations in our method.

Currently, the core and three-body equations are solved separately in
an iterative manner, which means the solutions for the two (core and
valence) equations are viewed independently.  At the moment this is
done as explained in Sect.~\ref{sec:3bEq} with the core solution found
in the ``external'' average field of a full three-body solution, including
angular and radial parts.  The resulting mean-field calculation
returns the input potentials for the three-body calculation and the
process is repeated until convergence is reached.

One interesting, and more correct, procedure is to extend the
hyperspheric adiabatic expansion, used to solve the three-body part, to
perform the mean-field calculation at each hyperradius, $\rho$.  More
precisely this means to calculate the input needed in the mean-field
calculation from the angular three-body wave function for a given
fixed $\rho$.  The core solution in turn provides the input needed for
the three-body calculation.  The process is repeated until convergence
is reached for each $\rho$.

After core and angular calculations are found the three-body radial
equation is solved as usual, only the equation itself is slightly
different now.  Our original radial equation from Eq.~(\ref{eq:rad3b})
is obtained by inserting the $\rho$ dependent three-body wave
function, expanded on the angular wave functions, see
Eq.(\ref{eq:3bExpanded}), into the Schr{\"o}dinger equation, and
using the solution from the angular part.  If the connection between
mean-field and three-body calculation is established at each $\rho$,
the core wave function would also depend on $\rho$, and the radial
equation would be slightly modified.  The radial equation would be
based on the full wave function expanded as usual
\begin{eqnarray}
\Psi =\psi_{3b}(\rho,\Omega)  \psi_c(\rho, r_1, \dots , r_A)=
\frac{1}{\rho^{5/2}} \sum_n f_n(\rho) \phi_{n}(\rho,\Omega) \psi_c(\rho, r_1, \dots , r_A) \;, \label{eq:AdEx_wave}
\end{eqnarray}
where this total many-body wave function should be fully
anti-symmetrized.

If the iterative process is conducted for fixed $\rho$, there is only
remaining free parameter in the space of ${r}_x$ and ${r}_y$, since,
$\rho^2 = |x|^2 + |y|^2$.  This affects the three-body aspect of the
calculation in that there would be one integral less. For example, the
three-body density in terms of traditional three-body coordinates is
\begin{eqnarray}
	n_3(\bm{r})
&=
\int | \psi_{3b}(\bm{r}_x = \bm{r}, \bm{r}_y)|^2 d^3 r_y
+ \int | \psi_{3b}(\bm{r}_x, \bm{r}_y =\bm{r})|^2 d^3 r_x
\nonumber
\\
&=
\int r_y^2 | \psi_{3b}(\bm{r}_x, \bm{r}_y) |^2 dr_y d \Omega_y
+ \int r_x^2 | \psi_{3b}(\bm{r}_x, \bm{r}_y) |^2 dr_x d \Omega_x,
\end{eqnarray}
for an infinitely heavy core and identical valence nucleons in the
Jacobi set where $\bm{x}$ is between the core and a valence
nucleon. If the iteration is carried out for a given $\rho$, one
coordinate integral has to be omitted.  However, the general procedure
would be the same as described in Sec.~\ref{sec:3bEq}.

The differences are related to the $\rho$ variation.  In the radial
differential three-body equation we can collect the new terms simply
by applying $T_{\rho}$ on the wave function in
Eq.~(\ref{eq:AdEx_wave}).  To distinguish the origin we denote the new
coupling terms, $\tilde{Q}$ and $\tilde{P}$.  They necessarily all
contain derivatives of $\psi_c$, which has to be added to the ordinary
and effective mass $Q$ and $P$-terms.  We get
\begin{eqnarray}
  \tilde{Q}_{nm} = \delta_{nm}\langle \psi_c | \frac{\partial^2}{\partial \rho^2}
  | \psi_c \rangle  + 
2 P_{nm} \langle \psi_c | \frac{\partial}{\partial \rho} | \psi_c \rangle 
 \\
  2  \tilde{P}_{nm} = \delta_{nm} \langle \psi_c | \frac{\partial}{\partial \rho}
  | \psi_c \rangle \; ,
\end{eqnarray}
where the explicit matrix elements are to be computed over the core
coordinates. The second of these $Q$-terms has a factor equal to the
ordinary $P$ coupling term.  These new terms constitute a very natural
extension of the previous radial equations, where we only added $P$-
and $Q$-like terms arising from $\rho$-derivatives of $\psi_{c}$.

Using this expanded adiabatic formulation would more accurately
connect the short-distance with the large-distance structure.
However, it is more time consuming, since angular solutions now have to
be found for each hyperradius, and the reduction due to fixed $\rho$
in the density calculations does not provide sufficient compensation.
Clearly both concept and formalism remain unchanged.  This improved
method provides the correct large-distance structure, since the large
$\rho$ implies isolated core structure surrounded by far away
non-interacting nucleons.  This is in contrast to the present
$\rho$-independent core structure arising from an average influence of
the valence particles.  This would open up for interesting
possibilities to study the evolution of the wave function, between
these extremes (small and large $\rho$), in ways not currently
possible with existing few- or many-body theories.

\subsection{Deformed and/or odd systems \label{subsec:deform}}

Before or after the improvements described in the previous
subsections, it is possible to extend the region of applicability in
several ways.  We shall here consider the two extensions of
deformation and odd numbers of neutrons and/or protons in the core.

\subsubsection{Deformed even-even core.}

Allowing axial deformations, for instance in the Skyrme
parameterization, would provide a more general core description,
which in turn would produce modified core-valence nucleon
interactions. This is a straight-forward extension of the core
Skyrme-Hartree-Fock description, and it was also the first improvement
of the original spherical Skyrme-Hartree-Fock method \cite{vau73}.
For systems with axially deformed core structures, which maintain an
even-even configuration, the assumption of time-reversal symmetry
still holds true. In the derivation contained in
\ref{appen:vari} no assumptions were made regarding the shape
of the core, so the fundamental equations from
Eqs.~(\ref{eq:sch3Appen}) and (\ref{eq:schCAppen}) are still
valid. However, it will be necessary to express the various densities
in appropriate cylindrical coordinates \cite{vau73}.

In the spherical case, as explained in Sec.~\ref{sec:coreEq}, the core
Hartree-Fock equations were solved in two iterative steps (not to be
confused with the iteration between our core and three-body
equations). First an ansatz to a radial wave function was produced
from harmonic oscillator potentials. This was used to produce first
approximations to the needed densities, which could then be used to
solve the Schr{\"o}dinger equation and calculate the effective mass,
and central and spin-orbit potentials. Using these interactions a new
radial wave function was produced, and the process continued until
convergence in total energy is reached. For spherical nuclei this was
all done in coordinate space, but solving a deformed Schr{\"o}dinger
equation in coordinate space is more difficult.  The deformed wave
function can be expanded and solved on a deformed harmonic oscillator
basis.  The evaluation of the potentials and the construction of the
densities subsequently done in coordinate space \cite{vau73}.

Incorporating an axially deformed core in the three-body part is
potentially more difficult. This is not something that has received
much attention in the community or been considered terribly important,
and as such there is not a well-established framework to draw
upon. Internal structure of the constituent particles is traditionally
not considered in few-body physics; the particles are viewed as
point-like. At best internal structure is simulated using the
characteristics of the effective potentials. Usually, this is not a
problem as the finer structure of the constituent particles is
unimportant at larger distances, in particular for halos \cite{mis97}.
The validity of the few-body results is generally related to the
accuracy of the large-distance asymptotic behavior. The short-range
structure is neglected as irrelevant and in any case previously not
included properly.

The method presented here consistently connects the long-distance
behavior with the short-range structure. As such the nature of the
constituent particle structures becomes much more important, and a
possibly deformed core potential is therefore important for the
coherence of the method. There are various possible procedures to
resolve this problem. The simplest and most obvious solution would be
to use the deformed mean-field formalism for the core calculations,
but averaging over angles when applying the potential in the
three-body description. This is equivalent to projection on zero
core-angular momentum.  In this way, the freedom allowed by the
deformed formalism is used by the core to assume the most favorable
configuration and produce the most realistic potential, while the
information lost at short-distances by averaging in the three-body
part should be minor even for moderately strong deformations. For
larger distances this should have no effect at all. For very strong
deformations the loss might be more significant, but situations with
very strong core deformations are unlikely in this scenario as the
most tempting candidates for this method are systems where the most
significant extension is at the valence level, not the core level.

Another option would be to use the deformed potential directly in the three-body calculation. Due to the direction associated with the deformation, angular momentum would no longer be a conserved quantity. This could be rectified by calculating a number of angular momentum states and coupling them appropriately. As long as all relevant states are calculated, there should be no loss of generality.

\subsubsection{Odd nucleons in the core.}

Unlike the expansion to include deformed core structures, the
expansion to include odd nucleon numbers in the core is relatively
difficult to incorporate in the mean-field part, but very simple to
incorporate in the three-body part.  The main problem from a
mean-field perspective, is that an unpaired nucleon will break the
crucial time-reversal symmetry. Without this symmetry, many of the
equalities from \ref{app:assump} no longer apply, and a number of
new terms appear beyond the even-even formalism.

Various methods have been developed to treat unpaired nucleons in
mean-field calculations.  One method is by the so-called exact
blocking of the odd state \cite{sch10}. The idea is to view the
unpaired nucleon as a quasi-particle excitation on top of the
even-even "vacuum" produced by the remaining even number of
nucleons. Done fully this is in principle exact. However, the
complications incurred by breaking time-reversal symmetry are so
considerable that other alternatives are often used instead.

One of the more popular alternatives is the equal filling approximation \cite{bon07}, where the central idea is to consider the unpaired nucleon as being half in one state and half in the time-reversed state simultaneously. This conserves time-reversal symmetry, and thereby simplifies the situation greatly. The simplicity of the approximation has caused it to be widely adopted and it has been used to study a great variety of odd nuclear systems \cite{rod10,rod10b,rod11}. The precision of the equal filling approximation, both in comparison with available experimental information and with the exact blocking method, is generally considered to be more than sufficient for most practical purposes \cite{sch10,rod10}.

The most common justification for the approximation is found in a statistical interpretation of quantum mechanics \cite{per08}. Another argument is found in the decomposition of the mean-field energy functional into time-even and time-odd parts. It can be shown that the time-even part of the energy density matrices are identical for the equal filling approximation and the blocking method \cite{sch10}. As such the average values of time-even observables are identical in the two methods, which means the average values of time-even observables such as radii and multipole moments are the same in the two methods, while time-odd observables such as spin alignments and magnetic moments will differ slightly \cite{sch10}.

For the three-body part there are no such complications, but the
valence nucleons would be allowed to occupy non-time reversed orbits.
There are no inherent assumptions about time-reversal symmetries or
pairing of nucleons. The three constituent particles just enter as a
mass and charge with a given two- and three-body potential acting
between a number of possible relative partial waves but coupled to the
given conserved total angular momentum.  The potential produced by the
core calculation, be it with exact blocking or with equal filling or
any other approximation, is adopted directly as the effective two-body
core-valence nucleon interaction.

\subsection{Applications and generalizations}\label{subsec:applic}

We shall first describe how to employ more elaborate many-body methods
on the core and second how to extend to more complicated systems of valence
particles. Then we shall suggest a number of immediate applications,
and finally we indicate how to transfer the method from nuclei to
other subfields of physics.

\subsubsection{Improved descriptions and complicated valence structure.}

The core and valence parts can each be changed or improved
independently although both structures would be affected through the
coupling.  The present choice of Skyrme Hartree-Fock mean-field
treatment of the core structure can be modified in several ways.  The
conceptually smallest change is to employ a finite-range two-body
interaction, while maintaining the self-consistent mean-field
approximation.  This is discussed in more details in
Sec.~\ref{subsec:valNN} for the Gogny-interactions, but a Gogny type
interaction is not the only possibility, and the discussion is valid
for any other finite-range mean-field calculation.

The core treatment can be substantially further improved by
implementing advanced methods from many-body physics either variants
of interacting shell models or directly ab-initio techniques, see
Sec.~\ref{sec:many} for options.  In general, any many-body
description of the core, which yields a potential to be incorporated
into the few-body equation could be used.  Implementing a more
sophisticated core description, such as for instance a (no-core) shell
model description (see Sec.~\ref{subsec:deform} and
\ref{subsec:applic}) is possible, but not necessarily worth the
effort.  As was seen in Sec.~\ref{sec:24O}, as long as the core
description is fairly decent, very accurate results can be obtained.
So the increased complexity might not produce a significantly better
result, but the computation time would drastically increase.  As such,
other improvements, such as the inclusion of deformed or non-even
cores, might be more worth the effort.

The immediate technical advantage of finite-range interactions is that
the connection between in-medium and in-vacuum nucleon-nucleon
interactions becomes much smoother. Furthermore, the transition is
obviously easier to simulate correctly when the scattering length of
the finite-range interaction is close to the free value.  However, the
more fundamental advantage is that the same interaction is applied to
both core and valence spaces, that means no new parameter.  The
valence particle few-body calculation would no doubt be more
complicated as consequence of an improved core treatment, but in most
cases probably practically doable with the same variational
formulation.

The valence part is at the moment formulated for two structureless
particles, that is directly valid for the three different combinations
of two nucleons.  However, it is tempting to include intrinsic
structure of the valence particles, that could be one or two
alpha-particles as simple examples.  Three larger nuclear clusters
each with intrinsic structure would be more complicated due to the
combination of intrinsic structure, the relative coordinates, and the
necessary more complicated handling of the Pauli principle.
Sec.~\ref{sec:GenTheo} contains an overall general formulation.

To be practical, before the complicated extension to larger chunks of nuclear
clusters, it is illuminating to start with only two clusters. Each can
first be treated in the mean-field approximation, and subsequently
coupled in the product wave function ansatz described by relative
coordinates.  For two clusters described in the intrinsic coordinate
system this amounts to one relative distance coordinate. Such a
formulation is compatible with the two-center (non-interacting) shell
model \cite{mar72}, which first of all is aimed at describing the
fission process and heavy-ion collisions \cite{ghe03,dia08}.  A
version with a two-cluster structure is to allow both a ground and an
excited state in one of the clusters, but this is effectively a
three-body problem.

The next extension towards three clusters with intrinsic structure is
to allow only two of these and one point-like valence particle.  The
final step could turn out to be rather complicated but at least by now
qualitatively formulated, see Sec.~\ref{sec:GenTheo}.  Division into
more than three clusters would not uncover new features because more
clusters do not have radial divergences at the threshold of zero
binding \cite{yam10a,yam10b}.  This implies that a tendency to form
correlated structures with more than three centers would be met by
coalescence into fewer centers \cite{jen04}.

\subsubsection{Application to correlated wave functions.}

All the numerical examples in the present paper have been on weakly
bound and simple dripline nuclei.  However, it is also possible
to study correlations in ordinary well bound nuclei.  The binding
between core and valence particles should then be comparable to the
binding between all other nuclear substructures.  The two-body
correlation between deeper-lying single-particle orbits in ordinary
well bound nuclei is then within reach.  The correlations under
investigation correspond to the chosen valence structure, which in
practice so far has been structureless nucleons. This restriction
could be lifted by extension to more complicated systems as described
in the previous subsection.

Another unexplored question is the interplay between short and
long-range interactions. In the present model this is made explicit
in the different treatments of the dense core and the spatially
extended and weakly bound valence particles.  Nevertheless, the
related two parts of the wave function are connected through the
quantum mechanical equation of motion.  However, the path connecting
short and long-distance properties is not an observable, and as such
only indirectly open to experimental investigations.  Crudely speaking
we can say that large-distance probabilities are measurable but not
uniquely related to short-distance properties of the wave function
\cite{jen10}.

The goal of understanding the structure, from which the measured
probability originates, is then left to theoretical interpretation.
In extreme cases the connection may appear as (almost) unique. This
would be the case when the wave function in the corresponding
parameter space is confined to a narrow path similar to a classical
deterministic orbit.  On the other hand a smeared-out wave function
can be very similar at small distances but evolving into completely
different structures at large distances, see an example in
Sec.~\ref{sec:70Ca}, which mostly resembles chaos conditions.  An
understanding of possibly systematic relations and quantitative
validity conditions are highly desirable, because measurements then
can be interpreted directly or at least the underlying uncertainties
extracted.  The present method is an invitation to study how far the
information from measurements can be extended through this dynamic
evolution of the wave functions.

The static properties of a (not necessarily stationary) wave function
should be supplemented by more explicit time dependent properties.
The simplest is probably decay mechanisms of resonances \cite{alv08} or
maybe decay of continuum states in general \cite{gar15,gar14}.  Reaction investigations
transferring one structure into another are readily formulated but a
lot more difficult to carry out \cite{rom11,tim99}.  A non-stationary state
describable by our method may be populated by decay of a complex
many-body state, perhaps beta-decay.  The subsequent time evolution
carries information about both decay process and initial population.
Again the short and long-distance interplay is important.  In general,
time evolution of a non-stationary initial state is a tempting
application.

\subsubsection{Transfer to other subfields of physics.}

The concepts of halos, Efimov states, and many-body structure are
general physics topics.  The most special is probably the inherent
properties from the short-range nuclear interaction.  When the Coulomb
force is dominating the present method is therefore most likely not
suitable.  When both short and long-range forces are active the
interplay of correlations between different length scales is delicate
but essential.  For many subfields of physics this issue is of great
interest.  The core-valence division seems in general able to deal
with these issues.  A direct application is on hypernuclei mixed with
nucleons and perhaps pions or other mesons \cite{cob97}.

On the other hand many subfields of physics, most prominently
condensed matter, molecular and cold atomic physics, exhibit features
with dominating short-range characteristics \cite{nie01,jen04,zin05}.
The connecting property is the existence of universal structures which by
definition is independent of details of the supporting potentials.
The meaning of this statement is understandable already from
``classical'' nuclear physics where low-energy nucleon-nucleus
scattering could be explained by disparate potentials with the same
$s$-wave scattering length.  From this originated the notion of
nuclear halos, that is weakly bound and spatially extended structures
describable by the scattering length \cite{han87}.  Also existence of
the extreme Efimov states have been (indirectly) established within
the fields of cold atomic and molecular physics \cite{kra06}.
Recently also direct evidence has been obtained \cite{kun15}.

These universal structures appear as correlated substructures within
the hosting many-body system.  It is therefore appropriate to ask how
they materialize in analogy to the present investigation of emergence
of halos and Efimov states from the background of an essentially
uncorrelated many-body system.  The first necessary ingredient is here
to add a genuinely external field to confine spatially all (core and
valence) particles.  The core treatment is then simple with rather
weak two-body interactions perhaps parameterized as a mean-field.  The
correlations producing the few-body universal structures are then
described by the valence particles.  The appearance of halos and
Efimov states can be studied with the present method.  The advantage
is that choices of systems are much more flexible and suitable mass
asymmetric systems can be studied.  In other words applications to
other subfields of physics is only a matter of changing interactions
while keeping the methodology.

\section{Summary and conclusions\label{sec:con}}

In this work we present the detailed derivation of a new method for
treating many-body nuclear systems with correlated substructures.
The report describes the method and is applied to three different nuclear
systems, each illustrating characteristic aspects of the method. In
this final section we first briefly survey important pieces of the
theoretical development, second we draw lessons learned from the
applications, and third we conclude with a general perspective derived
from the present work.

\subsection{Development of the method}

A survey of central details of existing models describing many-body
nuclear systems is first presented in Sec.~\ref{sec:many}.  The
intend is to explain the underlying philosophy and the fundamental
operating methods in an effort to highlight strengths and weaknesses
specific to each approach.  We conclude that shortcomings among
established methods are treatments of weakly bound and spatially
extended systems as well as the computational requirements when
applied to systems heavier than the very lightest nuclei.  This
expresses the need for an efficient method which incorporates these
correlations along with ordinary many-body features.  We present a new
method to remedy these shortcomings, that is both allow these
structures in the wave function and without loss of computational
efficiency for applications on heavier nuclei.

The underlying philosophy and the theoretical derivations are
described in Sec.~\ref{sec:GenTheo}, with assumptions added as they
become necessary. We present the idea behind the combination of few-
and many-body structures and derive the fundamental coupled equations
of motion.  At first, this formalism is independent of specific
correlations and in particular valid for all choices of
nucleon-nucleon interactions.  The few- and many-body treatments are
independently formulated although connected in the resulting coupled
equations.  Then we continue the derivation in Sec.~\ref{sec:TheoSky}
with specific choices of few-body and many-body formalisms and
subsequently corresponding selection of the Skyrme nucleon-nucleon
interaction.

The few-body formalism is chosen to be the hyperspherical adiabatic
expansion of the Faddeev equations in coordinate space.  This choice has
several advantages of particular importance for practical
implementation of the method.  All Jacobi coordinate systems are
treated equally, and the bound and continuum channels decouple
completely, which makes it much easier to account for the Pauli
exclusion principle between few- and many-body treated nucleons.
Finally, this part of the program is then as computationally efficient as
the same few-body problem.

The many-body formalism is here chosen to be the mean-field
Skyrme-Hartree-Fock description.  The corresponding many-body
treatment is related to the bulk part (core) of the nucleus.  The
advantages is again simplicity and efficiency of the method which
in particular makes it possible to treat heavy nuclear systems.  Our
present focus is on weakly bound systems, where most of the
characteristic behavior is dictated by the few-body treated (valence)
nucleons.  Efficiency is therefore a higher priority than
sophistication, as long as a decent core description is delivered.

It should be noted that despite the close connection between the core
and valence aspects of the system, it is still possible to account very
well for the Pauli principle, as described in Sec.~\ref{sec:pauli}.
To achieve this goal several procedures are available within the
framework of our hyperspherical adiabatic expansion method.  Phase
equivalent potentials can be constructed to exclude occupation of one
specific state, while still retaining precisely the same scattering
properties.  Due to the decoupling of many solutions in the spectrum
of hyperangular eigenvalues, direct removal of occupied states is
also possible. By employing a combination of these methods the Pauli
principle is very accurately accounted for.

At the heart of the method is how the core description provides a
potential to be used between the clusters in the three-body
calculations. Producing this potential self-consistently within the
framework of the method eliminates one of the main weaknesses of
traditional three-body approaches, as there is (almost) no freedom to
use for phenomenological adjustments.  The choice of a nucleon-nucleon
interaction containing derivatives leads to effective mass terms in
the three-body as well as in the core calculations. Such terms are
unusual in traditional three-body calculations, but in principle they
amount to additional couplings between the different subsystems, and
are as such not unfamiliar.

Another central aspect of the method is how the structure of the core
affects the behavior of the remaining valence nucleons through the
iterative procedure of the calculations. These iterations intimately
connect the short-distance bulk properties of the system with the
long-range behavior of the weakly bound valence nucleons. This
connection allows for more meaningful investigations of the spatial
evolution and reaction mechanisms of nuclear systems.

\subsection{Summarizing the applications}

Having presented all the details of the derivations along with a
discussion of the technical implementation, we first apply the
method on $^{26}\textnormal{O} \, (^{24}\textnormal{O}+n+n)$ as
described in Sec~\ref{sec:24O}.  This system is particularly
interesting as located at the edge of an abrupt discontinuity in the
neutron dripline following $Z=8$.  As such it has in recent years
received an increasing amount of attention, both theoretically and
experimentally.  We demonstrate the procedure with the effect of
the iterations and the speed of the convergence. Using three
different, unaltered Skyrme parameterizations, both the experimentally
known energy levels and the half-life is very accurately reproduced.
This is particularly striking considering the inaccuracy of the
predictions from traditional Skyrme-Hartree-Fock calculations. This
illustrates how well the method can work even with a very simple core
description.

To further demonstrate the flexibility and computational efficiency of
the method we examine the heavier nucleus, $^{72}\textnormal{Ca} \,
(^{70}\textnormal{Ca} + n + n)$.  Although this system is currently
outside experimental reach, it is still very interesting due to the
possibilities it presents. It is on the edge of the neutron dripline,
but more importantly one of the lowest allowed states for the valence
nucleons is an $s_{1/2}$ state. Having neither a Coulomb nor a
centrifugal barrier, very extended configurations are possible, which
relates the system to the field of halo physics, and it could
potentially also allow for the formation of Efimov states. All these
issues are discussed in Sec.~\ref{sec:70Ca}.

In the same section, Sec.~\ref{sec:70Ca}, we also discuss how the
method closely relates the short-distance, bulk properties with the
long-distance physical observables. We show how the fingerprints of
the bulk properties can be detected in physical observables such as
the energy distribution.  We further discuss when and under which
conditions halo formations are possible, and we demonstrate that the
mere availability of a spatially confined $d$-state is not enough to
suppress the formation of extended $s$-wave halo configurations.
Finally, we consider the possibility of detecting Efimov states in
this system, maybe as excited states.  Despite the fascinating avenues
of physics that would be available by such a possibility, we conclude
that a series of Efimov states is not possible in
$^{72}\textnormal{Ca}$ or in any other nuclear system.

We continue with an application on the proton dripline,
$^{70}\textnormal{Kr} \, (^{68}\textnormal{Se} + p +p)$, with
astrophysical consequences in mind.  The nucleus,
$^{68}\textnormal{Se}$, is considered to be one of the most important of the
heavier critical waiting points in the rp-process. The understanding
of the effective lifetime of this nucleus in a stellar environment is
central in determining the evolution of for instance close binary
systems containing neutron stars or white dwarfs. This effective
lifetime is affected significantly by the possibility of skipping the
mass gap in the proton dripline through a three-body reaction and
forming $^{70}\textnormal{Kr}$ by capturing two protons.

With that in mind the primary focus of the analysis is here on the
two-proton capture rate, and to that end the associated proton capture
cross section is an essential intermediate step. The analysis is
complicated by the fact that very little experimental information is
available in this region of the nuclear chart, and some uncertainty
remains as to the energy level of the three-body resonance. Despite
these uncertainties, it is still possible to provide rather narrow
limits for the reaction rate.  We find that the continuum (off
resonance) background contribution is significant, and
thereby establishing a scale for the reaction rate.

In addition to these practical predictions, the $^{70}\textnormal{Kr}$
nucleus is also used to examine the more subtle parts of the new
coupling terms, as well as fundamental properties of the reaction
mechanism. The behavior of the new coupling terms is understandable
when comparing to the already existing coupling terms. The capture reaction is
seen to be a very unambiguous sequential reaction, which can be
understood as proceeding through a resonance state and the related
specific geometric configurations of the system.

\subsection{Summary and perspective}

In evaluating these results a number of obvious strong points in the
method is seen. First and foremost is the ease with which weakly bound
and extended systems are treated. This is an area that is very
difficult to treat properly with existing methods, making this
application particularly valuable. In relation to this point is the
fact that the method intimately and self-consistently connect
long-distance observable structures and short-distance bulk
properties.  This makes it possible to extract reliable information
about the properties of the nuclear systems from the observable behavior.

A final noteworthy aspect of the method is its simplicity and
computational efficiency. While many of the more sophisticated,
existing many-body methods need supercomputer clusters and extremely
large memory arrays to complete a calculation within a realistic time
frame, this method can be implemented on a single, regular computer
core, with a runtime of several days. Due to the simplicity of the
mean-field method this is even true for heavy and complicated systems.

These obvious strengths of the method have to be further enhanced by
implementing two or three improvements on top of the present
formulation and applications.  The first lies in the interaction
between pairs of valence nucleons.  As discussed in
Secs.~\ref{sec:ValInt} and \ref{sec:V3b}, it is possible, in
principle, to use the same interaction between the valence nucleons as
between the core nucleons.  However, due to the zero-range nature of
the Skyrme force, this leads to spurious, unphysical solutions, which
strongly couple to all physical solutions.  This is not much of a
problem, as minor variations in the valence nucleon-nucleon
interaction only has an insignificant influence on the three-body wave
function, as long as this interaction is asymptotically correct.

The second improvement is related to the three-body potential in the
three-body equation.  The philosophy behind the Skyrme effective
interaction suggests parameterization in terms of a density dependent
two-body interaction.  This is in principle straightforward, but
complicated, using the nuclear core density folded with the
nucleon-nucleon interaction.  However, this is not meaningful at the
moment, since all present global many-body calculations are unable to
provide keV-level accuracy, as needed to calculate desired reliable
few-body properties.  This is a problematic issue for the simple
Skyrme force and for all other interactions as well. To be practical,
we use here a phenomenological three-body interaction between the core
and the two valence nucleons.  So far, we keep this potential for
fine-tuning the energy without affecting the three-body structure.
This is important in situations where the energy is decisive as for
tunneling or close to break-up thresholds.  This degree of flexibility
is unavoidable for precision calculations.

The third improvement is related to the average treatment of the
``external'' potential acting on the core nucleons and arising from
the valence particles. This treatment should be replaced by a detailed
hyper-radial dependence such that each average distance produces its
own potential.  The asymptotic limit of three free particles would
then be correctly reproduced.

One limitation discussed in Sec.~\ref{sec:expansion} is the structure
of the system, currently assumed to be a spherical, even-even core
surrounded by two identical valence point-like nucleons. This could be
improved in several directions. The valence nucleons could be replaced
by more complicated clusters with intrinsic structure, where the
simplest probably is $\alpha$-particles but larger compact nuclei
could also be considered. The simplest choice is obviously to add
structure on one valence particle at a time, and in particular start
with only two clusters simulating the asymmetric fission process.  The
core could also be allowed to deviate from spherical symmetry, be
deformed and in addition possibly containing odd numbers of
nucleons. This would open up for many new possibilities.  Beside these
mean-field generalizations, other improvements of the many-body
treatment are tempting.  In general this could be to replace the
mean-field core description with a more sophisticated
\textit{ab-initio} method, but this would sacrifice much of the
efficiency, while probably gaining very little accuracy.

In conclusion, we have presented a conceptually simple, but
technically advanced new method, which has unique application
possibilities within the areas of weakly bound and extended three-body
systems, while at the same time apparently exhibiting rather

impressive computational efficiency. These initial applications
demonstrate the accuracy and viability of the method. But a number of
future improvements and applications, also beyond nuclear physics, are
both tempting and possible.

\ack

This work was funded by the Danish Council for Independent Research DFF Natural Science and the DFF Sapere Aude program. This work has been partially supported by the Spanish Ministerio de Econom\'{\i}a y Competitividad under Project FIS2014-51971-P.

\appendix

\section{Skyrme interaction variation\label{appen:vari}}

This appendix contains details of the variations expressed in Eq.~(\ref{eq:vari}). The variation of Eq.~(\ref{eq:HamCore}) with respect to $\psi_i^*$ is a completely standard Skyrme mean-field variation only with a slightly different $t_3$ term. The focus is therefore on terms involving the valence nucleons, i.e. variations of $V_{iv_1}$ and $V_{iv_2}$ in Eq.~(\ref{eq:Ham3b}). This will be done explicitly for the variation with respect $\psi_{3b}^*$, while the variation with respect to $\psi_i^*$ follows by symmetry.  Initially, the derivation is divided into contributions from the various terms in the Skyrme force ($t_0$, $t_1$, $t_2$, $t_3$, and $W_0$), which are then combined.

The results are expressed in terms of the densities known from regular mean-field calculations \cite{vau72}, along with analog densities for the three-body part. Specifically, the interactions are expressed in terms of
\begin{eqnarray}
n_q(\bm{r}) &= \sum_{i \sigma} | \psi_{iq\sigma}(\bm{r}) |^{2}
= n_{q\downarrow}(\bm{r}) + n_{q\uparrow}(\bm{r}),
\label{eq:def_nc}
\\
n_3(\bm{r}) &= \int |\psi_{3b}(\bm{r}_{cv1},\bm{r})|^2 d \bm{r}_{cv_1} + \int |\psi_{3b}(\bm{r},\bm{r}_{cv2})|^2 d \bm{r}_{cv_2}, 
\label{eq:def_n3}
\\
\tau_q(\bm{r}) &= \sum_{i\sigma} | \bm{\nabla} \psi_{i q \sigma}(\bm{r}) |^{2},
\label{eq:def_tauc}
\\
\tau_3(\bm{r}) &= \int |\bm{\nabla}_r \psi_{3b}(\bm{r}_{cv1},\bm{r})|^{2} d \bm{r}_{cv_1} + \int | \bm{\nabla}_r \psi_{3b}(\bm{r},\bm{r}_{cv2}) |^2 d \bm{r}_{cv_2},
\label{eq:def_tau3}
\\
\bm{J}_q(\bm{r}) &= -i \sum_{i \sigma \sigma^{\prime}} \psi_{i q \sigma}^{\ast}(\bm{r}) \left( \bm{\nabla} \psi_{i q \sigma^{\prime}} (\bm{r}) \times \Braket{\sigma | \bm{\sigma} | \sigma^{\prime} } \right),
\label{eq:def_Jc}
\\
\bm{J}_3(\bm{r}) &= -i \sum_{\sigma \sigma^{\prime}} 
\left( 
\int \psi_{3b, \sigma}^{\ast}(\bm{r}_{cv_1}, \bm{r})  \bm{\nabla}_r \psi_{3b,\sigma^{\prime}}(\bm{r}_{cv_1}, \bm{r})  d \bm{r}_{cv_1}
\right.
\nonumber
\\
&+
\left.
\int \psi_{3b, \sigma}^{\ast}(\bm{r}, \bm{r}_{cv_2})  \bm{\nabla}_r \psi_{3b,\sigma^{\prime}}(\bm{r}, \bm{r}_{cv_2})  d \bm{r}_{cv_2} \right) \times \braket{\sigma | \bm{\sigma} | \sigma^{\prime} },
\label{eq:def_J3}
\end{eqnarray}
where the sum over $i$ is a sum over single particle states, and the sum over $\sigma$ is over spin components (also indicated by $\uparrow, \downarrow$), while $\bm{\sigma}$ are the Pauli matrices, and $q$ indicates the associated nucleon type. Here $q$ can be either core neutrons $(n)$, core protons $(p)$, core neutrons and protons combined $(c)$, core nucleons of the same type as the valence nucleons $(s)$, or core nucleons of opposite type to the valence nucleons $(d)$.

To further simplify the expressions the following shorthand notation is also introduced at the end
\begin{eqnarray}
t_{i1} = t_i \left( 1 + \frac{1}{2}x_i \right), 
&&
t_{i2} = t_i \left(\frac{1}{2} + x_i \right), 
\end{eqnarray}
where $t_i$ and $x_i$ are the parameters from the Skyrme force, with $i = 0,1,2$, and $3$.

\subsection{Assumptions and equalities\label{app:assump}}

Because of the spin-operators in the Skyrme force from Eq.~(\ref{eq:skyrme}), and to treat the exchange term explicitly, the wave functions are decomposed into their spin-components. It will be used, that
\begin{eqnarray}
\ket{S=0} = \frac{1}{\sqrt{2}} \left( \ket{\uparrow}_{1} \ket{\downarrow}_{2} - \ket{\downarrow}_{1} \ket{\uparrow}_{2} \right),
\\
\ket{S=1,s_z=0} = \frac{1}{\sqrt{2}} \left( \ket{\uparrow}_{1} \ket{\downarrow}_{2} + \ket{\downarrow}_{1} \ket{\uparrow}_{2} \right),
\\
\ket{S=1,s_z= 1} = \ket{\uparrow}_{1} \ket{\uparrow}_{2},
\\
\ket{S=1,s_z=-1} = \ket{\downarrow}_{1} \ket{\downarrow}_{2},
\end{eqnarray}
where $\ket{\uparrow}_i$ and $\ket{\downarrow}_i$ indicates the spin-up and down components of the $i$'th particle. It is also used that
\begin{eqnarray}
P_{\sigma} \ket{S=0} &= - \ket{S=0} \mbox{ and }
P_{\sigma} \ket{S=1} &= \ket{S=1}. 
\end{eqnarray}

The product of $\psi_i$ and $\psi_{3b}$ is written out as
\begin{eqnarray}
\psi_i(\bm{r}) \psi_{3b}(\bm{r}_{cv1},\bm{r}_{cv2})
&= \left( \psi_{i\uparrow}(\bm{r}) \ket{\uparrow}_{r} + \psi_{i\downarrow}(\bm{r}) \ket{\downarrow}_{r} \right)
\nonumber
\\
&\times
\left( \psi_{3b,\uparrow}(\bm{r}_{cv1},\bm{r}_{cv2}) \ket{\uparrow}_{r_1} + \psi_{3b,\downarrow}(\bm{r}_{cv1},\bm{r}_{cv2}) \ket{\downarrow}_{r_1} \right)
\nonumber
\\
&= 
\frac{\ket{S=0}}{\sqrt{2}}  
\left(\psi_{i\uparrow}(\bm{r}) \psi_{3b,\downarrow}(\bm{r}_{cv1},\bm{r}_{cv2}) - \psi_{i\downarrow}(\bm{r}) \psi_{3b,\uparrow}(\bm{r}_{cv1},\bm{r}_{cv2}) \right)
\nonumber \\
&+ \frac{\ket{S=1, s_z=0}}{\sqrt{2}} 
\nonumber
\\
&\times
\left(\psi_{i\uparrow}(\bm{r}) \psi_{3b,\downarrow}(\bm{r}_{cv1},\bm{r}_{cv2}) + \psi_{i\downarrow}(\bm{r}) \psi_{3b,\uparrow}(\bm{r}_{cv1},\bm{r}_{cv2}) \right)
\nonumber
\\
&+ \ket{S=1,s_z=-1} \psi_{i\downarrow}(\bm{r}) \psi_{3b,\downarrow}(\bm{r}_{cv1},\bm{r}_{cv2})
\nonumber
\\
&+\ket{S=1,s_z=1} \psi_{i\uparrow}(\bm{r}) \psi_{3b,\uparrow}(\bm{r}_{cv1},\bm{r}_{cv2})
\end{eqnarray}

As we are focusing on cores with an even-even spherical structure, time-reversal symmetry is implied. This allows us to use many of the traditional equalities known from spherical mean-field Skyrme calculations, seen for instance in Ref.~\cite{vau72}. We therefore have the following equalities for the core wave function
\begin{eqnarray}
\frac{1}{2} n_n
&= n_{n\uparrow} = n_{n\downarrow}, \label{eq:HFee}
\\
0 &=
\sum_{i \in N} \psi_{i\downarrow}^* \psi_{i\uparrow} + \psi_{i\uparrow}^* \psi_{i\downarrow},
\label{eq:dens_ud}
\\
\frac{1}{4} \bm{\nabla} n_n
&= \sum_{i \in N} \psi_{i\downarrow}^* \psi_{i\downarrow}' 
= \sum_{i \in N} \psi_{i\uparrow}^* \psi_{i\uparrow}' 
= \sum_{i \in N} \psi_{i\downarrow}'^* \psi_{i\downarrow} 
= \sum_{i \in N} \psi_{i\uparrow}'^* \psi_{i\uparrow},
\label{eq:grad_ip}
\\
\bm{\nabla}^2 n_n 
&= 2 \tau_n + 2 \sum_{i \in N} \psi_i^* \bm{\nabla}^2 \psi_i,
\label{eq:twoGrad_ip}
\end{eqnarray}
where a prime indicates a gradient. The same equalities hold true for core proton densities. For the three-body wave function it is only needed that
\begin{eqnarray}
\frac{1}{2} \bm{\nabla} |\psi_{3b}|^2
&= \psi_{3b,\uparrow}^* \psi_{3b,\uparrow}' + \psi_{3b,\downarrow}'^* \psi_{3b,\downarrow} 
= \psi_{3b,\uparrow}'^* \psi_{3b,\uparrow} + \psi_{3b,\uparrow}^* \psi_{3b,\uparrow}'
\label{eq:grad_phi3}
\\
\bm{\nabla}^2 |\psi_{3b}|^2
&= 2 |\bm{\nabla} \psi_{3b}|^2 + 2 \psi_{3b}^* \bm{\nabla}^2 \psi_{3b}
\label{eq:twoGrad_phi3}
\end{eqnarray}

Finally, by writing out the definitions from Eqs.~(\ref{eq:def_Jc}) and (\ref{eq:def_J3}) it is seen that
\begin{eqnarray}
\bm{J}_n(\bm{r}) \cdot \bm{J}_3(\bm{r})
=
&-2 \sum_{i \in N} 
\int d \bm{r}_{cv_1}
\left(
\psi_{3b,\uparrow}^*(\bm{r}_{cv_1},\bm{r}) \bm{\nabla} \psi_{3b,\downarrow}(\bm{r}_{cv_1},\bm{r}) \cdot \psi_{i\uparrow}^*(\bm{r}) \bm{\nabla} \psi_{i\downarrow}(\bm{r})
\right.
\nonumber
\\
&+ 
\left.
\psi_{3b,\downarrow}^*(\bm{r}_{cv_1},\bm{r}) \bm{\nabla} \psi_{3b,\uparrow}(\bm{r}_{cv_1},\bm{r}) \cdot \psi_{i\downarrow}^*(\bm{r}) \bm{\nabla} \psi_{i\uparrow}(\bm{r}) 
\right)
\nonumber
\\
&-2 \sum_{i \in N} 
\int d \bm{r}_{cv_2}
\left(
\psi_{3b,\uparrow}^*(\bm{r},\bm{r}_{cv_2}) \bm{\nabla} \psi_{3b,\downarrow}(\bm{r},\bm{r}_{cv_2}) \cdot \psi_{i\uparrow}^*(\bm{r}) \bm{\nabla} \psi_{i\downarrow}(\bm{r}) 
\right.
\nonumber
\\
&+ 
\left.
\psi_{3b,\downarrow}^*(\bm{r},\bm{r}_{cv_2}) \bm{\nabla} \psi_{3b,\uparrow}(\bm{r},\bm{r}_{cv_2}) \cdot \psi_{i\downarrow}^*(\bm{r}) \bm{\nabla} \psi_{i\uparrow}(\bm{r}) 
\right)
\label{eq:Jdots}
\end{eqnarray}

\subsection{$t_0$ and $t_3$ energy contribution\label{app:Et0}} 

Here the contribution to the energy from the $t_0$ part of the core-valence neutron interaction for variation with respect to $\psi_{3b}^*$ is derived. Only the contribution from the interaction between the core and one valence nucleon is calculated, i.e. the contribution from $V_{i v_1}$. The contribution from the other valence nucleon is the same with a change of coordinates.

The interaction is
\begin{eqnarray}
V_{i v_1}^{(t_0)} 
=
t_0 \left( 1 + x_0 P_{\sigma} \right) \delta(\bm{r} - \bm{r}_{cv_1}).
\end{eqnarray}
The energy is then calculated by multiplying by $\Psi$ from the left and the right, and integrating over all coordinates
\begin{eqnarray}
E_{t_0} 
&= t_0 \left( \sum_{i\in N} + \sum_{i \in Z} \right) 
\int d \bm{r} d \bm{r}_{cv_1} d \bm{r}_{cv_2}
\Big\{
\nonumber
\\
&
\psi_{i}^*(\bm{r}) \psi_{3b}^*(\bm{r}_{cv_1},\bm{r}_{cv_2}) 
\delta(\bm{r}-\bm{r}_{cv_1}) \left( 1 + x_0 P_{\sigma}(i,v_1) \right)
\nonumber 
\\
& \times \left[ 
\psi_{i}(\bm{r}) \psi_{3b}(\bm{r}_{cv1},\bm{r}_{cv2}) - \psi_{i}(\bm{r}_{cv_1}) \psi_{3b}(\bm{r},\bm{r}_{cv2}) \delta_{q_i,q_{v_1}}
\right]
\Big\},
\end{eqnarray}
where the last $\delta$ function is to ensure exchange is only included when core and valence nucleon is of the same type. The calculation is also split into the sum over neutrons and protons. Here it is assumed the valence nucleons are neutrons, but the derivation for protons is identical. The situation is simplified greatly by the $\delta$ function in the Skyrme force, as only $\ket{S=0}$ contributes when core and valence nucleon are the same type. 

So for neutrons, the contribution is
\begin{eqnarray}
E_{t_0}^{(N)} \left( \ket{S=0} \right)
&= t_0 \sum_{i \in N} \int d \bm{r}_{cv_1} d \bm{r}_{cv_2}
\left\{ \frac{(1-x_0)}{2} 
\right.
\nonumber
\\
&
\times \left[
\psi_{i\uparrow}^*(\bm{r}_{cv_1}) \psi_{3b,\downarrow}^*(\bm{r}_{cv_1},\bm{r}_{cv_2}) - \psi_{i\downarrow}^*(\bm{r}_{cv_1}) \psi_{3b,\uparrow}^*(\bm{r}_{cv_1},\bm{r}_{cv_2})
\right]
\nonumber \\
& \times 
\left. \vphantom{\frac{1}{1}}
2 \left[
\psi_{i\uparrow}(\bm{r}_{cv_1}) \psi_{3b,\downarrow}(\bm{r}_{cv1},\bm{r}_{cv2}) - \psi_{i\downarrow}(\bm{r}_{cv_1}) \psi_{3b,\uparrow}(\bm{r}_{cv1},\bm{r}_{cv2})
\right]
\right\}
\nonumber \\
&= t_0 \int d \bm{r}_{cv_1} d \bm{r}_{cv_2} \Big\{
(1 - x_0) 
\\
& \times
\left[
n_{n\uparrow}(\bm{r}_{cv_1}) |\psi_{3b,\downarrow}(\bm{r}_{cv1},\bm{r}_{cv2})|^2 + n_{n\downarrow}(\bm{r}_{cv_1}) |\psi_{3b,\uparrow}(\bm{r}_{cv1},\bm{r}_{cv2})|^2
\right]
\Big\}
\nonumber
\end{eqnarray}
With a neutron as valence nucleon, there is no exchange when interacting with a core proton. Therefore both $\ket{S=0}$ and $\ket{S=1}$ contributes. For $\ket{S=0}$ the contribution is
\begin{eqnarray}
E_{t_0}^{(Z)} \left( \ket{S=0} \right)
&= 
t_0 \int d \bm{r}_{cv_1} d \bm{r}_{cv_2} \frac{1 - x_0}{2} 
\left[ n_{p\uparrow}(\bm{r}_{cv_1}) |\psi_{3b,\downarrow}(\bm{r}_{cv1},\bm{r}_{cv2})|^2
\right.
\nonumber
\\
& +
\left. n_{p\downarrow}(\bm{r}_{cv_1}) |\psi_{3b,\uparrow}(\bm{r}_{cv1},\bm{r}_{cv2})|^2 \right],
\end{eqnarray}
while for $\ket{S=1}$ the contribution consists of three components
\begin{eqnarray}
E_{t_0}^{(Z)} \left( \ket{S=1,S_z=0} \right)
&= 
t_0 \int d \bm{r}_{cv_1} d \bm{r}_{cv_2} \frac{1 + x_0}{2} 
\left[ n_{p\uparrow}(\bm{r}_{cv_1}) |\psi_{3b,\downarrow}(\bm{r}_{cv1},\bm{r}_{cv2})|^2 \right.
\nonumber
\\
&+
\left. n_{p\downarrow}(\bm{r}_{cv_1}) |\psi_{3b,\uparrow}(\bm{r}_{cv1},\bm{r}_{cv2})|^2 \right],
\\
E_{t_0}^{(Z)} \left( \ket{S=1, S_z=1} \right)
&= 
t_0 \int d \bm{r}_{cv_1} d \bm{r}_{cv_2} (1 + x_0)
n_{p\uparrow}(\bm{r}_{cv_1}) |\psi_{3b,\uparrow}(\bm{r}_{cv1},\bm{r}_{cv2})|^2,
\\
E_{t_0}^{(Z)} \left( \ket{S=1, S_z=-1} \right)
&= 
t_0 \int d \bm{r}_{cv_1} d \bm{r}_{cv_2} (1 + x_0)
n_{p\downarrow}(\bm{r}_{cv_1}) |\psi_{3b,\downarrow}(\bm{r}_{cv1},\bm{r}_{cv2})|^2
\end{eqnarray}

Combining both neutron and proton contributions and using Eq.~(\ref{eq:HFee}) the contribution to the energy is
\begin{eqnarray}
E_{t_0}
&=
t_0 \int d \bm{r}_{cv_1} d \bm{r}_{cv_2} 
\Big\{
\nonumber
\\
&
|\psi_{3b,\uparrow}(\bm{r}_{cv1},\bm{r}_{cv2})|^2 
\left[ 
(1-x_0) n_{n\downarrow}(\bm{r}_{cv_1}) + n_{p\downarrow}(\bm{r}_{cv_1}) + (1+x_0) n_{p\uparrow}(\bm{r}_{cv_1})
\right]
\nonumber 
\\
&+
|\psi_{3b,\downarrow}(\bm{r},\bm{r}_{cv_2})|^2 
\left[
(1-x_0) n_{n\uparrow}(\bm{r}_{cv_1}) + n_{p\uparrow}(\bm{r}_{cv_1}) + (1+x_0) n_{p\downarrow}(\bm{r}_{cv_1})
\right]
\Big\}
\nonumber 
\\
&= t_0 \int d \bm{r}_{cv_1} d \bm{r}_{cv_2}
\vphantom{\frac{1}{1}}
\nonumber
\\
& 
|\psi_{3b}(\bm{r}_{cv1},\bm{r}_{cv2})|^2 \left[
\left(1+\frac{1}{2} x_0 \right) n_c(\bm{r}_{cv_1}) - \left( \frac{1}{2} + x_0 \right) n_n(\bm{r}_{cv_1})
\right]
\nonumber \\
&= \int d \bm{r}_{cv_1} d \bm{r}_{cv_2} 
|\psi_{3b}(\bm{r}_{cv1},\bm{r}_{cv2})|^2 \left[ t_{01} n_c(\bm{r}_{cv_1}) - t_{02} n_n(\bm{r}_{cv_1}) \right].
\label{eq:Et0}
\end{eqnarray}

An identical $t_0$ contribution would arise from the interaction between the second valence nucleon and the core, only with a change of coordinates. If the valence nucleon is a proton the only difference would be a change of subscript, $n_n \leftrightarrow n_p$.

The calculation of the $t_3$ contribution to the energy is identical only including a factor of $(n_c + n_3)^\alpha/6$ from the interaction. The result is
\begin{eqnarray}
E_{t_3} 
&=
\frac{1}{6} \int d \bm{r}_{cv_1} d \bm{r}_{cv_2}  
\label{eq:Et3}
\\
& |\psi_{3b}(\bm{r}_{cv1},\bm{r}_{cv2})|^2 \left(n_c(\bm{r}_{cv_1}) + n_3(\bm{r}_{cv_1}) \right)^{\alpha}
\left[ t_{31} n_c(\bm{r}_{cv_1}) - t_{32} n_n(\bm{r}_{cv_1}) \right].
\nonumber
\end{eqnarray}

It should be noted that the $t_3$ term in $V_{ij}$ from Eq.~(\ref{eq:HamCore}) also depends on $n_3$. However, this is just the regular $t_3$ contribution \cite{vau72} with $n_c^{\alpha} \rightarrow \left( n_c + n_3\right)^{\alpha}$. This contribution is then
\begin{eqnarray}
E_{t_3}^{(c)}
&=
\frac{1}{12} \int d\bm{r} \left( n_c(\bm{r}) + n_3(\bm{r})\right)^{\alpha} 
\left[
t_{31} n_c(\bm{r})^2 - t_{32} \left( n_n(\bm{r})^2 + n_p(\bm{r})^2 \right)
\right]\!.
\label{eq:Et3C}
\end{eqnarray}

\subsection{\texorpdfstring{$t_1$}{t1} energy contribution\label{app:Et1}}

The contribution to the energy from the $t_1$ term is slightly more complicated because of the derivatives, but the procedure is the same. The $t_1$ interaction is 
\begin{eqnarray}
V_{i v_1}^{(t_1)} 
&= \frac{1}{2} t_1 \left( 1 + x_1 P_{\sigma} \right) 
\left( \bm{k}'^{2} \delta(\bm{r}_{cv_1} - \bm{r}) + \delta(\bm{r}_{cv_1} - \bm{r}) \bm{k}^{2} \right) 
\label{eq:Vt1}
\\
&= \frac{-t_1}{8} \left( 1 + x_1 P_{\sigma} \right) 
\left( \left( \bm{\nabla}'_{v_1} - \bm{\nabla}' \right)^2 \delta(\bm{r}_{cv_1} - \bm{r}) 
+ \delta(\bm{r}_{cv_1} - \bm{r}) \left( \bm{\nabla}_{v_1} - \bm{\nabla} \right)^2 \right),
\nonumber
\end{eqnarray}
and the contribution to the energy is then
\begin{eqnarray}
E_{t_1}
&= - \frac{1}{8} t_1 \left( \sum_{i \in N} + \sum_{i \in Z} \right) \int d \bm{r} d \bm{r}_{cv_1} d \bm{r}_{cv_2}
\psi_{i}^*(\bm{r}) \psi_{3b}^*(\bm{r}_{cv_1},\bm{r}_{cv_2}) 
\left( 1 + x_1 P_{\sigma}(i,v_1) \right) 
\nonumber \\
&\times
\left( \left( \bm{\nabla}'_{v_1} - \bm{\nabla}' \right)^2 \delta(\bm{r}_{cv_1} - \bm{r}) 
+ \delta(\bm{r}_{cv_1} - \bm{r}) \left( \bm{\nabla}_{v_1} - \bm{\nabla} \right)^2 \right)
\nonumber
\\
&\times
\left[ \psi_{i}(\bm{r}) \psi_{3b}(\bm{r}_{cv1},\bm{r}_{cv2}) - \psi_{3b}(\bm{r},\bm{r}_{cv_2}) \psi_{i}(\bm{r}_{cv_1}) \delta_{q_i,q_{v_1}} \right].
\end{eqnarray}

The general procedure is the same as for $t_0$. That is; divide the derivation into the sum over neutrons and the sum over protons, separate into $S=1$ and $S=0$ components, calculate all the derivatives, restructure using partial integration, apply the assumptions from  \ref{app:assump}, regroup to comply with definitions from the beginning of \ref{appen:vari}, and finally simplify the expressions. 

Using Eqs.~(\ref{eq:HFee}) to (\ref{eq:twoGrad_phi3})  the neutron contribution to the energy from the second term in Eq.~(\ref{eq:Vt1}) is then

\begin{eqnarray}
E^{(N)}_{t_1}
&= - \frac{1-x_1}{8} t_1 \sum_{i \in N} \int d \bm{r}_{cv_1} d \bm{r}_{cv_2}
\Big\{
\left[ \psi_{i\uparrow}^* \psi_{3b,\downarrow}^* - \psi_{i\downarrow}^* \psi_{3b,\uparrow}^*  \right]
\nonumber
\\
&\times
\left[
\psi_{i\uparrow} \psi_{3b,\downarrow}'' - \psi_{i\downarrow} \psi_{3b,\uparrow}'' + \psi_{i\uparrow}'' \psi_{3b,\downarrow} - \psi_{i\downarrow}'' \psi_{3b,\uparrow} + 2 \left( \psi_{i\downarrow}' \psi_{3b,\uparrow}' - \psi_{i\uparrow}' \psi_{3b,\downarrow}' \right)
\right]
\Big\}
\nonumber
\\
&= - \frac{1-x_1}{8} t_1 \int d \bm{r}_{cv_1} d \bm{r}_{cv_2}
\left\{
\frac{3}{4} \bm{\nabla}^2 n_n |\psi_{3b}|^2 
- \frac{1}{2} n_n |\bm{\nabla} \psi_{3b}|^2
- \frac{1}{2} \tau_n |\psi_{3b}|^2
\vphantom{\sum_{\sigma^{\prime}}}
\right.
\nonumber 
\\
& \left.
+i \bm{J}_n \cdot  \sum_{\sigma \sigma^{\prime}} \psi_{3b, \sigma}^{\ast} \bm{\nabla} \psi_{3b,\sigma^{\prime}} \times \braket{\sigma | \bm{\sigma} | \sigma^{\prime} }
\right\}.
\label{eq:t1_n}
\end{eqnarray}

The same is done for the proton contribution. This is again divided into $\ket{S=0}$ and $\ket{S=1}$. The $\ket{S=0}$ contribution is half what it was for the neutron contribution because of the missing exchange term, and the $\ket{S=1}$ contribution is calculated similarly. The total contribution from the proton is then
\begin{eqnarray}
E_{t_1}^{(Z)}
&= - \frac{1 + x_1}{8} t_1 \int d \bm{r}_{cv_1} d \bm{r}_{cv_2} \left\{ 
\frac{9}{8} \bm{\nabla}^2 n_p |\psi_{3b}|^2 - \frac{3}{4} \tau_p |\psi_{3b}|^2 
\vphantom{\sum_{\sigma^{\prime}}}
\right.
\nonumber
\\
&-
\left.
\frac{3}{4} n_p |\bm{\nabla} \psi_{3b}|^2 
-i \frac{1}{2} \bm{J}_p \cdot  \sum_{\sigma \sigma^{\prime}} \psi_{3b, \sigma}^{\ast} \bm{\nabla} \psi_{3b,\sigma^{\prime}} \times \braket{\sigma | \bm{\sigma} | \sigma^{\prime} }
\right\}.
\label{eq:t1_p_s1}
\end{eqnarray}

Combining Eqs.~(\ref{eq:t1_n}) and (\ref{eq:t1_p_s1}) gives the energy contribution from the second term in Eq.~(\ref{eq:Vt1}). The first term in Eq.~(\ref{eq:Vt1}) results in the same contribution because of symmetry. The total energy contribution from the $t_1$ term for the interaction between the core and one of the valence neutrons is then
\begin{eqnarray}
E_{t_1} 
&= 
\int d \bm{r}_{cv_1} d \bm{r}_{cv_2} \left\{
i \frac{t_1}{8}
\right.
\nonumber
\\
&
\left( x_1 \bm{J}_c({\bm{r}_{cv_1}}) - \bm{J}_n({\bm{r}_{cv_1}}) \right) 
\cdot  \sum_{\sigma \sigma^{\prime}} \psi_{3b, \sigma}^{\ast}(\bm{r}_{cv_1},\bm{r}_{cv_2}) \bm{\nabla}_{v_1} \psi_{3b,\sigma^{\prime}}(\bm{r}_{cv_1},\bm{r}_{cv_2}) \times \braket{\sigma | \bm{\sigma} | \sigma^{\prime} } 
\nonumber
\\
&+
\frac{t_{11}}{4} \left( 
-\psi_{3b}^*(\bm{r}_{cv_1},\bm{r}_{cv_2}) \bm{\nabla}_{v_1} \left(  n_c(\bm{r}_{cv_1}) \bm{\nabla}_{v_1} \psi_{3b}(\bm{r}_{cv1},\bm{r}_{cv2}) \right) 
\vphantom{\frac{1}{1}}
\right.
\nonumber
\\
&+
\left. \tau_c(\bm{r}_{cv_1}) |\psi_{3b}(\bm{r}_{cv1},\bm{r}_{cv2})|^2 - \frac{3}{2} \bm{\nabla}_{v_1}^2 n_c(\bm{r}_{cv_1}) |\psi_{3b}(\bm{r}_{cv1},\bm{r}_{cv2})|^2 \right)
\nonumber
\\
&- 
\frac{t_{12}}{4} 
\left( -\psi_{3b}^*(\bm{r}_{cv_1},\bm{r}_{cv_2}) \bm{\nabla}_{v_1} \left( n_n(\bm{r}_{cv_1}) \bm{\nabla}_{v_1} \psi_{3b}(\bm{r}_{cv1},\bm{r}_{cv2}) \right) 
\vphantom{\frac{1}{1}}
\right.
\nonumber
\\
&+
\left.
\left. \tau_n(\bm{r}_{cv_1}) |\psi_{3b}(\bm{r}_{cv1},\bm{r}_{cv2})|^2 - \frac{3}{2} \bm{\nabla}_{v_1}^2 n_n(\bm{r}_{cv_1}) |\psi_{3b}(\bm{r}_{cv1},\bm{r}_{cv2})|^2 \right) \right\}.
\label{eq:Et1}
\end{eqnarray}

There is an identical contribution from the interaction with the other valence neutron, only where the coordinate dependence is $\bm{r}_{cv_2}$ instead of $\bm{r}_{cv_1}$ in 
$n_i$ and $\tau_i$.

\subsection{$t_2$ energy contribution\label{app:Et2}}

The $t_2$ interaction is
\begin{eqnarray}
V_{i v_1}^{(t_2)}
&= 
 t_2 \left(1 + x_2 P_{\sigma} \right) \bm{k}' \delta(\bm{r}_{cv_1} - \bm{r} ) \bm{k}
\nonumber \\
&= \frac{1}{4} t_2 \left( 1 + x_2 P_{\sigma} \right) \left( \bm{\nabla}'_{v_1} - \bm{\nabla}' \right) \delta(\bm{r}_{cv_1} - \bm{r}) \left( \bm{\nabla}_{v_1} - \bm{\nabla} \right).
\end{eqnarray}

The contribution to the energy is therefore
\begin{eqnarray}
E_{t_2} 
&= \frac{1}{4} t_2 \left( \sum_{i \in N} + \sum_{i \in Z} \right) \int d \bm{r} d \bm{r}_{cv_1} d \bm{r}_{cv_2}
\Big\{
\nonumber
\\
&
\psi_{i}^*(\bm{r}) \psi_{3b}^*(\bm{r}_{cv_1},\bm{r}_{cv_2}) 
\left( \bm{\nabla}'_{v_1} - \bm{\nabla}' \right) 
\left[1 + x_2 P_{\sigma}(i,v_1) \right] 
\delta(\bm{r}_{cv_1} - \bm{r}) 
\nonumber \\
& \times \left( \bm{\nabla}_{v_1} - \bm{\nabla} \right) 
\left[ \psi_{i}(\bm{r}) \psi_{3b}(\bm{r}_{cv1},\bm{r}_{cv2}) - \psi_{3b}(\bm{r},\bm{r}_{cv_2}) \psi_{i}(\bm{r}_{cv_1}) \delta_{q_i,q_{v_1}} \right]
\Big\}.
\end{eqnarray}

The derivation of the $t_2$ contribution to the energy proceeds exactly as in the previous sections, and the final result is 
\begin{eqnarray}
E_{t_2} 
&= 
\int d \bm{r}_{cv_1} d \bm{r}_{cv_2} \left\{
 i \frac{t_2}{8} \right.
 \nonumber
 \\
 &\left( \bm{J}_n({\bm{r}_{cv_1}}) + x_2 \bm{J}_c({\bm{r}_{cv_1}}) \right)
\cdot  \sum_{\sigma \sigma^{\prime}} \psi_{3b, \sigma}^{\ast}(\bm{r}_{cv_1},\bm{r}_{cv_2}) \bm{\nabla}_{v_1} \psi_{3b,\sigma^{\prime}}(\bm{r}_{cv_1},\bm{r}_{cv_2}) \times \braket{\sigma | \bm{\sigma} | \sigma^{\prime} } 
\nonumber
\\
&+
\frac{t_{21}}{8}
\Big( 
-2 \psi_{3b}^*(\bm{r}_{cv_1},\bm{r}_{cv_2}) \bm{\nabla}_{v_1} \left( n_c(\bm{r}_{cv_1}) \bm{\nabla}_{v_1} \psi_{3b}(\bm{r}_{cv1},\bm{r}_{cv2}) \right) 
\nonumber
\\
&+ 
2 |\psi_{3b}(\bm{r}_{cv1},\bm{r}_{cv2})|^2 \tau_c(\bm{r}_{cv_1}) 
+ |\psi_{3b}(\bm{r}_{cv1},\bm{r}_{cv2})|^2 \bm{\nabla}_{v_1}^2 n_c(\bm{r}_{cv_1}) 
\Big) 
\nonumber
\\
&+ \frac{t_{22}}{8}
\Big(
-2  \psi_{3b}^*(\bm{r}_{cv_1},\bm{r}_{cv_2}) \bm{\nabla}_{v_1} \left( n_n(\bm{r}_{cv_1}) \bm{\nabla}_{v_1} \psi_{3b}(\bm{r}_{cv1},\bm{r}_{cv2}) \right)
\nonumber
\\
&\left.+ 2 |\psi_{3b}(\bm{r}_{cv1},\bm{r}_{cv2})|^2 \tau_n(\bm{r}_{cv_1}) 
+ |\psi_{3b}(\bm{r}_{cv1},\bm{r}_{cv2})|^2 \bm{\nabla}_{v_1}^2 n_n(\bm{r}_{cv_1}) 
\Big)\right\}.
\label{eq:Et2}
\end{eqnarray}

Again there is an identical contribution from the interaction with the other nucleon, only where $n_c, \, n_n$, $\bm{J}_c$, and $\bm{J}_n$ depends $\bm{r}_{cv_2}$ instead of $\bm{r}_{cv_1}$, and the gradients are also with respect to $\bm{r}_{cv_2}$.

\subsection{$W_0$ energy contribution\label{app:EW0}}

As in a regular Skyrme-Hartree-Fock calculation the spin-orbit part is much simpler to calculate. In this case, using a prime notation to indicate a complex conjugated operator acting to the right, makes the calculation slightly more intuitive. The interaction can be written out explicitly as
\begin{eqnarray}
V_{jv_1}^{(SO)}
&= \frac{i}{4} W_0
\left( \bm{\sigma}_{j} + \bm{\sigma}_{v_1} \right) \cdot \left[ \left( \bm{\nabla}'_{j} - \bm{\nabla}'_{v_1} \right) \times \delta(\bm{r}_j - \bm{r}_{cv_1} ) \left( \bm{\nabla}_j - \bm{\nabla}_{v_1} \right) \right]
\nonumber
\\
&=
\frac{i}{4} W_0 \left[
\bm{\sigma}_{j} \cdot \bm{\nabla}'_{j} \times \delta \bm{\nabla}_j 
- \bm{\sigma}_{j} \cdot \bm{\nabla}'_{v_1} \times \delta \bm{\nabla}_j 
- \bm{\sigma}_{j} \cdot \bm{\nabla}'_{j} \times \delta \bm{\nabla}_{v_1}
\right.
\nonumber
\\
&+
\bm{\sigma}_{j} \cdot \bm{\nabla}'_{v_1} \times \delta \bm{\nabla}_{v_1}
+ \bm{\sigma}_{v_1} \cdot \bm{\nabla}'_{j} \times \delta \bm{\nabla}_j 
- \bm{\sigma}_{v_1} \cdot \bm{\nabla}'_{v_1} \times \delta \bm{\nabla}_j 
\nonumber
\\
&- \left.
\bm{\sigma}_{v_1} \cdot \bm{\nabla}'_{j} \times \delta \bm{\nabla}_{v_1}
+ \bm{\sigma}_{v_1} \cdot \bm{\nabla}'_{v_1} \times \delta \bm{\nabla}_{v_1}
\right],
\label{eq:VSO}
\end{eqnarray}
To rewrite this expression a number of common vector identities will be used, along with the assumption of time-reversal symmetry, and the fact that (see Ref.~\cite{vau72})
\begin{eqnarray}
\sum_{i \sigma \sigma^{\prime}}
\psi_{i \sigma}^{\ast}(\bm{r}) 
\Braket{\sigma | \bm{\sigma} | \sigma^{\prime} }
\psi_{i \sigma^{\prime}} (\bm{r}) 
=
\sum_i \psi_{i}^*(\bm{r}) \bm{\sigma} \psi_{i}(\bm{r})
= 
0,
\label{eq:sumSpinor}
\end{eqnarray}
where the second equality is just a shorthand notation. The idea is then to eliminate any $\bm{\nabla}'$ using partial integration, and rewrite the terms into the form $\bm{\nabla}_A \cdot \bm{\nabla}_B \times \bm{\sigma}_{B}$.

The first term can be rewritten as
\begin{eqnarray}
\bm{\sigma}_{j} \cdot \bm{\nabla}'_{j} \times \delta \bm{\nabla}_j
&=
-\bm{\sigma}_{j} \cdot \bm{\nabla}'_{v_1} \times \delta \bm{\nabla}_j
-\bm{\sigma}_{j} \cdot \bm{\nabla}_j \times \delta \bm{\nabla}_j
-\bm{\sigma}_{j} \cdot \bm{\nabla}_{v_1} \times \delta \bm{\nabla}_j
\nonumber
\\
&=
-2 \delta \bm{\nabla}_{v_1} \cdot \bm{\nabla}_j \times \bm{\sigma}_{j},
\end{eqnarray}
as $ \bm{\nabla}_j \cdot \bm{\nabla}_j \times \bm{\sigma}_{j} = 0$, and $-\bm{\sigma}_{j} \cdot \bm{\nabla}'_{v_1} \times \delta \bm{\nabla}_j = -\bm{\sigma}_{j} \cdot \bm{\nabla}_{v_1} \times \delta \bm{\nabla}_j$ because of time-reversal symmetry. The second term in Eq.~(\ref{eq:VSO}) follows by the same time-reversal argument.

The third term is rewritten using Eq.~(\ref{eq:sumSpinor}) as
\begin{eqnarray}
- \bm{\sigma}_{j} \cdot \bm{\nabla}'_{j} \times \delta \bm{\nabla}_{v_1}
&=
\bm{\sigma}_{j} \cdot \bm{\nabla}'_{v_1} \times \delta \bm{\nabla}_{v_1}
+\bm{\sigma}_{j} \cdot \bm{\nabla}_j \times \delta \bm{\nabla}_{v_1}
+\bm{\sigma}_{j} \cdot \bm{\nabla}_{v_1} \times \delta \bm{\nabla}_{v_1}
\nonumber
\\
&= - \delta \bm{\nabla}_{v_1} \cdot \bm{\nabla}_j \times \bm{\sigma}_{j}.
\end{eqnarray}
Finally, the fourth term is just zero because of Eq.~(\ref{eq:sumSpinor}) . The next four terms follow by symmetry. 

After integration over $\bm{r}$, $\psi_{3b}$ depends on $(\bm{r}_{cv_1},\bm{r}_{cv_2})$ and $\psi_i$ depends on $(\bm{r}_{cv_1})$. Using the definitions from Eqs.~(\ref{eq:def_nc}), and (\ref{eq:def_Jc}), along with Eqs.~(\ref{eq:grad_ip}) and (\ref{eq:grad_phi3}) the contribution to the energy from the spin-orbit term becomes
\begin{eqnarray}
E_{SO} 
&= -i W_0 \sum_i \int d \bm{r}_{cv_1} d \bm{r}_{cv_2} \Big\{
\nonumber
\\
& \psi_i^* \psi_{3b}^* \left( \bm{\nabla}_{v_1} \cdot \bm{\nabla}_j \times \bm{\sigma}_{j} + \bm{\nabla}_j \cdot \bm{\nabla}_{v_1} \times \bm{\sigma}_{v_1} \right) \psi_i \psi_{3b} \left(1 + \delta_{q_i,n} \right)
\Big\}
\label{eq:EW0}
\\
&= - \frac{1}{2} W_0 \int d \bm{r}_{cv_1} d \bm{r}_{cv_2} \Big\{
|\psi_{3b}(\bm{r}_{cv1},\bm{r}_{cv2})|^2 \left[  \bm{\nabla}_{v_1} \cdot \bm{J}_c({\bm{r}_{cv_1}}) 
+  \bm{\nabla}_{v_1} \cdot \bm{J}_n({\bm{r}_{cv_1}}) \right]
\nonumber
\\
&
+i \left[ 
\bm{\nabla}_{v_1} n_c(\bm{r}_{cv_1}) +  \bm{\nabla}_{v_1} n_n(\bm{r}_{cv_1}) \right]
\cdot \psi_{3b}^*(\bm{r}_{cv_1},\bm{r}_{cv_2}) \bm{\nabla}_{v_1} \psi_{3b}(\bm{r}_{cv1},\bm{r}_{cv2}) \times \bm{\sigma}_{v_1}
\Big\},
\nonumber
\end{eqnarray}
with an identical contribution from the other valence neutron only where $n_c$, $n_n$, $\bm{J}_c$, and $\bm{J}_n$ depends on $\bm{r}_{cv_2}$ instead of $\bm{r}_{cv_1}$, and the gradients are also with respect to $\bm{r}_{cv_2}$.

\subsection{Three-body Schr{\"o}dinger equation}

Having derived the energy contribution from the various terms in the Skyrme interaction in the previous sections, and given that the energy must be stationary under individual variation of both the core and three-body wave function, the Schr{\"o}dinger equation for the three-body part follows from Eqs.~(\ref{eq:Et0},\ref{eq:Et3},\ref{eq:Et3C},\ref{eq:Et1},\ref{eq:Et2},\ref{eq:EW0}). Given that $\psi_{3b}^*$ intentionally has been isolated to the left in all the equations, the variation is trivial. 

The three-body Schr{\"o}dinger equation becomes
\begin{eqnarray}
E_3 \psi_{3b}(\bm{r}_{cv1},\bm{r}_{cv2})
&= \Big[
T_x + T_y
+ V_{cv}(\bm{r}_{cv_1}) + V_{cv}(\bm{r}_{cv_2})
+ V_{v_1 v_2}(\bm{r}_{cv_1} , \bm{r}_{cv_2})
\nonumber
\\
&+
V_{3}(\bm{r}_{cv_1} , \bm{r}_{cv_2})
\Big] \psi_{3b}(\bm{r}_{cv1},\bm{r}_{cv2}), \label{eq:sch3Appen}
\end{eqnarray}
where $T_x$ and $T_y$ are the kinetic energy operators related to the three-body equation, $V_{cv}$ are the core - valence neutron interaction, $V_{v_1v_2}$ is the interaction between the two valence neutrons, and $V_3$ is the three-body interaction. The core - valence neutron interaction contains ordinary central, $V_{cen}$ and spin-orbit, $V_{SO}$ terms as well as effective masses, $V_{grad}$ in the form of gradient terms
\begin{eqnarray}
V_{cv}(\bm{r}) &= V_{cen}(\bm{r}) + V_{SO}(\bm{r}) + V_{grad}(\bm{r})
\label{vcore}
\\
V_{cen}(\bm{r}) 
&= n_{eff}(\bm{r}) + n_{H_c}(\bm{r}) + V^C(\bm{r}),
\label{vcen}
\\
V_{SO}(\bm{r}) 
&= - i \, \bm{n}_J(\bm{r}) \cdot \bm{\nabla}_r \times \bm{\sigma} \; ,
\label{vso}
\\
V_{grad}(\bm{r}) 
&= \bm{\nabla}_r \cdot \left( n_a(\bm{r}) \bm{\nabla}_r \right),
\label{vgrad}
\end{eqnarray}
where $n_{H_c}$ is specifically from the variation of Eq.~(\ref{eq:Et3C}). The parameters entering 
in the three-body equation are
\begin{eqnarray}
n_{eff}(\bm{r}) 
&=
 t_{01} n_c(\bm{r}) 
+ \frac{t_{31}}{6}  n_c(\bm{r}) \left( (n_c(\bm{r}) + n_3(\bm{r}))^{\alpha} + \alpha n_3(\bm{r}) (n_c(\bm{r}) + n_3(\bm{r}) )^{\alpha-1} \right)
\nonumber 
\\
&- 
t_{02} n_n(\bm{r}) - \frac{t_{32}}{6}  n_n(\bm{r}) \left( (n_c(\bm{r}) + n_3(\bm{r}))^{\alpha} + \alpha n_3(\bm{r}) (n_c(\bm{r}) + n_3(\bm{r}) )^{\alpha-1} \right)
\nonumber 
\\
&+ 
\frac{1}{8} \bm{\nabla}^{2} n_c(\bm{r}) \left( -3 t_{11} + t_{21} \right)
+ 
\frac{1}{8} \bm{\nabla}^{2} n_n(\bm{r}) \left( 3 t_{12} + t_{22} \right)
\nonumber 
\\
&+
\frac{\tau_c(\bm{r})}{4}  \left( t_{11} + t_{21} \right)
+ 
\frac{\tau_n(\bm{r})}{4}  \left( t_{22} - t_{12} \right)
-\frac{W_0}{2}  \left[ \bm{\nabla} \cdot \bm{J}_c(\bm{r}) + \bm{\nabla} \cdot \bm{J}_n(\bm{r}) \right] ,
\nonumber 
\\
n_{H_c}(\bm{r}) 
&= \frac{\alpha}{12} (n_c(\bm{r}) + n_3(\bm{r}))^{\alpha-1} \left[ t_{31} n_c(\bm{r})^2 - t_{32} \left( n_p(\bm{r})^2 + n_n(\bm{r})^2 \right) \right],
\nonumber 
\\
n_a(\bm{r}) 
&= 
\frac{1}{4} n_c(\bm{r}) \left( -t_{11} - t_{21} \right) + \frac{1}{4}
n_n(\bm{r}) \left( t_{12} - t_{22} \right),
\label{densities} \\
\bm{n}_J(\bm{r}) 
&= 
\frac{1}{8} \left[  (t_1-t_2) \bm{J}_n(\bm{r}) - (t_1 x_1 + t_2 x_2) \bm{J}_c(\bm{r})
+4 W_0 \left( \bm{\nabla} n_c(\bm{r}) +  \bm{\nabla} n_n(\bm{r}) \right) \right].
\nonumber
\end{eqnarray}
In the case of valence protons the expressions are similar after exchanging neutrons by protons ($n \leftrightarrow p$).

\subsubsection{Core Schr{\"o}dinger equation.}

The previous sections focused on the variation with respect to $\psi_{3b}^*$. From Eq.~(\ref{eq:vari}) it is clear that another variation with respect to $\psi_i^*$ is needed. Fortunately, this is fairly straightforward given the previous derivations. The variation will contain two parts; one part from Eq.~(\ref{eq:HamCore}) and one part from Eq.~(\ref{eq:Ham3b}). 

The part from Eq.~(\ref{eq:HamCore}) is identical to regular Skyrme-Hartree-Fock \cite{vau72}, only $n_c^{\alpha} \rightarrow (n_c + n_3 )^{\alpha}$ in the $t_3$ in the Skyrme interaction. The part from Eq.~(\ref{eq:Ham3b}) could be calculated exactly as was done in \ref{app:Et0} to \ref{app:EW0}. However, the only difference would be that the surviving integral would be with respect to $\bm{r}$ instead of $\bm{r}_{cv_1}$ and $\bm{r}_{cv_2}$, which means that $|\psi_{3b}|^2 \rightarrow n_3$ and likewise for $\tau_3$ and $\bm{J}_3$. The core Schr{\"o}dinger equation then becomes 
\begin{eqnarray}
\epsilon_{iq} \psi_{iq}(\bm{r})
&= 
\left[- \bm{\nabla} \cdot \frac{\hbar^{2}}{2m^{\ast}_q(\bm{r})}  \bm{\nabla}
+ U_q(\bm{r}) 
- i \bm{W}_q(\bm{r}) \cdot ( \bm{\nabla} \times \bm{\sigma} )
\right.
\nonumber
\\
&- \left.
\bm{\nabla} \cdot \frac{1}{m_q^{\prime \ast}(\bm{r})}  \bm{\nabla}
+
U^{\prime}_q(\bm{r}) 
- i \bm{W}^{\prime}_q(\bm{r}) \cdot ( \bm{\nabla} \times \bm{\sigma} )
\right] \psi_{iq} (r), \label{eq:schCAppen}
\end{eqnarray}
where a prime indicates the interaction is due to the valence nucleons. The specific interactions entering in the core equation are
\begin{eqnarray}
\frac{\hbar^{2}}{2m_q^{\ast}(\bm{r}) }
&= \frac{\hbar^{2}}{2m_q} 
+ \frac{1}{4} 
\left[ 
\left( t_{11} + t_{21} \right) n_c(\bm{r}) 
+ \left( t_{22} - t_{12} \right) n_q(\bm{r}) 
\right],
\nonumber 
\\
U_q(\bm{r})  
&=
t_{01} n_c(\bm{r}) - t_{02} n_q(\bm{r})
+ \frac{1}{4} \tau_c(\bm{r}) \left[ t_{11} + t_{21} \right]
+ \frac{1}{4} \tau_q(\bm{r}) \left[ t_{22} - t_{12} \right]
\nonumber 
\\
&+ \frac{t_{31}}{12} 
\left( 
2 (n_c(\bm{r}) + n_3(\bm{r}))^{\alpha} n_c(\bm{r}) 
+ \alpha n_c(\bm{r})^2 (n_c(\bm{r}) + n_3(\bm{r}))^{\alpha-1} 
\right) 
\nonumber
\\
&- \frac{t_{32}}{6}(n_c(\bm{r}) + n_3(\bm{r}))^{\alpha} n_q(\bm{r}) 
\nonumber
\\
&- \frac{t_{32}}{12} \alpha (n_c(\bm{r}) + n_3(\bm{r}) )^{\alpha-1} 
\left( n_n(\bm{r})^{2} + n_p(\bm{r})^{2} \right) 
\nonumber 
\\
&+ 
\frac{\bm{\nabla}^{2} n_c(\bm{r})}{8}  
\left( t_{21} - 3 t_{11} \right)
+ \frac{\bm{\nabla}^{2} n_q(\bm{r})}{8}  
\left[ 3 t_{12} + t_{22} \right]
\nonumber
\\
&- 
\frac{W_0}{2} 
\left( 
\bm{\nabla} \cdot \bm{J}_c(\bm{r})
+ \bm{\nabla} \cdot \bm{J}_q(\bm{r})
\right) 
+ V_c^C
\label{eq cOrd par}
\\
\bm{W}_q(\bm{r}) 
&= \frac{1}{2} W_0 \left( \bm{\nabla} n_c(\bm{r}) + \bm{\nabla} n_q(\bm{r}) \right)
+ \frac{1}{8} \left( (t_1 - t_2) \bm{J}_q - (t_1 x_1 + t_2 x_2 ) \bm{J}_c \right), 
\nonumber
\end{eqnarray}
The core-valence neutron interaction basically amounts to adding $n_3$ to $n_n$ (only some care has to be taken with the $t_3$ term), if the valence nucleons are neutrons. So, for core nucleons of the same type as the valence neutrons ($s \in N$), the addition to the interaction becomes
\begin{eqnarray}
\frac{1}{m_s^{\prime \ast}(\bm{r})}
&= 
\frac{1}{4} n_3(\bm{r}) \left[ t_{11} - t_{12} + t_{21} + t_{22} \right],
\nonumber 
\\
U_s^{\prime}(\bm{r})
&=  
n_3(\bm{r}) 
\left( t_{01} - t_{02} \right)
+  \frac{1}{4} \tau_3(\bm{r}) \left[ t_{11} - t_{12} + t_{21} + t_{22} \right]
\nonumber
\\
&+ 
\frac{n_3(\bm{r}) }{6}  
\Big\{ 
\alpha ( n_c(\bm{r}) + n_3(\bm{r}) )^{\alpha-1} 
\left( n_c(\bm{r}) t_{31} - n_n(\bm{r}) t_{32} \right)
\nonumber
\\
&
+ ( n_c(\bm{r}) + n_3(\bm{r}) )^{\alpha} ( t_{31} - t_{32})
\Big\} 
\nonumber 
\\
&+ \frac{1}{8} \bm{\nabla}^{2} n_3(\bm{r}) 
\left[ t_{21} + t_{22} + 3 ( t_{12} - t_{11} ) \right]
- W_0  \bm{\nabla} \cdot \bm{J}_3(\bm{r}) ,
\nonumber
\\
\bm{W}_s^{\prime}(\bm{r})
&=  W_0 \bm{\nabla} n_3(\bm{r}) 
+\frac{1}{8} \bm{J}_3(\bm{r}) \left[ t_1 - t_2 - (t_1 x_1 + t_2 x_2) \right].
\label{eq nval para}
\end{eqnarray}
Likewise, for core nucleons of a different type than the valence neutrons ($d\in Z$) the interaction simplifies to 
\begin{eqnarray}
\frac{1}{m_d^{\prime \ast}(\bm{r})}
&= \frac{1}{4} n_3(\bm{r}) \left[ t_{11} + t_{21} \right],
\nonumber 
\\
U_d^{\prime}(\bm{r})
&= 
t_{01} n_3(\bm{r})
+ \frac{1}{4} \tau_3(\bm{r})
\left[ t_{11} + t_{21} \right]
+ \frac{1}{8} \bm{\nabla}^{2} n_3(\bm{r}) 
\left[ t_{21} - 3 t_{11} \right]
\nonumber
\\
&+ \frac{n_3(\bm{r})}{6}
\alpha ( n_c(\bm{r}) + n_3(\bm{r}) )^{\alpha-1} 
\left( t_{31} n_c(\bm{r})  - t_{32} n_n(\bm{r}) \right) 
\nonumber
\\
&+ \frac{n_3(\bm{r})}{6} t_{31} ( n_c(\bm{r}) + n_3(\bm{r}) )^{\alpha}
- \frac{1}{2} W_0 \bm{\nabla} \cdot \bm{J}_3(\bm{r}),
\nonumber 
\\
\bm{W}_d^{\prime}(\bm{r})
&= \frac{1}{2} W_0 \bm{\nabla} n_3(\bm{r})
- \frac{1}{8} \bm{J}_3(\bm{r}) \left[  t_1 x_1 + t_2 x_2 \right] . \label{eq cval para}
\end{eqnarray}

In the case of valence protons the expressions are similar after exchanging neutrons by protons ($n \leftrightarrow p$). In the same way $s\in Z$ and $d \in N$. Also, the Coulomb
term should be added in this case to $U'_s$. The Coulomb interaction within the Slater approximation for the exchange part takes the form \cite{cha98,erl11,gu13}
\begin{eqnarray}
V^{C}_{c}(\bm{r})
= \frac{e^2}{2} \int \frac{n_p(\bm{r}^{\prime})}{|\bm{r} - \bm{r}^{\prime}|} d \bm{r}^{\prime} - \frac{e^2}{2} \left( \frac{3}{\pi} \right)^{1/3} n_p(\bm{r})^{1/3},
\label{eq_app:coul}
\end{eqnarray}
and similarly for the Coulomb interaction of the valence protons.

\section{New three-body equations\label{appen:3bEq}}

Due to the almost-local, effective masses in Eq.~(\ref{eq:sch3}) the usual three-body 
equations \cite{nie01} change slightly. This is due to the appearance of the $V_{grad}(\bm{r})$
term (\ref{vgrad}) into the nucleon-core interaction $V_{cv}(\bm{r})$ in Eq.(\ref{vcore}).
Also, since the core is assumed to be spherical all the density functions in Eq.(\ref{densities}),
and therefore also the potential $V_{grad}$, do actually depend only on the distance $r$.

Taking this into account, the almost-local terms are expressed in terms of the Jacobi coordinates from Eq.~(\ref{eq:JacCoorx})
\begin{equation}
\bm{\nabla} \cdot (n_a(r) \bm{\nabla} )
= \frac{\mu_x}{m} \bm{\nabla}_x \cdot (n_a(r) \bm{\nabla}_x )
= \frac{\mu_x}{m} \left\{ \bm{\nabla}_x \right\},
\end{equation}
where $ \left\{ \bm{\nabla}_x \right\}$ can also be written as
\begin{eqnarray}
\left\{ \bm{\nabla}_x \right\}
&= \bm{\nabla}_x \cdot (n_a(r) \bm{\nabla}_x )
= n_a(r) \bm{\nabla}_x^2 + \left( \bm{\nabla}_x n_a(r) \right) \cdot \bm{\nabla}_x
\nonumber
\\
&=
\left\{ \bm{\nabla}_x \right\}_1 + \left\{ \bm{\nabla}_x \right\}_2.
\label{eq:EffMass}
\end{eqnarray}

Let us focus now on the first part in the equation above, $\left\{ \bm{\nabla}_x \right\}_1$, which
in spherical coordinates takes the form
\begin{eqnarray}
\left\{ \bm{\nabla}_x \right\}_1 
=
n_a(r)
\left(
\frac{1}{x^2} \frac{\partial}{\partial x} 
\left( x^2 \frac{\partial}{\partial x} \right) 
- \frac{\hat{\ell}_x^2 }{x^2} 
\right),
\label{eq:grx1}
\end{eqnarray}
where $\hat{\ell}_x$ is the usual angular momentum operator associated with the $\bm{x}$ Jacobi coordinate. Translation of this into hyperspherical coordinates leads to
\begin{eqnarray}
\left\{ \bm{\nabla}_x \right\}_1
&=
n_a(r)
\left( \vphantom{\frac{\hat{\ell}_x^2 }{\rho^2\sin^2\alpha} }
\sin^2\alpha \frac{\partial^2}{\partial \rho^2} + \frac{2+\cos^2\alpha}{\rho} \frac{\partial}{\partial \rho}
+\frac{\cos^2\alpha}{\rho^2} \frac{\partial^2}{\partial \alpha^2}
+\frac{2\cos^3\alpha}{\rho^2\sin\alpha} \frac{\partial}{\partial \alpha}
\right.
\nonumber
\\
&+
\left.
\frac{2\sin\alpha\cos\alpha}{\rho} \frac{\partial}{\partial \rho} \frac{\partial}{\partial \alpha}
-\frac{\hat{\ell}_x^2 }{\rho^2\sin^2\alpha} 
\right).
\label{eq:grx1Reduced}
\end{eqnarray}

The second term in Eq.~(\ref{eq:EffMass}), $\left\{ \bm{\nabla}_x \right\}_2$, is much simpler, as $n_a$ only depends on $r$, and it can be written as
\begin{eqnarray}
\left\{ \bm{\nabla}_x \right\}_2
= \left( \bm{\nabla}_x n_a(r) \right) \cdot \bm{\nabla}_x
= \frac{dn_a}{dx} \frac{\partial}{\partial x}
= \frac{dn_a}{dx} \left(
\sin\alpha \frac{\partial}{\partial \rho}
+\frac{\cos \alpha}{\rho} \frac{\partial}{\partial \alpha} \right).
\label{eq:grx2Reduced}
\end{eqnarray}

Combining Eqs.~(\ref{eq:grx1Reduced}) and (\ref{eq:grx2Reduced}) the full non-local part of the two-body potential given in Eq.~(\ref{eq:EffMass}) can be written as
\begin{eqnarray}
\bm{\nabla}_x \cdot (n_a(r) \bm{\nabla}_x ) 
&=
n_a \sin^2\alpha \frac{\partial^2}{\partial \rho^2}  
+ 
\left(
n_a \frac{2+\cos^2\alpha}{\rho} + \frac{dn_a}{dx} \sin\alpha 
\right)
\frac{\partial}{\partial \rho}
\nonumber
\\
&
+ n_a \frac{\sin2\alpha}{\rho} \frac{\partial}{\partial \rho} \frac{\partial}{\partial \alpha} 
+ n_a \frac{\cos^2\alpha}{\rho^2} \frac{\partial^2}{\partial \alpha^2}
\nonumber 
\\ 
&+ \left( 
n_a \frac{2\cos^3\alpha}{\rho^2\sin\alpha} + \frac{dn_a}{dx} \frac{\cos \alpha}{\rho}
\right) 
\frac{\partial}{\partial \alpha}
- n_a \frac{\hat{\ell}_x^2 }{\rho^2\sin^2\alpha}.
\label{eq:NonLocal}
\end{eqnarray}

The expression in Eq.~(\ref{eq:NonLocal}) can be divided in two parts, $\left\{ \bm{\nabla}_x \right\}_{pot}$ and $\left\{ \bm{\nabla}_x \right\}_{coup}$, given by
\begin{eqnarray}
\left\{ \bm{\nabla}_x \right\}_{pot}
&= n_a \frac{\cos^2\alpha}{\rho^2} \frac{\partial^2}{\partial \alpha^2}
+ \left( n_a \frac{2\cos^3\alpha}{\rho^2\sin\alpha} + \frac{dn_a}{dx} \frac{\cos \alpha}{\rho}\right) \frac{\partial}{\partial \alpha}
-n_a \frac{\hat{\ell}_x^2 }{\rho^2\sin^2\alpha},
\label{eq:grPot}
\\
\left\{ \bm{\nabla}_x \right\}_{coup}
&=
 n_a \sin^2\alpha \frac{\partial^2}{\partial \rho^2}  
+\left( n_a \frac{2+\cos^2\alpha}{\rho} + \frac{dn_a}{dx} \sin\alpha \right) \frac{\partial}{\partial \rho}
\nonumber
\\
&+
n_a \frac{\sin2\alpha}{\rho} \frac{\partial}{\partial \rho} \frac{\partial}{\partial \alpha}
\nonumber
\\
&= g_1(\rho,\alpha)\frac{\partial^2}{\partial \rho^2}
+ g_2(\rho,\alpha)\frac{\partial}{\partial \rho}
+ g_3(\rho,\alpha)\frac{\partial}{\partial \rho}\frac{\partial}{\partial \alpha},
\label{eq:grCoup}
\end{eqnarray}
where $g_1$, $g_2$, and $g_3$ have been introduced to simplify the notation.

The first term, $\left\{ \bm{\nabla}_x \right\}_{pot}$, does not depend on the derivatives of the hyperradius, and it can be fully included in the two-body potential and treated when solving the  angular part of the Faddeev equations. The second term, $\left\{ \bm{\nabla}_x \right\}_{coup}$, is more complicated due to the derivatives of $\rho$, as they give rise to new coupling terms between the radial equations, similar to what the usual $P$'s and $Q$'s do \cite{nie01}.

Inserting the expanded three-body wave function from Eq.~(\ref{eq:3bExpanded}) into Eq.~(\ref{eq:3Bsch}) and using Eqs.~(\ref{eq:grPot}) and (\ref{eq:grCoup}) the three-body Schr\"{o}dinger equation becomes
\begin{eqnarray}
0
&=
\sum_m \left[
-\frac{15}{4} \frac{1}{\rho^2} f_m \phi_m
+ \frac{\partial^2 f_m}{\partial \rho^2} \phi_m
+ 2 \frac{\partial f_m}{\partial \rho} \frac{\partial \phi_m}{\partial \rho}
+ f_m \frac{\partial^2 \phi_m}{\partial \rho^2} 
\right.
\nonumber
\\
&
-\frac{2\mu_{cv_1} \rho^{5/2}}{\hbar^2} 
\left\{\bm{\nabla}_{x_{cv_1}} \right\}_{coup} \frac{1}{\rho^{5/2}} f_m \phi_m
-\frac{2\mu_{cv_2}\rho^{5/2}}{\hbar^2}  
\left\{\bm{\nabla}_{x_{cv_2}}\right\}_{coup} \frac{1}{\rho^{5/2}} f_m \phi_m
\nonumber
\\
&
\left.
+ \frac{2m(E-V_3)}{\hbar^2} f_m \phi_m
- \frac{f_m}{\rho^2} 
\left(
\hat{\Lambda}^2 
+ \frac{2m\rho^2}{\hbar^2} (\tilde{V}_{12}+\tilde{V}_{13}+\tilde{V}_{23})
\right) \phi_m
\right],
\label{eq:3BFullSch}
\end{eqnarray}
where $\mu_{cv_i}$ is the reduced mass between the core and the $i$'th valence nucleon, $\left\{\bm{\nabla}_{x_{cv_1}} \right\}_{coup}$ is the related new coupling, and the two-body potentials $\tilde{V}_{ij}$ contain the usual central and spin orbit terms as well as the $\left\{ \bm{\nabla}_x \right\}_{pot}$ term discussed above.

It should be noted that the angular functions $\phi_n$ are defined to be the eigenfunctions of the angular part of the Faddeev equations
\begin{eqnarray}
\left(\hat{\Lambda}^2 
+ \frac{2m\rho^2}{\hbar^2}(\tilde{V}_{12} + \tilde{V}_{13} + \tilde{V}_{23})
\right) \phi_n(\rho,\Omega) 
=
\lambda_n(\rho) \phi_n(\rho,\Omega),
\label{b10}
\end{eqnarray}
in such a way that multiplying Eq.~(\ref{eq:3BFullSch}) from the left by $\phi_n^*(\rho,\Omega)$ and integrating over $\Omega$  Eq.~(\ref{eq:3BFullSch}) results in
\begin{eqnarray}
0
&=
\frac{\partial^2 f_n}{\partial \rho^2} -\frac{\lambda_n+\frac{15}{4}}{\rho^2} f_n + \frac{2m(E_3 - V_3)}{\hbar^2}f_n
+2\sum_m P_{nm} \frac{\partial f_m}{\partial \rho}
+\sum_m Q_{nm} f_m  
\nonumber
\\ 
&- \sum_m 
\Braket{ \Phi_n | 
\frac{2\mu_{12}\rho^{5/2}}{\hbar^2} \left\{\bm{\nabla}_{x_{12}}\right\}_{coup} 
+\frac{2\mu_{13}\rho^{5/2}}{\hbar^2} \left\{\bm{\nabla}_{x_{13}}\right\}_{coup}
| \frac{f_m \Phi_m}{\rho^{5/2}} }_\Omega,
\label{eq:schCoup}
\end{eqnarray}
which except for the last term is the usual coupled set of radial equations \cite{nie01}.

The last line in Eq.~(\ref{eq:schCoup}) gives rise to quite a few additional couplings. In the case of two identical valence particles they are the same for both valence nucleons, resulting in a factor of 2. For this particular case the radial equations are
\begin{eqnarray}
0 &= 
\frac{\partial^2 f_n}{\partial \rho^2} 
- \frac{\lambda_n+\frac{15}{4}}{\rho^2} f_n 
+ \frac{2m(E_3 - V_3)}{\hbar^2}f_n
+ 2 \sum_m P_{nm} \frac{\partial f_m}{\partial \rho}
+ \sum_m Q_{nm} f_m
\nonumber
\\
&
- \sum_m (2 C_{nm}^{(12)}+C_{nm}^{(21)}+C_{nm}^{(31)} - \frac{5}{\rho} C_{nm}) \frac{\partial f_m}{\partial \rho}
- \sum_m C_{nm} \frac{\partial^2 f_m}{\partial \rho^2}
\\ 
&-\sum_m (C_{nm}^{(13)}
+C_{nm}^{(22)}
+C_{nm}^{(32)}
+\frac{35}{4\rho^2} C_{nm}
-\frac{5}{\rho} C_{nm}^{(12)}
-\frac{5}{2\rho} C_{nm}^{(21)}
-\frac{5}{2\rho} C_{nm}^{(31)}) f_m,
\nonumber
\end{eqnarray}
where
\begin{eqnarray}
C_{nm}
&= 2\frac{2\mu_x}{\hbar^2} 
\Braket{\phi_n | 
g_1(\rho,\alpha) 
| \phi_m }_\Omega,
\nonumber
\\
C_{nm}^{(12)} 
&= 2\frac{2\mu_x}{\hbar^2} 
\Braket{ \phi_n| 
g_1(\rho,\alpha)  \frac{\partial}{\partial \rho} 
| \phi_m }_\Omega, 
\nonumber
\\
C_{nm}^{(13)}
&= 2\frac{2\mu_x}{\hbar^2} 
\Braket{ \phi_n | 
g_1(\rho,\alpha) \frac{\partial^2}{\partial \rho^2} 
| \phi_m }_\Omega,
\nonumber
\\
C_{nm}^{(21)}
&= 2 \frac{2\mu_x}{\hbar^2} 
\Braket{\phi_n | 
g_2(\rho,\alpha) 
| \phi_m }_\Omega, 
\nonumber
\\
C_{nm}^{(22)}
&= 2 \frac{2\mu_x}{\hbar^2} 
\Braket{\phi_n | 
g_2(\rho,\alpha) \frac{\partial}{\partial \rho}
| \phi_m }_\Omega,
\nonumber
\\
C_{nm}^{(31)} 
&= 2 \frac{2\mu_x}{\hbar^2} 
\Braket{ \phi_n | 
g_3(\rho,\alpha) 
| \frac{\partial\phi_m}{\partial\alpha} }_\Omega ,
\nonumber
\\
C_{nm}^{(32)} 
&= 2 \frac{2\mu_x}{\hbar^2} 
\Braket{ \phi_n | 
g_3(\rho,\alpha) \frac{\partial}{\partial \rho}
| \frac{\partial \phi_m}{\partial \alpha} }_\Omega. \label{eq:Cnm}
\end{eqnarray}

These new coupling terms can be easily computed. The only problem comes from the $C_{nm}$, which mixes the second derivatives of the radial functions. However, the non-diagonal $C_{nm}$ terms are very small, and the simplest is to neglect them, in such a way that the coupled set of radial equations to be solved are
\begin{eqnarray}
0 
&= (1-C_{nn}) \frac{\partial^2 f_n}{\partial \rho^2} 
- \frac{\lambda_n+\frac{15}{4}}{\rho^2} f_n + \frac{2m(E - V_3)}{\hbar^2}f_n
\nonumber
\\
&+ 2 \sum_m \left( P_{nm} + P_{nm}^{\prime} \right) \frac{\partial f_m}{\partial \rho}
+\sum_m \left( Q_{nm} + Q_{nm}^{\prime} \right) f_m,
\label{eq:NewRad}
\\
Q_{nm}^{\prime}
&= \frac{5}{\rho} C_{nm}^{(12)} + \frac{5}{2}\frac{1}{\rho} C_{nm}^{(21)} + \frac{5}{2}\frac{1}{\rho} C_{nm}^{(31)} - C_{nm}^{(13)} - C_{nm}^{(22)} - C_{nm}^{(32)} - \frac{35}{4}\frac{1}{\rho^2} C_{nm},
\nonumber
\\
P_{nm}^{\prime}
&=
\frac{5}{2} \frac{1}{\rho} C_{nm} - C_{nm}^{(12)} - \frac{1}{2} C_{nm}^{(21)} - \frac{1}{2} C_{nm}^{(31)}.
\label{eq:PpQp}
\end{eqnarray}

\section*{References}


\end{document}